\documentclass[11pt,a4paper]{article}
\pdfoutput=1
\usepackage{amsfonts,amssymb,amsmath,epsfig, graphicx,yfonts,jheppub}
\usepackage[utf8]{inputenc}
\usepackage{lineno}

\title{Melting holographic mesons by cooling a magnetized quark gluon plasma}

\author{Daniel \'Avila,}
\author{Leonardo Pati\~no,}
\affiliation{Departamento de F\'isica, Facultad de Ciencias, Universidad Nacional Aut\'onoma de M\'exico, \\  A.P. 70-542, M\'exico D.F. 04510, Mexico} 

\abstract{We extend our analysis of holographic meson dissociation in the presence of an intense magnetic field. In addition to the previously known critical temperature above which the mesons melt, we found that for certain magnetic field intensities there exists a second lower critical temperature, below which stable mesons cease to exist. While we showed before that there is a range of high temperatures for which mesons can be melted by changing the magnetic field intensity, here we show that, as a consequence of the second critical point, there is also a range of low temperatures for which this phenomenom, which we term Magnetic Meson Melting (MMM), can be triggered. Additionaly, we also show that the magnetic field decreases the mass gap of the meson spectrum along with their masses. We are able to observe this by constructing a configuration that makes it possible to apply gauge/gravity methods to study fundamental degrees of freedom in a quark-gluon plasma subject to a magnetic field as intense as that expected in high energy collisions. This is achieved by the confection of a ten-dimensional background that is dual to the magnetized plasma and nonetheless permits the embedding of D7-branes in it. The main difference with previous approaches, which in consequence gives the novel results, is that the magnetic field retroacts in the geometry itself, as opposed to be confined to the world volume of the probe D7-branes.}


\emailAdd{davhdz06@ciencias.unam.mx} 
\emailAdd{leopj@ciencias.unam.mx}

\begin{document} 

\maketitle
\setlength{\parskip}{3pt}

\section{Introduction}
It has become increasingly accepted that an intense magnetic field is produced in high energy collisions and that understanding its effects is relevant to properly analyze experimental observations 
\cite{Skokov:2009qp,Wilde:2012wc,Basar:2012bp,Andersen:2014xxa,Ayala:2018wux}. Two of the properties of QCD matter that are of interest are the quark condensate and meson masses, along with their behavior as functions of the magnetic field. 

The quark condensate is an order parameter for chiral symmetry breaking. It vanishes at high temperatures, where chiral symmetry is restored, but it is non-zero on the hadronic phase. In this context the term `magnetic catalysis' (MC) refers to an increase in the magnitude of the quark condensate as a response to the magnetic field \cite{Gusynin:1994re,Shovkovy:2012zn}. However, using lattice calculations \cite{Bali:2011qj,Bali:2012zg,Endrodi:2019zrl} and the Nambu-Jana-Lasinio model \cite{Farias:2016gmy}, it has been found that for light quarks and sufficiently high temperatures the opposite behavior is observed, and the magnitude of the quark condensate decreases as the magnetic field intensity increases. This phenomenom is now known as `inverse magnetic catalysis' (IMC). 

The dependence of the meson spectrum on the magnetic field has been studied in different frameworks. Chiral perturbation theory \cite{Andersen:2012dz} as well as linear sigma models \cite{Ayala:2018zat} show that charged and neutral pions respond oppositely to the magnetic field. While the former increase their masses in the presence of the magnetic field, the latter find their masses reduced. However, this doesn't means that all neutral mesons behave the same. Using lattice QCD it has been shown \cite{Bali:2017ian} that the neutral pion meson mass decreases monotonically as the magnetic field grows, while for the $\rho$-meson the oposite behavior is observed. Different results are also obtained for neutral heavy mesons using a potential model with constituent quarks \cite{Yoshida:2016xgm}.

The non-perturbative regime of the physics just described has been studied using the gauge/gravity correspondence \cite{Maldacena:1997re}. It should be noted however, that the theory that can be studied through holographic methods is not properly quantum chromodynamics (QCD), because the exact gravity dual to this theory has not been found yet. Instead, what has been done is to consider theories similar to QCD, such as finite temperature $\mathcal{N}=4$ super Yang-Mills (SYM) with gauge group $SU(N_{c})$, and modify it to bring it as close to QCD as possible. 

A relevant step in this direction was the inclusion of a small number of flavor degrees of freedom $N_{f}$, or in other words, fields in the fundamental representation of the gauge group $SU(N_{c})$. In an abuse of language, we are going to refer to these fields generically as `quarks' even if bosonic and fermionic degrees of freedom are included. Holographically, this is done by adding $N_{f}$ probe D-branes to the desired gravitational background \cite{Karch:2002sh}, keeping $N_{c}\gg N_{f}$ to avoid backreaction (see \cite{Erdmenger:2007cm} for a review). In particular, in order to add fundamental matter to SYM $\mathcal{N}=4$ it is necessary to study the dynamics of probe D7-branes on $AdS_{5}\times S^{5}$. The configuration is such that the D7-branes extend along the $AdS_{5}$ directions while wrapping a $S^{3}\subset S^{5}$. 
 
In \cite{Kruczenski:2003be,Babington:2003vm,Mateos:2006nu,Mateos:2007vn} it was found that if the background was deformed to allow a finite temperature, the system showed an interesting thermodynamic behavior. It was discovered that, for a critical value of the temperature $T_{fun}$, the D7-branes undergoes a first order phase transition, taking place between a phase where they lie completely outside the horizon, in what it is called a Minkowski embedding, and a phase in which they fall into it, the black hole embedding. From the dual gauge theory perspective, this corresponds to a transition between a discrete meson spectrum (quark-antiquark bound states) and a gapless distribution of excitations. It is important to remark that what we have just described is a `melting' or dissociation phase transition, not a confinement-deconfinement one. The holographic description of a confining phase involves a horizon-free geometry. At $T_{deconf}$ the gluons and adjoint matter become deconfined, at which point the dual geometry develops a black hole horizon. However, if the quark mass is large enough it is possible for the branes to lie outside the horizon and thus mesons are stable (to leading order within the approximations of large $N_{c}$ and strong coupling) for $T_{deconf}<T<T_{c}$, behavior consistent with lattice simulations \cite{Asakawa:2003re,Datta:2003ww,Hatsuda:2005nw}.

The incorporation of a magnetic field in this setup was done in \cite{Filev:2007gb,Albash:2007bk} and independently in \cite{Erdmenger:2007bn}. In these works the authors considered a pure gauge $B$-field along the gauge theory directions, which then enters as an excitation in the DBI action of the probe branes and thus it is equivalent to a magnetic field on the world-volume. The results obtained in \cite{Albash:2007bk,Erdmenger:2007bn} showed that the effect of the magnetic field was to increase the melting temperature of the mesons, to the point of not having dissociation at all for field intensities above a certain critical value. Not only that, but it was also found that the presence of the magnetic field induces chiral symmetry breaking, exhibited by a non-zero quark condesate even at zero bare quark mass. Regarding the meson spectrum, it was found that the magnetic field breaks the degeneracy of the levels given in \cite{Kruczenski:2003be} displaying a Zeeman like effect. Using a similar setup, in \cite{Herzog:2010uh} it was studied how the magnetic field affects the spectrum of heavy-light mesons, showing that a Zeeman like effect is also induced in this case. It is important to remark that the excited modes on the $S^{3}$ were not considered in \cite{Filev:2007gb,Albash:2007bk,Erdmenger:2007bn}. Given that the magnetic field lives in the probe branes world volume, it couples to all the other fields livig there. As a consequence, in the dual gauge theory the magnetic field under consideration couples to the baryon current.

Another holographic setup for the magnetized quark gluon plasma was given in \cite{DHoker:2009mmn}, where all the matter fields were in the adjoint representation. The five-dimensional theory presented there, that considers the full backreaction of the magnetic field, is dual to the desired gauge theory because it is a solution to a consistent truncation of supergravity (SUGRA) IIB \cite{Cvetic:1999xp}. The magnetic field is introduced by factorizing a $U(1)$ from the $SO(6)$ symmetry of the compact space and changing it to a gauge symmetry. Hence, from the dual gauge theory perspective, the magnetic field in this case couples to the conserved current associated with a $U(1)$ subgroup of the $SU(4)$ $R-$symmetry \cite{Arciniega:2013dqa}. While this five-dimensional truncation is enough in many scenarios, in order to add flavor degrees of freedom to the theory we require the ten-dimensional uplift. Embedding D7-branes into this background has proved difficult, because the compact part of the resulting geometry warps in a way that prevents an easy identification of the right 3-cycle that the D7-brane must wrap (see App. \ref{App10D}).

We followed a different approach in \cite{Avila:2018hsi}, where we implemented a new consistent five-dimensional truncation. The family of solutions that we found features the full backreaction of a constant magnetic field and a scalar field dual to an operator of scaling dimension 2. Given that this solutions are part of the same general truncation ansatz \cite{Cvetic:1999xp}, from the gauge theory perspective the magnetic field couples to the $R-$current. From the five-dimensional perspective, we found that the presence of this scalar field induces the existence of a critical intensity for the magnetic field $b_{c}$, above which the system becomes unstable. For intensities below $b_{c}$, two branches of solutions exists, with one of them being thermodynamically preferred over the other. From the ten-dimensional perspective, the inclusion of this scalar field allows the compact space of the uplifted geometry to factorize as a warped 3-sphere, a 1-cycle and an angular coordinate $\theta$ that distributes the volume between these two spaces. Given this factorization, the probe D7-brane naturally wraps the 3-sphere while its embedding is described by the radial dependence of $\theta$.

In \cite{Avila:2019pua} we presented our first results using the construction just described. We found that, in contrast to the results obtained with the setup from \cite{Albash:2007bk,Erdmenger:2007bn}, the effect of the magnetic field is to decrease the critical temperature at which the mesons melt, that is, we observed IMC for meson dissociation. We also showed that the melting can be triggered at fixed temperature by adjusting the magnetic field intensity, but only for a certain range of temperatures. We named this phenomenom \textit{magnetic meson melting} (MMM) in \cite{Avila:2019pua}. 

The main objective of this manuscript is to provide additional details and extend the analysis that we presented in \cite{Avila:2019pua}. Concretely, here we present a more in depth thermodynamic analysis and widen our study of the spectrum of mesons by considering excitations over the 3-sphere. We also give more details on the numerical construction of the D7-brane embeddings and the holographic renormalization of its action. Our main new result is that for a certain range of magnetic field intensities, in addition to the previously known critical temperature above which the mesons melt, there is a second critical temperature \textit{below} which stable mesons cease to exist. While for the first hot temperature the effect of the magnetic field was the one of IMC, we observe MC for the new cold critical temperature\footnote{We use cold and hot to differentiate both temperatures, even if both are high enough for the system to be in a plasma state.}. As we will see, the transition is of first order in both cases. We will also show that, in addition to decrease the mass of the mesons, the magnetic field impose restrictions over the quantum numbers of the mesons over the 3-sphere.

The manuscript is organized as follows. In Sec. \ref{Background} we review the gravitational background, both from the $5D$ and $10D$ perspectives, while in Sec. \ref{Brane_embeddings} we introduce the probe D7-branes. In Sec. \ref{Ph_D} we show the phase diagram of the flavor degrees of freedom, and in Sec. \ref{Thermo} we present the thermodynamic analysis. In Sec. \ref{spectrum} we compute the meson spectrum and we close in Sec. \ref{discussion} with a discussion of our results. Some of the more technical details of our computations are contained in a series of appendices.

\section{The gravitational background}
\label{Background}

The gravitational background is a solution to ten-dimensional SUGRA IIB that assymptotes $AdS_{5}\times S^{5}$ and admits a deformation that encodes the dual of a magnetic field in the gauge theory. Additionally, in order to be able to add flavor degrees of freedom it is crucial that the geometry its such that it permits to factorize the compact part of the space in the way described in the introduction. 

It turns out that the line element that achieves this is of the form
\begin{equation}
ds_{10}^{2}=\Delta^{\frac{1}{2}}ds_{5}^{2}+\frac{L^{2}}{\Delta^{\frac{1}{2}}}\left[X\Delta d\theta^{2}+X^{2}\sin^{2}\theta d\phi^{2}+X^{-1}\cos^{2}\theta d\Sigma_{3}^{2}(A)\right],
\label{metric_10}
\end{equation}
where $\Delta$ is a wrapping factor given by
\begin{equation}
\Delta=X^{-2}\sin^{2}\theta+X\cos^{2}\theta, \qquad X=e^{\frac{1}{\sqrt{6}}\varphi(r)},
\label{Delta}
\end{equation}
$ds_{5}^{2}$ is the line element of a non-compact five-dimensional space that can be written as
\begin{equation}
ds_{5}^{2}=\frac{L^{2}}{U(r)}dr^{2}+\frac{1}{L^{2}}(-U(r)dt^{2}+V(r)(dx^{2}+dy^{2})+W(r)dz^{2}),
\label{metric}
\end{equation}
while $d\Sigma_{3}^{2}(A)$ is the line element of a 3-cycle given by
\begin{equation}
d\Sigma_{3}^{2}(A)=d\psi^{2}+\sin^{2}\psi\left(d\vartheta_{1}+\sqrt{2}\frac{A}{L}\right)^{2} +\cos^{2}\psi\left(d\vartheta_{2}+\sqrt{2}\frac{A}{L}\right)^{2},
\label{3_esfera}
\end{equation}
which for vanishing $A$ reduces to the line element of a unit 3-sphere given in Hopf (toroidal) coordinates \cite{LachiezeRey:2005hs,Achour:2015zpa}. In what follows we will take, without loss of generality, $L=1$. This implies that t'Hooft coupling $\lambda$ is related to the string length $l_{s}$ by
\begin{equation}
\lambda=\frac{1}{2l_{s}^{4}}.
\end{equation}

The $r$ coordinate appearing in \eqref{metric} meassures the radial distance such that we expect that the geometry assymptotes $AdS_{5}\times S^{5}$ as $r\rightarrow\infty$. The 1-form $A$ parametrizes an infinitesimal rotation along a periodic direction of the compact manifold that, in turn, codifies the internal degrees of freedom of the dual gauge theory. By keeping $A$ and its exterior derivative $F=dA$ in the cotangent space to the directions dual to those of the gauge theory, it will represent a U(1) vector potential. In order to introduce the desired magnetic field in the dual gauge theory we will look for solutions that admit
\begin{equation}
F=b\,dx\wedge dy,
\end{equation} 
and thus $F$ will respresent a constant magnetic field in the $z$ direction with intensity $b$.

The two properties of \eqref{metric_10} that permit an easy embedding of a D7-brane are that the metric does not depend on the angular coordinate $\phi$ and that, according to \eqref{metric_10}, the direction that it represents remains orthogonal to the rest of the spacetime. Notice that the inclusion of $\varphi$ was crucial in order to achieve this orthogonality. It is bacause of this that the brane can be placed at a fixed location in $\phi$. Concerning the 3-cycle in \eqref{metric_10}, we notice that its volume depends on the position $\theta$ and, regardless of the intensity of the magnetic field $b$, this cycle becomes maximal at $\theta=0$, while for $b=0$ it reduces to $S^3$. For non-vanishing $b$, the 3-cycle gets tilted towards the five dimensional non-compact part of the spacetime in a manner that is volume preserving within the eight dimensions of this two spaces together.

In App. \ref{App10D} we show that the configuration just described is part of the general truncation anzats presented in \cite{Cvetic:1999xp} and thus it constitutes a solution to type IIB supergravity, as long as the self-dual five-form is given by \eqref{5form} while the five-dimensional metric and fields are a solution to five-dimensional gauged supergravity. In \cite{Avila:2018hsi} we found a family of such solutions to the effective five-dimensional theory governed by the action
\begin{equation}
S_{5}=\frac{1}{16\pi G_{5}}\int d^{5}x \sqrt{-g}\left[R-\frac{1}{2}(\partial\varphi)^{2}+4\left(X^{2}+2X^{-1}\right)-X^{-2}(F)^{2}\right],
\label{eff-action}
\end{equation}
where $g$ is the determinant of the five-dimensional metric and $R$ is the five-dimensional Ricci scalar. Notice that $G_{5}=\pi/2N_{c}^{2}$ because we set $L=1$. The equations of motion that can be derived from this effective action are
\begin{eqnarray}
&& R_{\mu\nu}-\frac{1}{2}\partial_{\mu}\varphi\partial_{\nu}\varphi-2X^{-2}F_{\mu\sigma}{F_{\nu}}^{\sigma}+g_{\mu\nu}\left[\frac{4}{3}\left(X^{2}+2X^{-1}\right)+\frac{1}{3}X^{-2}F_{\rho\sigma}F^{\rho\sigma}\right]=0,
\cr
&& \frac{1}{\sqrt{-g}}\partial_{\mu}(\sqrt{-g}g^{\mu\nu}\partial_{\nu}\varphi)+4\sqrt{\frac{2}{3}}(X^{2}-X^{-1})+\sqrt{\frac{2}{3}}X^{-2}F_{\mu\nu}F^{\mu\nu}=0,
\cr
&& \partial_{\mu}(\sqrt{-g}X^{-2}F^{\mu\nu})=0.
\label{EOM_fondo}
\end{eqnarray}
Additionally, in order for the solutions to be part of the truncation ansatz it is necessary to impose the constriction $F\wedge F=0$, which our family of solutions satisfies inmediately. It is also important to note that a non-vanishing magnetic field implies a non-constant scalar field $\varphi(r)$ in order for it to be a solution to \eqref{EOM_fondo}. This means that the gravitational model found in \cite{DHoker:2009mmn} cannot be recovered from ours for $b$ other than zero. 

Given that the equations of motion \eqref{EOM_fondo} are highly non-linear, we resorted to numerical methods to solve them for non-vanishing intensity of the magnetic field. The general procedure is described in detail in \cite{Avila:2018hsi} and here we will just review some steps that will be relevant to the following calculations. The near-boundary behavior of the solutions is presented in App. \ref{AppB}.

To solve \eqref{EOM_fondo} we perform a numerical integration from the location of the horizon $r_{h}$, at which the metric function $U(r)$ vanishes, up to the boundary at $r\rightarrow\infty$ where the geometry asymptotes $AdS_{5}$. The symmetries of the equations of motion \eqref{EOM_fondo} allow us to completely specify any member of the family of solutions by the position of the horizon $r_{h}$, the value that the scalar field takes at $r_{h}$, and a parameter $B$. This three quantities are parameters over which we have computational control. However, what we want is to specify any member of the family of solutions by the values that the different physical quantities take in the dual gauge theory.  

The parameter $B$ is related to the magnetic field intensity by
\begin{equation}
b=\frac{B}{V_{\infty}}, \qquad V_{\infty}=\lim_{r\rightarrow\infty}\frac{V(r)}{r^{2}},
\end{equation}
while $r_{h}$ is related to the temperature by 
\begin{equation}
T=\frac{3r_{h}}{2\pi}.
\label{Temperature}
\end{equation}
Note that althought $T$ only depends on $r_{h}$, the magnetic field $b$ is a function of the position of the horizon $r_{h}$ and the parameter $B$. Thus, in order to vary $T$ while keeping $b$ constant, both $r_{h}$ and $B$ need to be fine-tuned.

Regarding $\varphi$, the general behavior of the scalar field build by the procedure just described near the boundary is
\begin{equation}
\varphi\rightarrow\frac{1}{r^{2}}\left(\varphi_{0}+\psi_{0}\log{r}\right),\label{phiasy}
\end{equation}
where $\varphi_{0}$ and $\psi_{0}$ are coefficients that can be read from the asymptotics of a specific solution. This particular behavior implies that $\varphi$ is dual to an operator $\mathcal{O}_{\varphi}$ of dimension 2, and thus it saturates the BF bound \cite{Breitenlohner:1982jf,Bianchi:2001kw}. In consequence, $\psi_{0}$ is dual to the source of the operator and $\varphi_{0}$ to its vacuum expectation value $\langle \mathcal{O}_{\varphi}\rangle$ \cite{Bianchi:2001kw}. From the gauge theory perspective, it makes sense to specify the source of the operator and then compute the vacuum expectation value that it generates in response to said source. Thus, we use the freedom to choose $\varphi(r_{h})$ in order to fix $\psi_{0}$ to a given value. 

It was found in \cite{Avila:2018hsi} that for any given $\psi_{0}$ there exists a critical value for the intensity of the magnetic field $b_{c}/T^{2}$ above which a singularity develops outside the horizon, indicating that the dual state in the gauge theory is unstable. For a given magnetic field intensity $b/T^{2}$ below this critical value, two different solutions exist, each one corresponding to a different value for the dimensionless ratio $\langle \mathcal{O}_{\varphi}\rangle/T^{2}$. The one with the higher value for $\langle \mathcal{O}_{\varphi}\rangle/T^{2}$ corresponds to a state with negative specific heat, higher free energy and lower entropy than the other, showing that the solutions with smaller $\langle \mathcal{O}_{\varphi}\rangle/T^{2}$ are thermodynamically preferred. 

Finally, in \cite{Avila:2018hsi} it was shown that because of the presence of the magnetic field the dual theory features a conformal anomaly, given by the fact that the trace of the stress-energy tensor does not vanish for a non-zero magnetic field. This causes some dimensionless observables not to depend only on dimensionless quantities such as $b/T^{2}$ and introduces a scheme dependent energy scale. This will have consequences on some of our numerical results, as some of them will be scheme dependent. The physical conclusion however, will be true regarding of the chosen scheme.

\section{Flavor D7-brane embeddings}
\label{Brane_embeddings}

In order to add flavor degrees of freedom to the theory, it is necessary to add probe D7-branes to the ten-dimensional gravitational backgrounds just described. This system can be thought of as a D3/D7 system with $N_{c}$ D3-branes and $N_{f}$ flavor D7-branes, with $N_{c}\gg N_{f}$ to be able to describe the latter as probe objects moving in the fixed geometry generated by the former. These objects extend in the following directions:
\begin{eqnarray}
\begin{array}{l| cccc|c|ccc}
& t & x & y & z & r & \theta & \phi & \Sigma_{3} \\
\hline 
N_{c} ~~ \mbox{  D3 } & \times & \times & \times & \times & & & & \\
N_{f} ~~\mbox{ D7 } &  \times &  \times &  \times & \times & \times &  & & \times 
\end{array}\,.
\label{extend}
\end{eqnarray}

In general, the position of the D7-branes is described by $\phi$ and $\theta$ as functions of the eight world-volume coordinates. This functions are in turn determinated by the equations of motion derived from the DBI action\footnote{There is no Wess-Zumino term in the action because the only SUGRA IIB form that is turned on is the self-dual five-form, and it doesn't couple to the D7-branes.}
\begin{equation}
S_{DBI}=-T_{D7}N_{f}\int d^{8}x\sqrt{-\text{det}(g_{D7})},
\label{DBI}
\end{equation}
where $T_{D7}$ is the D7-brane tension given by
\begin{equation}
T_{D7}=\frac{1}{(2\pi l_{s})^{7}l_{s}g_{s}}=\frac{1}{16\pi^{6}}\lambda N_{c},
\end{equation}
$g_{D7}$ denotes the induced metric on the world-volume, and the integration is performed over the eight directions on which the D7-branes extends \eqref{extend}. However, it turns out that it is possible to set 
\begin{equation}
\phi=0, \qquad \chi(r)=\sin\theta(r).
\label{phi_constant}
\end{equation}
consistently with the equations of motion, meaning that this choice automatically satisfy the equation for $\phi$ and leaves a second order ordinary differential equation for the profile function $\chi$ which can be solved after suitable initial conditions are given. With this choice the induced metric over the D7-brane world-volume, which can be computed from \eqref{metric_10}, is given by
\begin{equation}
ds^{2}_{D7}=\Delta^{\frac{1}{2}}\left[-Udt^{2}+V(dx^{2}+dy^{2})+Wdz^{2}+\frac{1-\chi^{2}+UX\chi'^{2}}{U(1-\chi^{2})}dr^{2}\right]+\frac{1-\chi^{2}}{\Delta^{\frac{1}{2}}X}d\Sigma_{3}^{2}(A),
\label{metric_D7}
\end{equation}
where the prime denotes the derivative with respect of $r$ and the profile function $\chi$ also appears in the wrapping factor
\begin{equation}
\Delta=X+\chi^{2}(X^{-2}-X).
\end{equation}

The remaining equation of motion for $\chi$ can also be derived from the general DBI action \eqref{DBI} after substitution of \eqref{metric_D7} and integration along the compact directions
\begin{equation}
S_{DBI}=-2\pi^{2}T_{D7}N_{f}\int d^{5}x\mathcal{L},
\end{equation} 
where
\begin{equation}
\mathcal{L}=V(1-\chi^{2})\sqrt{W\Delta X^{-3}(1-\chi^{2}+XU\chi'^{2})},
\label{Lagrangian}
\end{equation}
is the Lagrangian density. The equation of motion for $\chi$ is given by
\begin{equation}
\begin{aligned}
0&=\partial_{r}\frac{\partial\mathcal{L}}{\partial\chi'}-\frac{\partial\mathcal{L}}{\partial\chi}\\&=\partial_{r}\left(P(r)\frac{\sqrt{\Delta}(1-\chi^{2})XU\chi'}{\sqrt{1-\chi^{2}+XU\chi'^{2}}}\right)+P(r)\sqrt{\Delta}\frac{3\chi(1-\chi^{2})+2XU\chi\chi'^{2}}{\sqrt{1-\chi^{2}+XU\chi'^{2}}}\\& -P(r)\frac{X^{-2}-X}{\sqrt{\Delta}}\chi(1-\chi^{2})\sqrt{1-\chi^{2}+XU\chi'^{2}},
\end{aligned}
\label{EOM_chi}
\end{equation}
where
\begin{equation}
P(r)=V\sqrt{W X^{-3}}.
\end{equation}
The near-boundary behavior of the embedding profile $\chi$ can be obtained by solving \eqref{EOM_chi} for large $r$, which can be acomplished by using the asymptotic expansions for the metric functions and the scalar field given in App. \ref{AppB}. The result is 
\begin{equation}
\chi=\frac{m}{r}-\frac{U_{1}m}{2r^{2}}+\frac{c}{r^{3}}+\mathcal{O}\left(\frac{1}{r^{4}}\right),
\label{chi_boundary}
\end{equation}
where $U_{1}$ is one of the coefficients appearing in the expansions \eqref{r_expansions} for the background metric, $m$ is related to the bare quark mass by\footnote{Given our choice of radial coordinate, the parameter $m$ used here differs from the $m$ used in \cite{Mateos:2007vn} by a factor of $\pi T/\sqrt{2}$.}
\begin{equation}
M_{q}=\frac{1}{\pi}\sqrt{\frac{\lambda}{2}}m,
\label{quark_mass}
\end{equation}
and as we will show below, $c$ is related to the quark condensate. In order to directly compare our results to the ones presented in \cite{Mateos:2007vn} we will report them in terms of the rescaled mass
\begin{equation}
\bar{M}=\frac{2M_{q}}{\sqrt{\lambda}}=\frac{\sqrt{2}}{\pi}m,
\label{barM}
\end{equation}
which is related to the mass gap of the meson spectrum \cite{Kruczenski:2003be,Arean:2006pk,Ramallo:2006et,Myers:2006qr}.

By solving \eqref{EOM_chi} across all the bulk, there are two different kinds of embeddings that can be obtained: one where the branes lie completely outside the horizon, called Minkowski embedding, and another one where they fall through it, called black hole embedding. The general solution to \eqref{EOM_chi} for different values for the magnetic field needs to be computed numerically, given that the background itself is not known analytically. We now procced to explain how to perform said numerical integration to compute both types of embeddings.

Given that in the case of black hole embeddings the D7-branes fall into the horizon, the numerical integration is performed from there to the boundary. In practice, we solved the equation of motion \eqref{EOM_chi} as a power series of $r$ around $r_{h}$
\begin{equation}
\chi(r)=\chi_{h}+\sum_{i=1}^{\infty}\chi_{i}(r-r_{h})^{i}.
\label{chi_horizon}
\end{equation} 
With this procedure it is possible to write all the undeterminated coefficients up to any desired order in terms of $\chi_{h}$ and the values that the metric functions and the scalar field take at the horizon. Even if the equation of motion is of second order, our choice of coordinates is such that the value of $\chi'$ at the horizon is fixed by requiring the solution to be regular at $r=r_{h}$. Then \eqref{chi_horizon} is used to provide initial data for the numerical integration, starting at $r=r_{h}+\epsilon$ with $\epsilon\ll r_{h}$ all the way to the boundary at $r\rightarrow\infty$.

For the case of Minkowski embeddings the D7-branes lie completely outside the horizon, localized at $\theta=\pi/2$ for a given $r_{i}>r_{h}$. We then solve \eqref{EOM_chi} as a power series of $r$ around $r_{i}$
\begin{equation}
\chi(r)=1+\sum_{j=1}^{\infty}\chi_{j}(r-r_{i})^{j},
\label{chi_eje}
\end{equation}
and write all the undeterminated coefficients in terms of the values that the metric functions take at $r_{i}$. Again, the value for $\chi'(r_{i})$ is fixed by requiring the solution to be regular at $r=r_{i}$. By using \eqref{chi_eje} as initial conditions it is possible to numerically integrate \eqref{EOM_chi} from $r=r_{i}+\epsilon$ to the boundary $r\rightarrow\infty$. 

Finally, given that we wish to study the effect of the magnetic field and the temperature on the quarks, we need to work at fixed $\bar{M}$. In practice, this is done by choosing the value of $\chi_{h}$ or $r_{i}$ that gives the desired value of $m$ (and in consequence of $\bar{M}$ by \eqref{barM}) for any given values of $T$ and $b$.

\section{Phase diagram}
\label{Ph_D}

Following the procedure described in the previous section, it is possible to compute the D7-brane embeddings for any $b/T^{2}$ below its critical value.  In the following we fix the source term for the scalar operator $\psi_{0}$ equal to zero, which implies that the maximum intensity for the magnetic field that the background can hold is $b_{c}/T^{2}\approx 11.24$.  

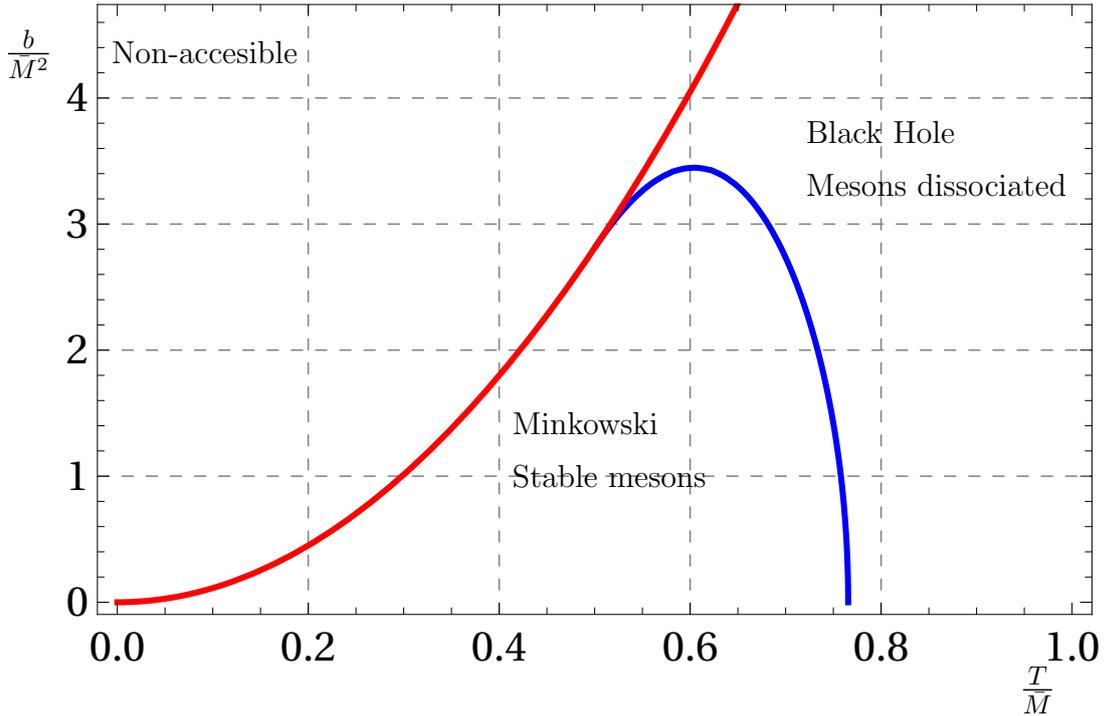
\begin{figure}[ht!]
 \centering
 \includegraphics[width=0.9\textwidth]{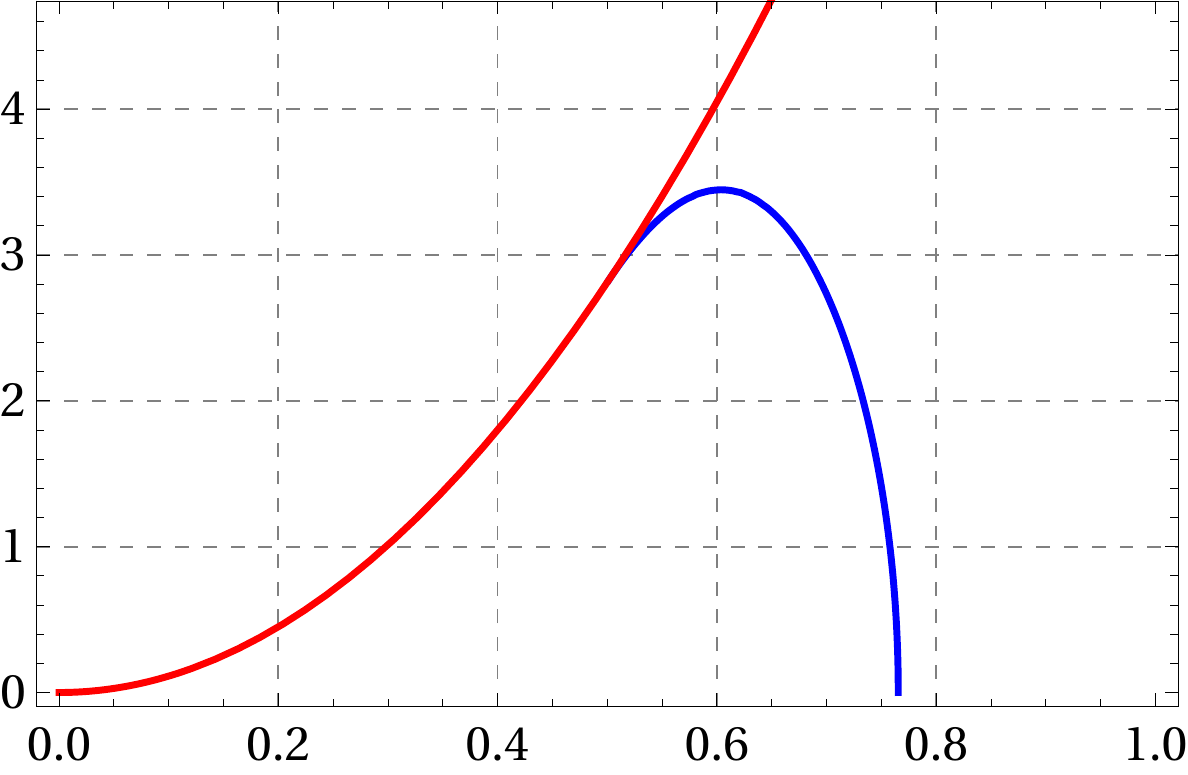}
 \put(-30,-10){\Large $\frac{T}{\bar{M}}$}
 \put(-410,230){\Large $\frac{b}{\bar{M}^{2}}$}
 \put(-110,200){\large Black Hole}
 \put(-110,180){\large Mesons dissociated}
 \put(-220,90){\large Minkowski}
 \put(-220,70){\large Stable mesons}
 \put(-370,230){\large Non-accesible}
\caption{\small Phase diagram of the flavor D7-branes. The red curve indicates the critical value $b/T^{2}=11.24$, thus the region to its left it's not accesible with our gravitational backgrounds. The blue
curve denotes the interphase between the two types of embeddings. The maximum of the blue curve is located at $b/\bar{M}^{2}=3.44$ and $T/\bar{M}=0.604$. The intersection of both curves is located at $b/\bar{M}^{2}=2.82$ and $T/\bar{M}=0.501$.}
\label{Phase_Diagram}
\end{figure}

We present the phase diagram of the system in Fig. (\ref{Phase_Diagram}), which visually summarizes many of our results. In practice we generated the phase diagram fixing $\bar{M}=1$, but we explicitly checked that it is independent of the chosen value for $\bar{M}$. The red curve indicates the critical value $b/T^{2}=11.24$, thus the region to its left it's not accesible with our current gravitational backgrounds. Note that we cannot explore the region given by $T/\bar{M}=0$ and $b/\bar{M}^{2}\neq 0$, which is exactly the physical scenario considered in \cite{Filev:2007gb}. Nonetheless, we can study the effect of any value of $b/\bar{M}^{2}$ and $T/\bar{M}$ to the right of the red curve, which is included in the region considered in \cite{Albash:2007bk,Erdmenger:2007bn}. The blue curve denotes the interphase between the two types of embeddings. 

Many interesting phenomena can be read from the phase diagram. The first can be seen by keeping $b/\bar{M}^{2}$ fixed. For $b/\bar{M}^{2}=0$ we recover the familiar results from \cite{Mateos:2007vn}, where from $T/\bar{M}=0$ up to the melting temperature $T_{hot}/\bar{M}=0.765$, the D7-branes are in the Minkowski phase, while for higher temperatures the D7-branes are in the black hole phase. As long as the intensity of the magnetic field is such that the value of $b/\bar{M}^2$ remains smaller than approximately $2.82$, the behavior does not change much with respect to the $b/\bar{M}^2=0$ case, in the sense that from the minimum $T/\bar{M}$ for that given $b/\bar{M}^{2}$ up to a melting temperature $T_{hot}/\bar{M}$ the branes are in the Minkowski phase, while for higher temperatures the branes are in the black hole phase. From the gauge theory perspective this means that, for $0<b/\bar{M}^{2}<2.82$, stable mesons states exist for low temperatures while for high temperatures the mesons melt.

However, for $2.82<b/\bar{M}^{2}<3.44$ something interesting happens: black hole embeddings exist for $T/\bar{M}$ near the minimum temperature. This new \textit{cold black hole phase}, which is a consequence of the presence of the background magnetic field, is dual to a low temperature phase in which stable mesons also cease to exist, and hence their melting can alternatively be achieved by lowering the temperature. Thus, for $2.82<b/\bar{M}^{2}<3.44$, in addition to the previous high melting temperature $T_{hot}/\bar{M}$, there exist now a low melting temperature $T_{cold}/\bar{M}$. For $b/\bar{M}^{2}=3.44$ both temperatures coincide, and for $b/\bar{M}^{2}>3.44$ no Minkowski embedding exist, meaning that stable meson states don't exist for high intensities of the magnetic field.

We would like to remark that our findings are very different, and even completely opposite when comparable, to the ones obtained using the setup from \cite{Albash:2007bk,Erdmenger:2007bn}. In particular, the cold phase transition is a novelty of our construction. While for vanishing magnetic field \cite{Albash:2007bk,Erdmenger:2007bn} also recovered the results from \cite{Mateos:2007vn}, they found that the melting temperature $T_{hot}/\bar{M}$ increases with the magnetic field. This effect is such that, above a certain magnetic field intensity, there was no meson melting at all and the D7-branes were in the Minkowski phase regardless of the value of $T/\bar{M}$. Thus while \cite{Albash:2007bk,Erdmenger:2007bn} found that the magnetic field prevents the melting of the mesons, here we found that it facilitates it.

The last affirmation is more evident if we return to the phase diagram in Fig. (\ref{Phase_Diagram}), but now exploring what happens if we keep the temperature fixed while changing the magnetic field. For temperatures below the intersection of the red and blue curves, that is $T/\bar{M}<0.501$, no black hole embeddings are permited and the branes are in the Minkowski phase for all intensities of the magnetic field up to its maximum value. The situation changes for $0.501<T/\bar{M}<0.765$, since for this temperatures the D7-branes transition from Minkowski to black hole embeddings at a melting magnetic field intensity, which we denote $b_{MMM}/\bar{M}^{2}$. We named this phenomenom \textit{magnetic meson melting} (MMM) in \cite{Avila:2019pua}, because what happens from the dual theory perspective is that the mesons are being melted at a fixed temperature by increasing the magnetic field only. For higher values of the temperature, namely $T/\bar{M}>0.765$, only black hole embeddings exist regardless of the magnetic field intensity. From the gauge theory perspective, what happens is that the mesons are already melted because of the high temperature.

\begin{figure}[ht!]
\begin{center}
\begin{tabular}{cc}
 \includegraphics[width=0.8\textwidth]{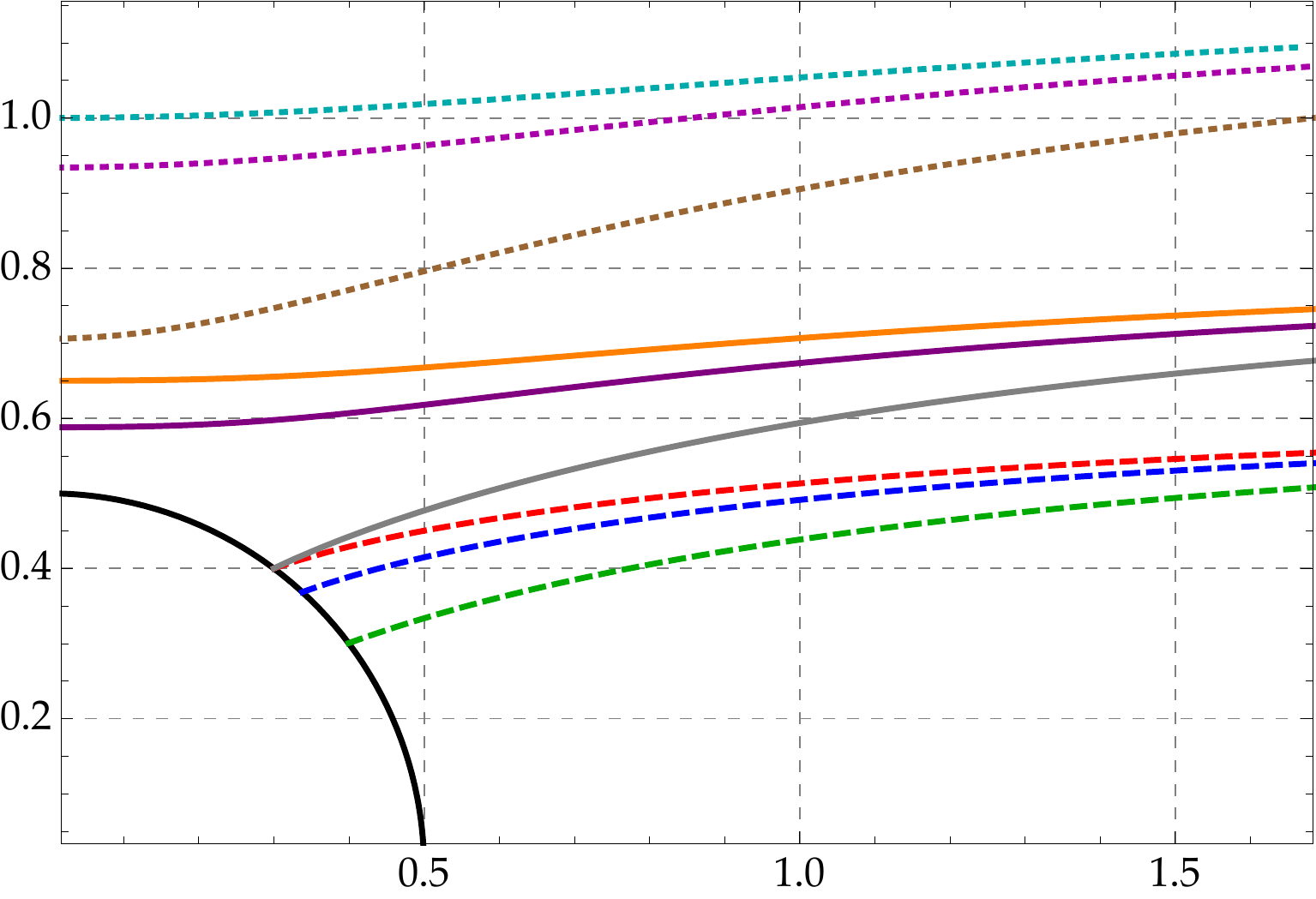}
 \includegraphics[width=0.24\textwidth]{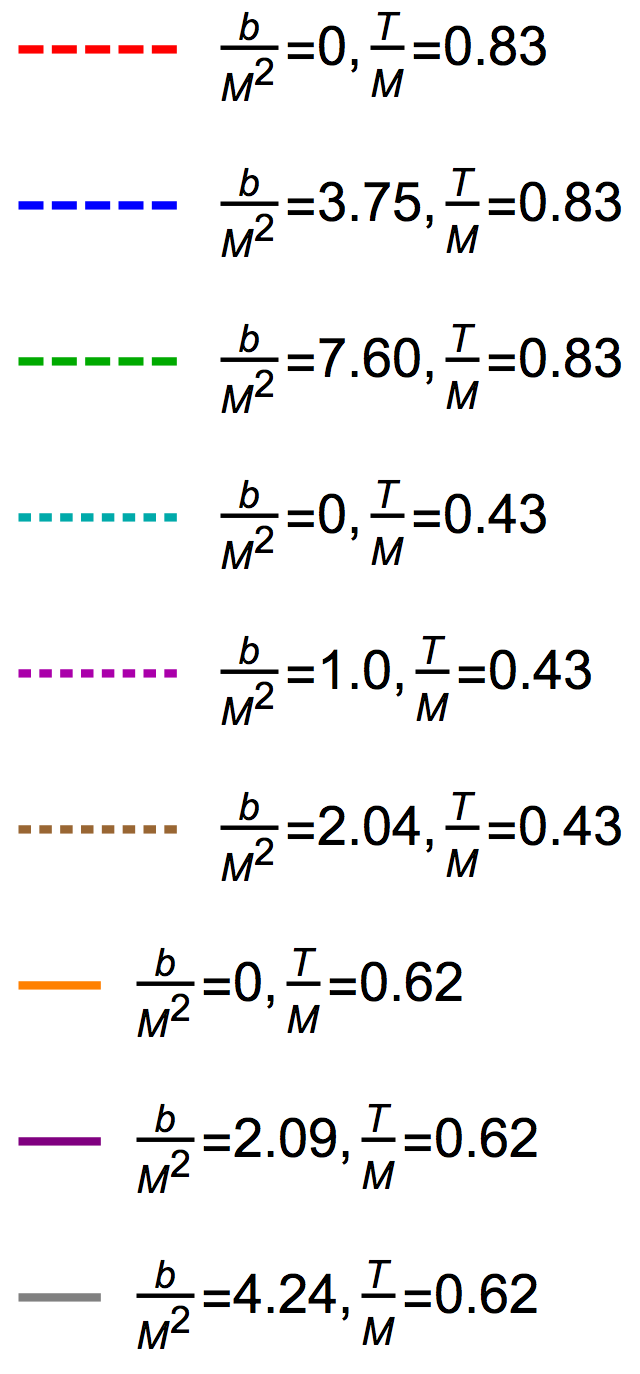}
 \put(-150,-10){\Large $r\cos\theta$}
 \put(-480,220){\Large $r\sin\theta$}
\end{tabular}
\end{center}
\caption{\small Profiles for various D7-brane embeddings in the $(r\cos\theta,r\sin\theta)$-plane. The black circle represents the horizon at $r_{h}=1/2$.}
\label{Embeddings_plots}
\end{figure}

To close this section, in Fig. (\ref{Embeddings_plots}) we present the profiles of some characteristic embeddings as parametric plots in the $(r\cos\theta,r\sin\theta)$-plane, since this presentation makes it easier to visualize how the transition between both types of embeddings takes place. For this plot we fixed the position of the horizon at $r_{h}=1/2$, meaning that the temperature of the system is fixed at $T=3/4\pi$. This is represented in Fig. (\ref{Embeddings_plots}) with a black line. However, each embedding in Fig. (\ref{Embeddings_plots}) is at a different value of $T/\bar{M}$ because each has a different value of $\bar{M}$.

We can see from Fig. (\ref{Embeddings_plots}) how the magnetic field tends to bend the branes towards the horizon, both for Minkowski and black hole embeddings, displaying how the magnetic field facilitates the phase transition. The dotted upper curves share $T/\bar{M}=0.43$ and are of Minkowski type for all three values of $b/\bar{M}^{2}$, as would be the case for any other embedding with $b/\bar{M}^{2}$ below the maximum intensity $2.04$ allowed at this $T/\bar{M}$. Meanwhile, the dashed lower curves share $T/\bar{M}=0.83$, which is a high enough value to make any embedding of black hole type regardless of the intensity of the magnetic field. The continuous curves in the middle are profiles for fixed $T/\bar{M}=0.62$. We can see that in this case, as we increase the magnetic field intensity, the brane is pulled towards the horizon until it falls through it for a certain critical magnetic field $b_{MMM}/\bar{M}^2$.

\section{Thermodynamic analysis}
\label{Thermo}

In this section we will perform a thermodynamic analysis of the phase transition between Minkowski and black hole embeddings described above. This will include the computation of the quark condensate, free energy, entropy, and energy densities of the branes. According to the holographic dictionary, all of this quantities can be computed from the on-shell D7-brane Euclidean action
\begin{equation}
S_{E}=2\pi^{2}T_{D7}N_{f}\int d^{5}x\mathcal{L},
\label{Euclidean_DBI}
\end{equation}
where the integration is performed over one period $\beta=1/T$ of Euclidean time $t_{E}=it$ and the metric is replaced by its Euclidean version. However, the result of this direct evaluation diverges as the integration is taken all the way up to the boundary at $r\rightarrow\infty$. To deal with this issue it is necessary to substract this divergent behavior by adding covariant boundary terms to the action, in a process known as holographic renormalization \cite{Skenderis:2002wp,Bianchi:2001kw}. We include the actual computation of the counterterm action in App. \ref{AppA}, while here we just present the result expressed in the Fefferman-Graham coordinate $u$ defined in \eqref{metric_5_FG}, in terms of which the boundary is located at $u=0$:
\begin{equation}
S_{ct}=-2\pi^{2}T_{D7}N_{f}\int d^{4}x \sqrt{\gamma}\left(\frac{1}{4}-\frac{1}{2}\chi^{2}-\frac{1}{8}F_{ij}F^{ij}\log{\epsilon}-\frac{1}{2\sqrt{6}}\varphi+\frac{1}{12}\varphi^{2}\right),
\label{counter_action}
\end{equation}
where $\gamma$ is the determinant of the induced metric at the boundary located at a radial cut-off $u=\epsilon$. The action \eqref{counter_action} contains the minimum terms that need to be added to the action in order to render it finite. However, we still have the freedom to add terms whose contributions by themselves are finite. In the current case the only ones are
\begin{equation}
S_{f}=-2\pi^{2}T_{D7}N_{f}\int d^{4}x \sqrt{\gamma}\left(C_{1}\chi^{4}+C_{2}F_{ij}F^{ij}\right).
\label{finite_term}
\end{equation}

This freedom is associated with the existence of a conformal anomaly in the gravitational background related to the magnetic field \cite{Avila:2018hsi}. As a consequence of this conformal anomaly, an independent arbitrary energy scale $\mu$ is introduced. This means that some physical quantities, such as the free energy of the system, do not depend only on the two dimensionless ratios that can be made out of $b$, $T$ and $\bar{M}$, but on the three independent ratios that can be built from $b$, $T$, $\bar{M}$ and $\mu$. We want to meassure everything in units of $\bar{M}$, hence we will work with $b/\bar{M}^{2}$, $T/\bar{M}$ and $\mu/\bar{M}$. 

On the other hand, different choices for the free coefficients $C_{1}$ and $C_{2}$ lead to different values in observables such as the free energy. We can use this dependence to reverse the change that would be introduced in physical quantities as the regularization scale is modified in the quest to fix a particular renormalization scheme, achieving in this way the scheme independence of relevant results. In that sense, $C_{1}$ and $C_{2}$ are scheme-dependent quantities. As explained in \cite{Mateos:2007vn}, we can fix $C_{1}=1/4$ by demanding that the on-shell action vanishes for the supersymmetric embedding. Because we don't have a physical reason to choose a specific value for $C_{2}$, some of our next results will dependend on its value. 

Having said that, the full renormalized action, defined as
\begin{equation}
S_{D7}=S_{E}+S_{ct}+S_{f},
\label{renormalized_action}
\end{equation}
is then finite in the limit $\epsilon\rightarrow 0$. We would like to remark that in order to properly evaluate \eqref{renormalized_action} both the DBI action and boundary terms need to be expressed in the same radial coordinate. Given that our numerical solutions are naturally computed using the $r$ coordinate, we will work using it in what follows.
\subsection{Quark condensate}
With the renormalized action at hand, we can compute the quark condensate by taking the variation of \eqref{renormalized_action} with respect to the quark mass. The computation is performed explicitly in App. \ref{AppC}, while here we just present the end result
\begin{equation}
\frac{\langle\bar{q}q \rangle}{\kappa}=c-\frac{1}{4}mU_{1}^{2},
\label{condensate}
\end{equation}
where
\begin{equation}
\kappa=-\frac{1}{2^{3/2}\pi^{3}}\sqrt{\lambda}N_{c}N_{f}.
\end{equation}

Once a given numerical embedding is known, it is possible to extract the value of the coefficients $m$ and $c$ by analyzing its near-boundar behavior. Thus, in practice $m$ and $c$ are both functions of $b$, $T$ and either $\chi_{h}$ or $r_{i}$ depending on the type of embedding. By inverting this relations we can eliminate the latter in favor of $m$ and then express the quark condensate as a function of $b/\bar{M}^{2}$ and $T/\bar{M}$. We explicitly checked that the quark condensate is independent of $\mu/\bar{M}$, rendering it a function only of the two dimensionless quantities just listed, or those constructed exclusively out of them. Note that, consistently, the quark condensate does not depend on the free coefficient $C_{2}$, as can be seen from the fact that the variation of the $F^{2}$ finite term with respect to the quark mass is zero. In practice the plots in this subsection were all produced at $\mu/\bar{M}=1$. Notice that, given that $U_{1}$ is a function of the magnetic field and the temperature, said term will affect the value of the quark condensate in a non-trivial way. 

\begin{figure}[ht!]
 \centering
 \includegraphics[width=0.85\textwidth]{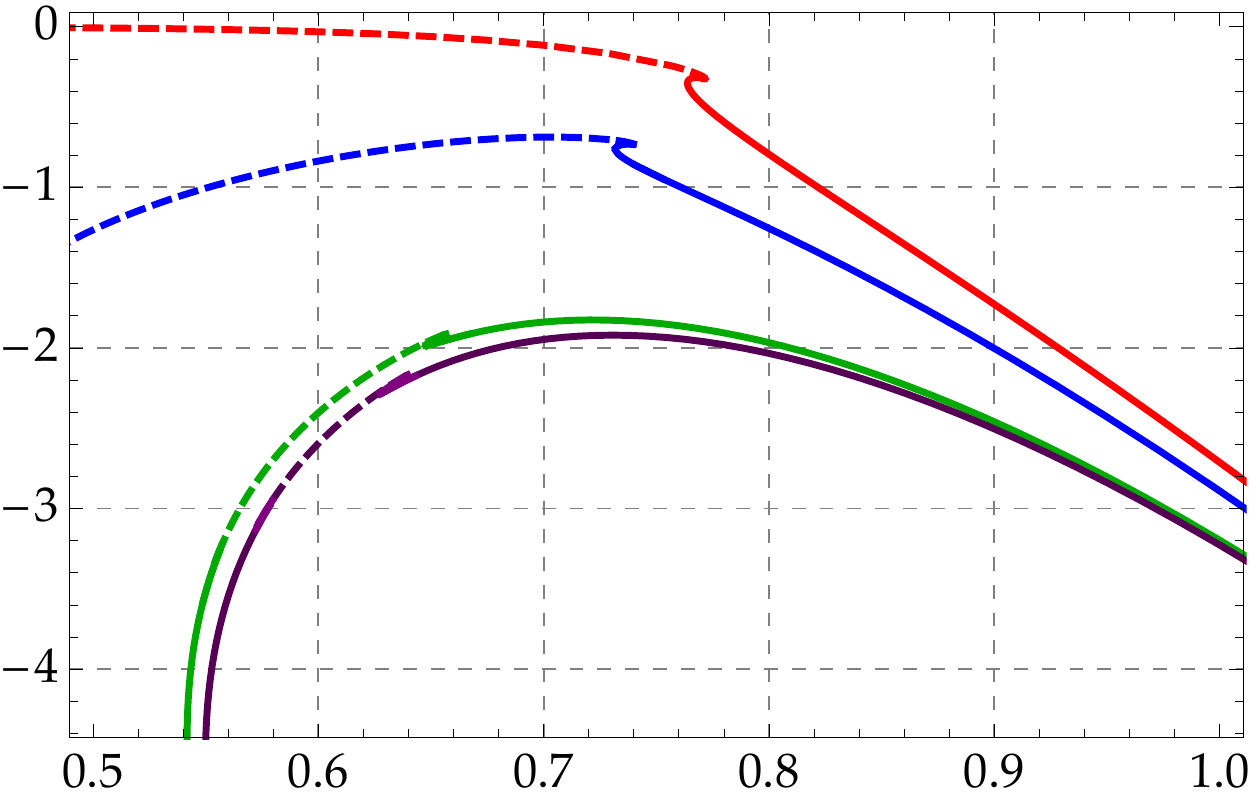}
 \put(-30,-10){\Large $\frac{T}{\bar{M}}$}
 \put(-400,220){\Large $\frac{\langle \bar{q}q\rangle}{\kappa\bar{M}^{3}}$}
\caption{\small Quark condensate $\langle \bar{q}q\rangle/\kappa\bar{M}^{3}$ as a function of $T/\bar{M}$. Red, blue, green, and purple curves (top to bottom) correspond to $b/\bar{M}^{2}=\lbrace0,2,3.3,3.4\rbrace$ respectively. The dashed segments correspond to Minkowski embeddings, while the continuous segments correspond to the black hole embeddings.}
\label{Condensate_bM2}
\end{figure}

In Fig. (\ref{Condensate_bM2}) we show the quark condensate in units of $\bar{M}$ as a function of $T/\bar{M}$ for different values of $b/\bar{M}^{2}$, ranging between zero and the maximum of the phase transition curve in the phase diagram Fig. (\ref{Phase_Diagram}). For $b/\bar{M}^{2}=0$ (red curve) we recover the results from \cite{Mateos:2007vn}, while for $b/\bar{M}^{2}$ other than zero the general effect of the magnetic field is to increase the magnitude of the condensate. 

For any magnetic field intensity other than zero, the quark condensate never vanishes for the temperatures that we are allowed to explore with our gravity construction, which signalizes that chiral symmetry is broken. Additionally, the condensate diverges as $T/\bar{M}$ aproaches its minimum value for the given magnetic field\footnote{As we will see in the following, this behavior is shared among all thermodynamic quantities.}. Note that this happens whether the low temperature phase is Minkowski or black hole. This behavior is consistent with the fact that for that temperature the background itself is expected to undergo a phase transition, as explained in detail in \cite{Avila:2018hsi}. A consequence of this is that we cannot check if the quark condensate vanishes at $T/\bar{M}=0$ for $b/\bar{M}^{2}\neq 0$.

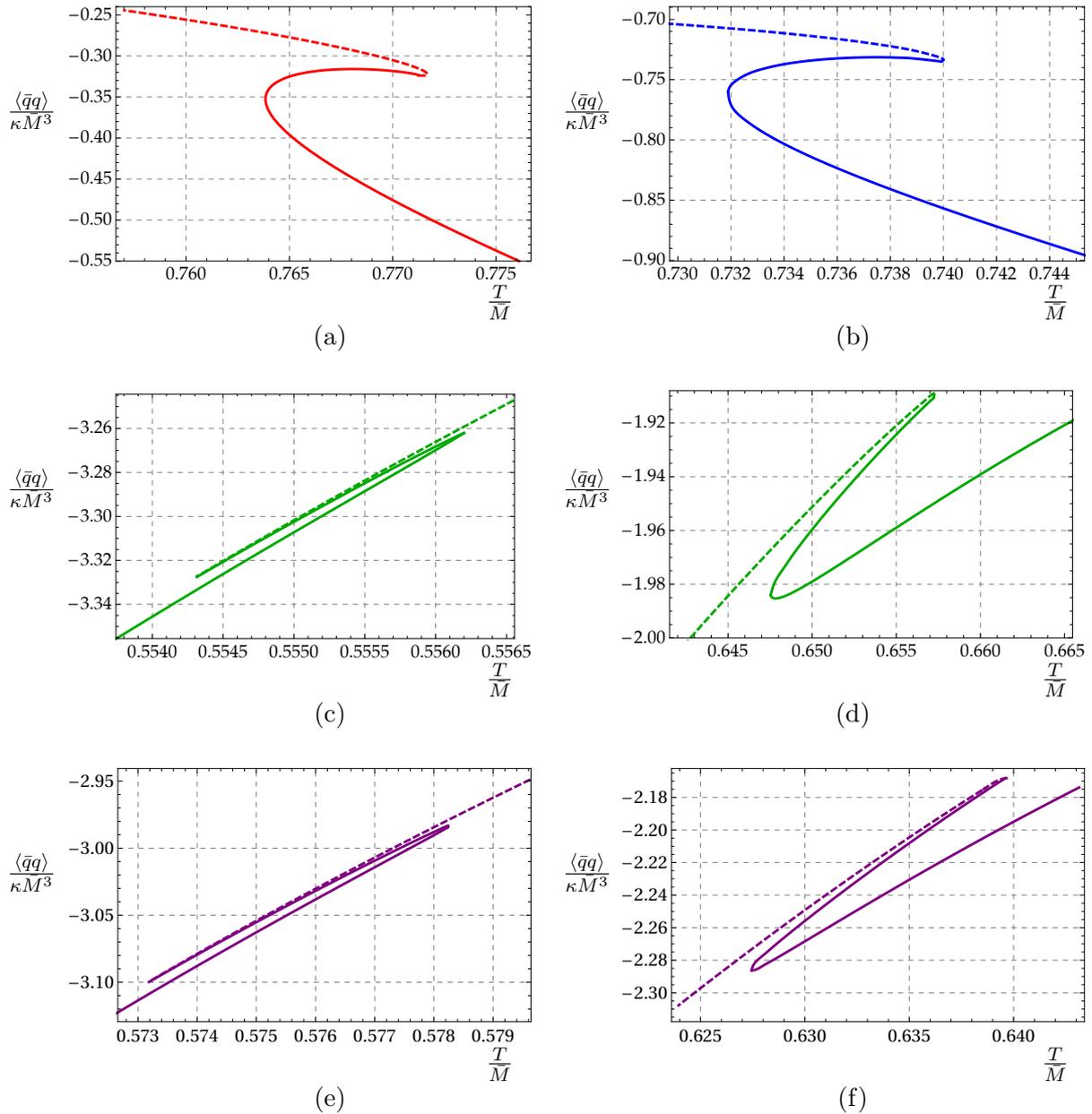
\begin{figure}
\begin{center}
\begin{tabular}{cc}
\includegraphics[width=0.45\textwidth]{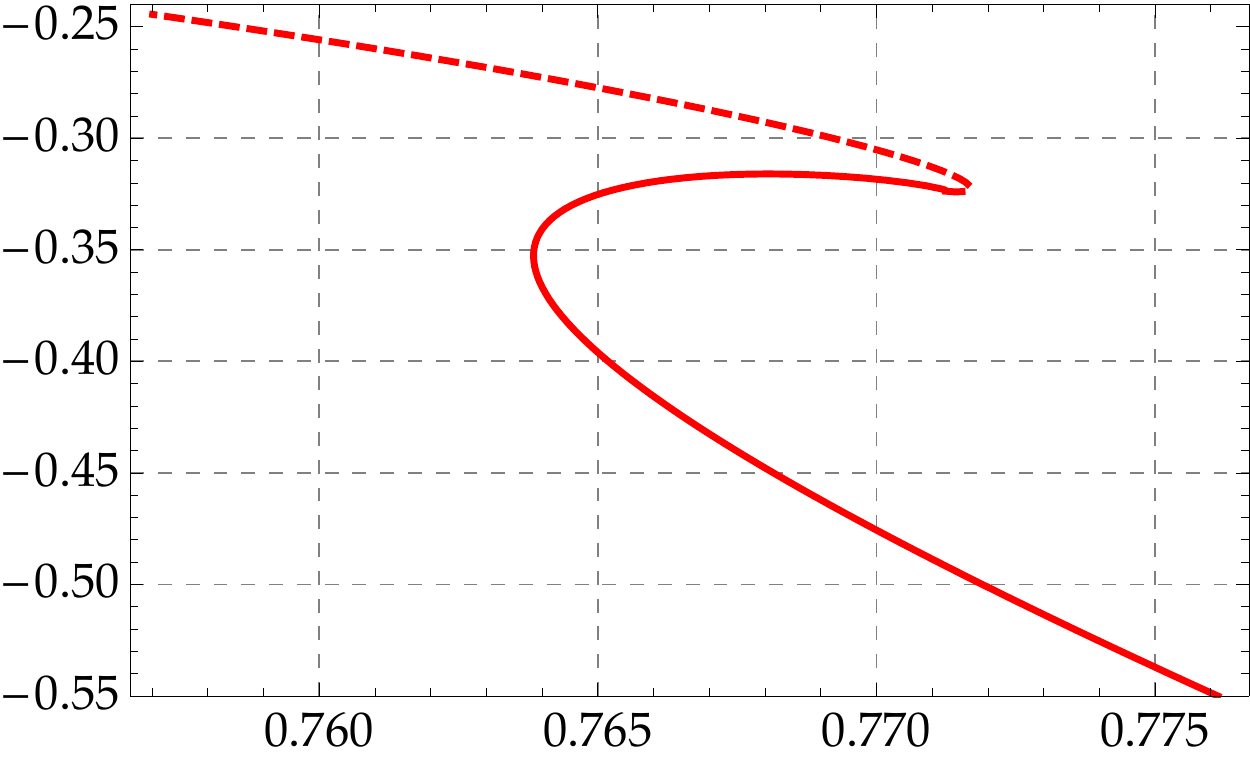} 
\qquad\qquad & 
\includegraphics[width=0.45\textwidth]{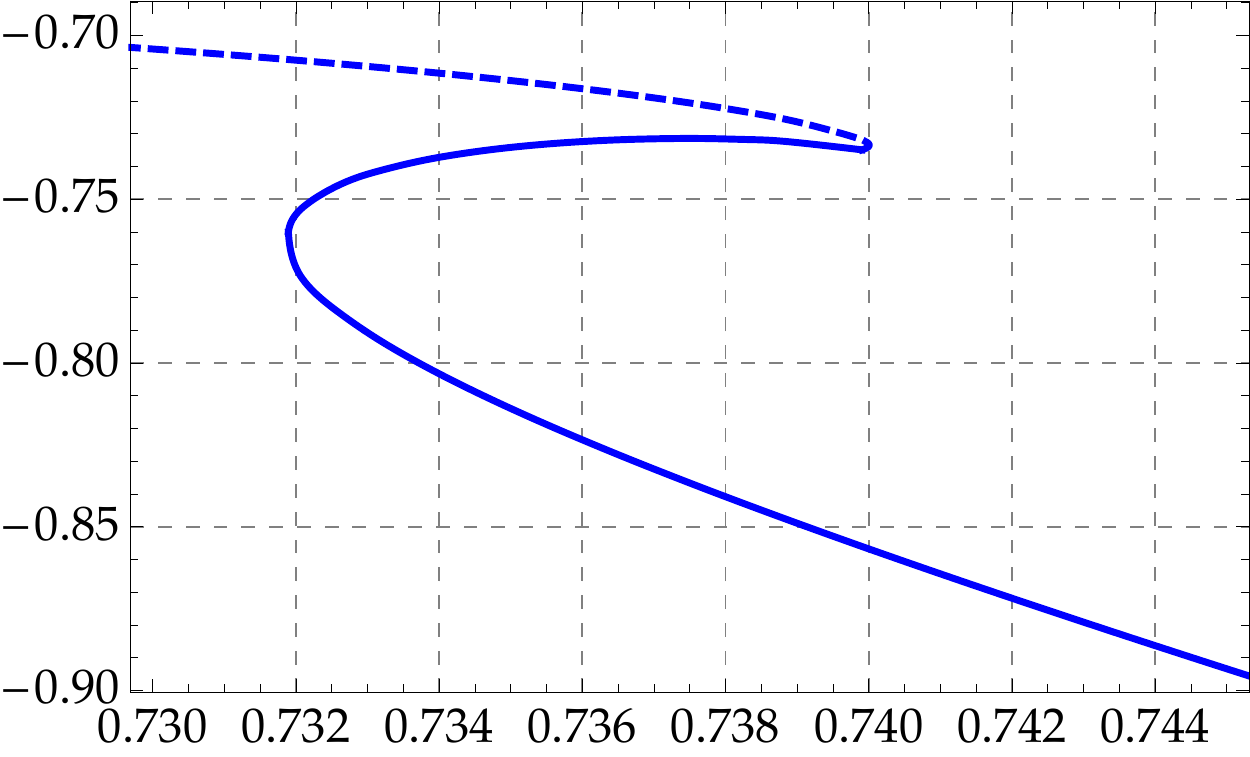}
\qquad
 \put(-450,70){$\frac{\langle \bar{q}q\rangle}{\kappa\bar{M}^{3}}$}
   \put(-250,-10){$\frac{T}{\bar{M}}$}
    \put(-218,70){$\frac{\langle \bar{q}q\rangle}{\kappa\bar{M}^{3}}$}
   \put(-18,-10){$\frac{T}{\bar{M}}$}
 \\
(a) & (b)\\
& \\
\includegraphics[width=0.45\textwidth]{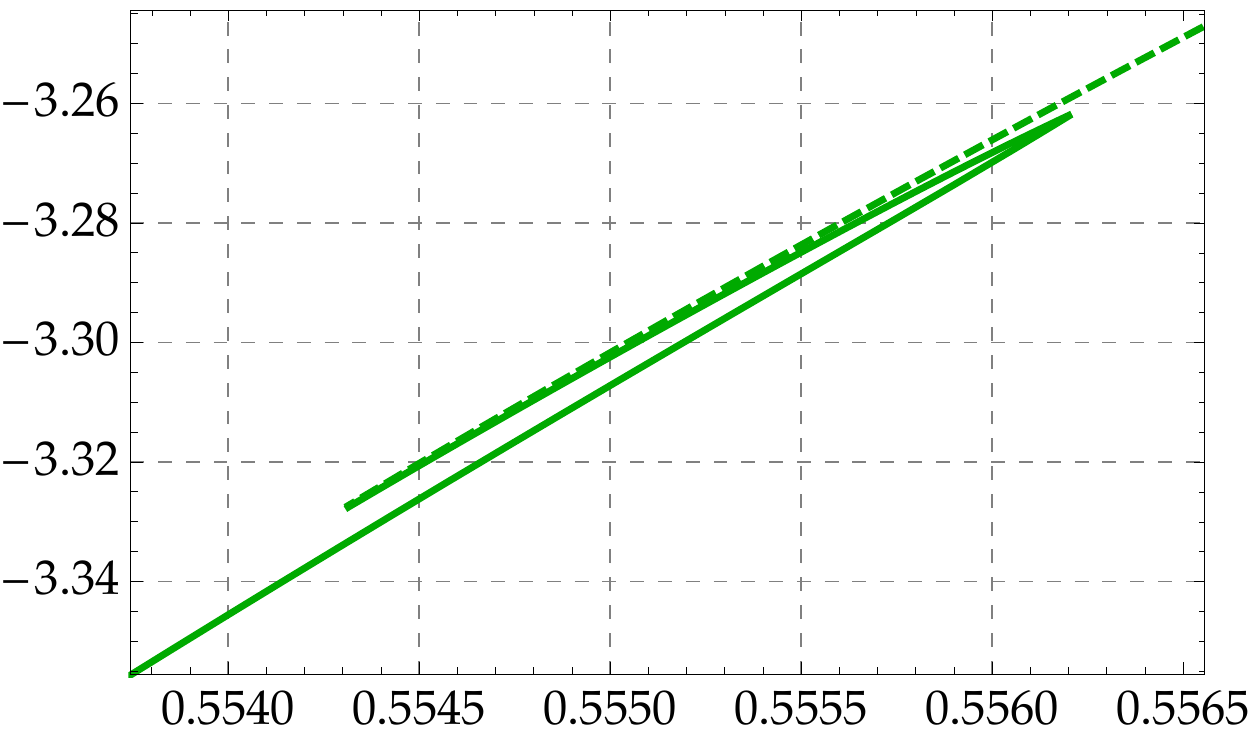} 
\qquad\qquad & 
\includegraphics[width=0.45\textwidth]{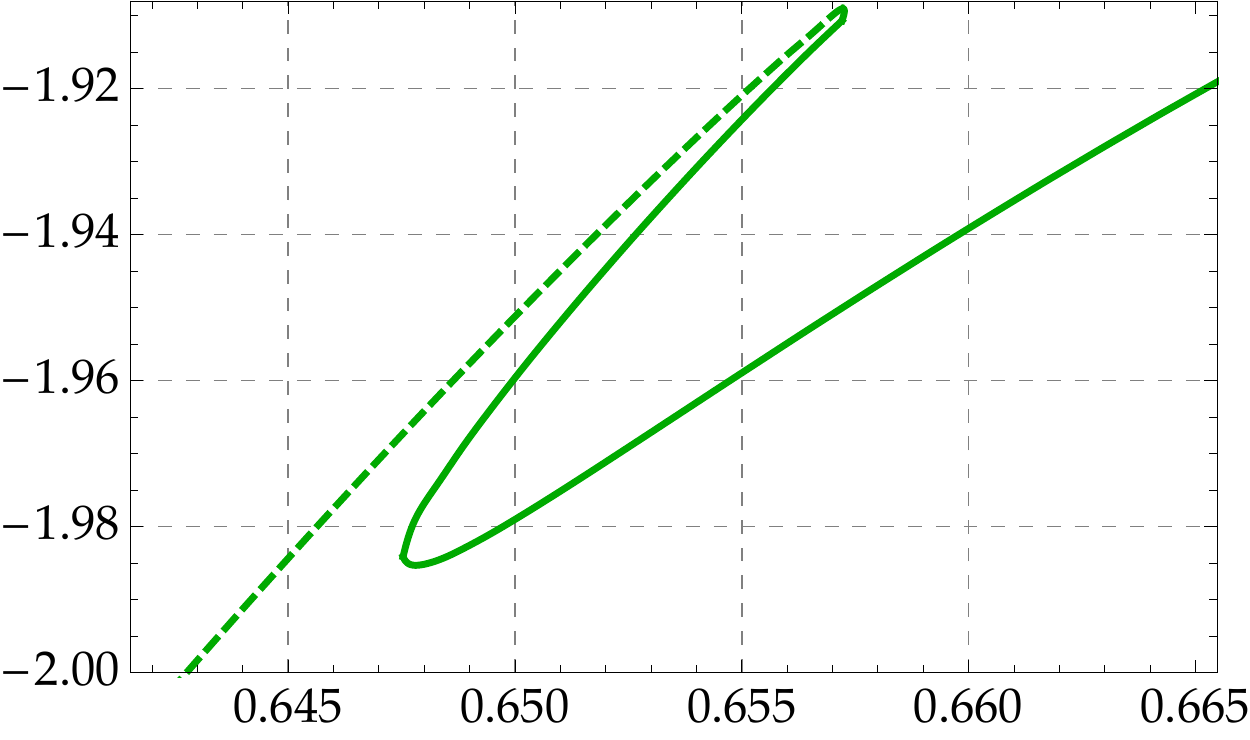}
\qquad
 \put(-450,70){$\frac{\langle \bar{q}q\rangle}{\kappa\bar{M}^{3}}$}
   \put(-250,-10){$\frac{T}{\bar{M}}$}
   \put(-218,70){$\frac{\langle \bar{q}q\rangle}{\kappa\bar{M}^{3}}$}
   \put(-18,-10){$\frac{T}{\bar{M}}$}
         \\
(c) & (d)\\
& \\
\includegraphics[width=0.45\textwidth]{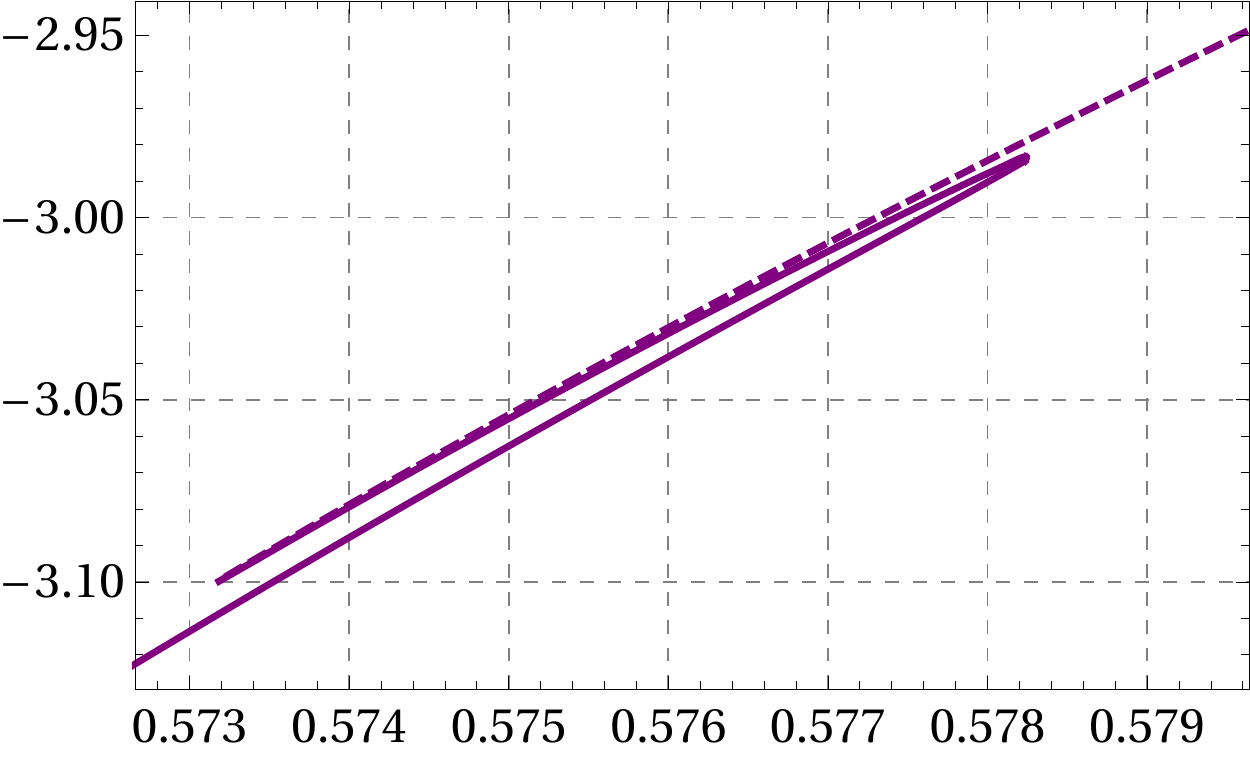} 
\qquad\qquad & 
\includegraphics[width=0.45\textwidth]{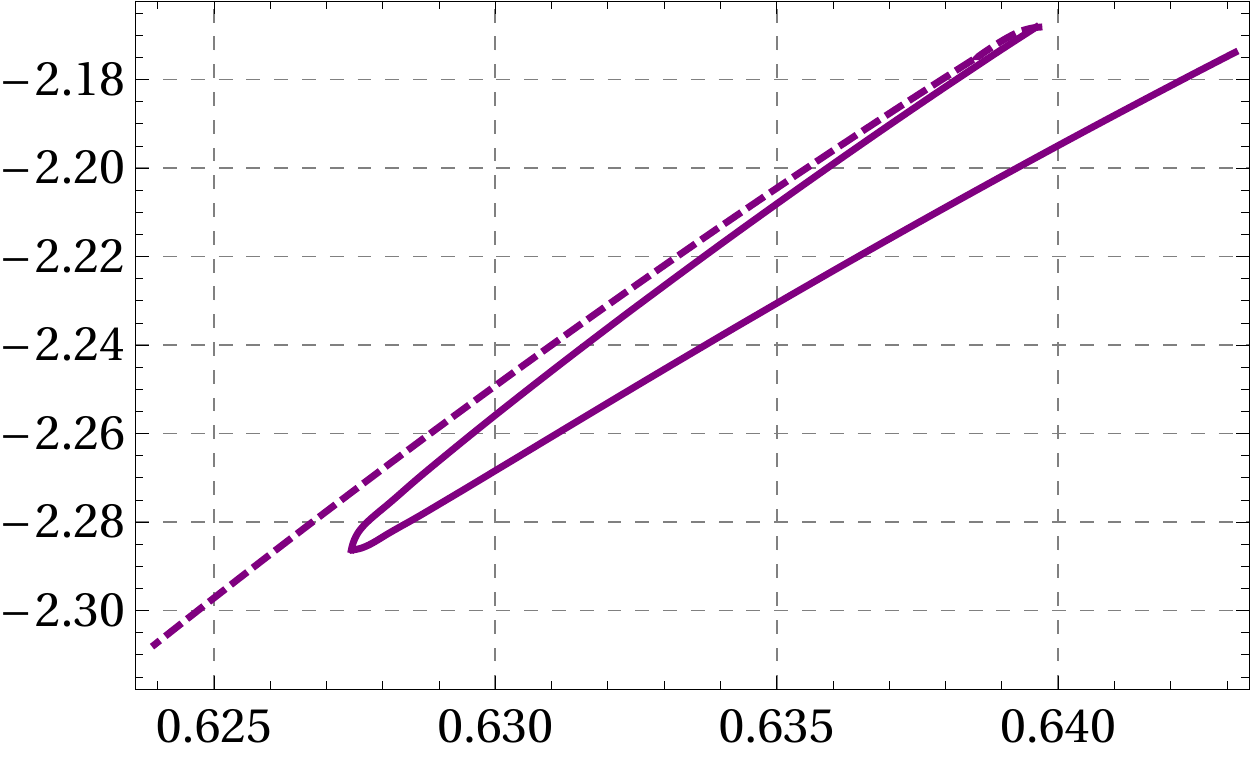}
\qquad
 \put(-450,70){$\frac{\langle \bar{q}q\rangle}{\kappa\bar{M}^{3}}$}
 \put(-250,-10){$\frac{T}{\bar{M}}$}
 \put(-218,70){$\frac{\langle \bar{q}q\rangle}{\kappa\bar{M}^{3}}$}
 \put(-18,-10){$\frac{T}{\bar{M}}$}
\\
(e) & (f) 
\end{tabular}
\end{center}
\caption{\small Zoom into the transition region between Minkowski and black hole embeddings for the quark condensate. The values of the magnetic field are $b/\bar{M}^{2}=$ 0 (a), 2 (b), 3.3 (c) and (d), and 3.4 (e) and (f). Dashed lines represent Minkowski embeddings, whereas solid lines represent black hole embedding.}
\label{Condensate_bM2_zoom}
\end{figure}

In Fig. (\ref{Condensate_bM2_zoom}) we zoom-in near the region where the phase transition takes place. From Fig. (\ref{Condensate_bM2_zoom}) (a) we see that we recover the results from \cite{Mateos:2007vn} when the magnetic field vanishes. For non-vanishing magnetic field intensities the curves retain the same shape, which signalizes that the phase transition remains first order regardless of the magnetic field. This is also true for the new cold phase transition, as it can be seen from Fig. (\ref{Condensate_bM2_zoom}) (c) and (e).  

\begin{figure}[ht!]
 \centering
 \includegraphics[width=0.85\textwidth]{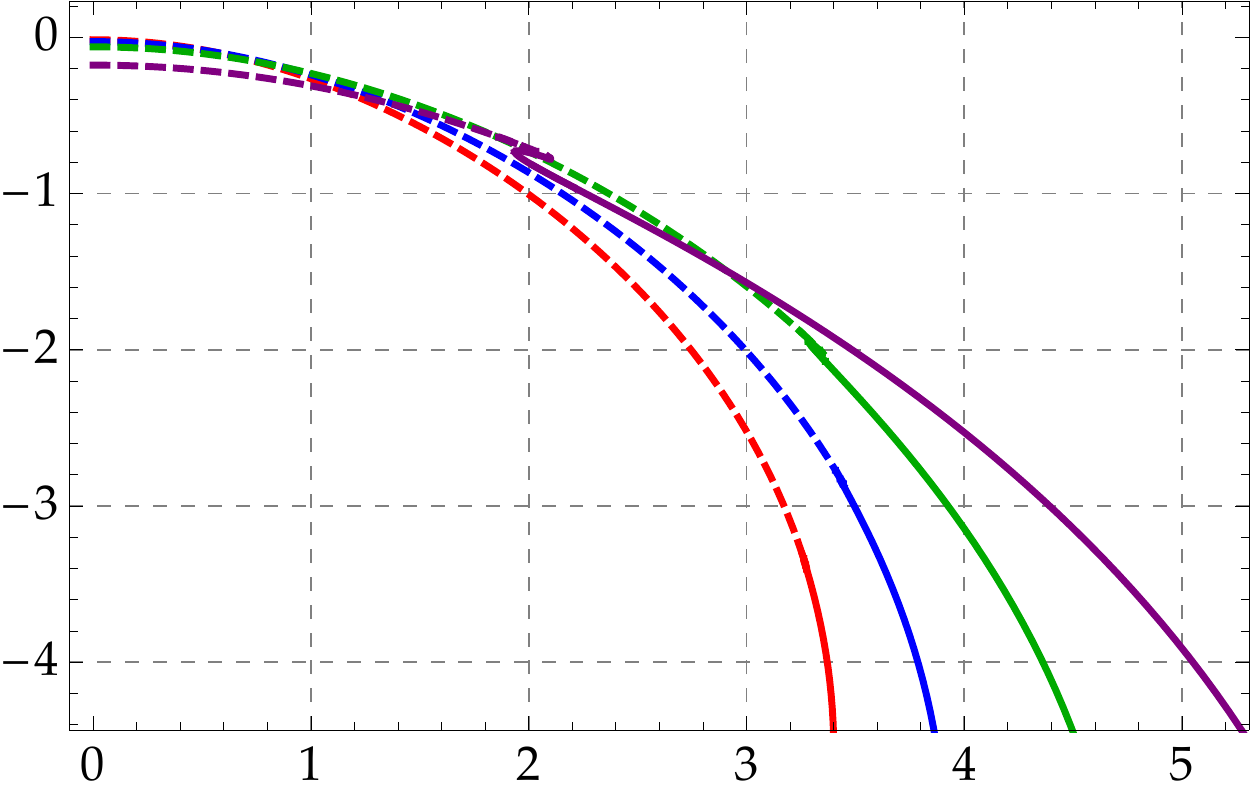}
 \put(0,0){\Large $\frac{b}{\bar{M}^{2}}$}
 \put(-410,210){\Large $\frac{\langle \bar{q}q\rangle}{\kappa\bar{M}^{3}}$}
\caption{\small Quark condensate $\langle \bar{q}q\rangle/\kappa$ as a function of $b/\bar{M}^{2}$. Red, purple, green, and blue curves (left to right at the bottom) correspond to $T/\bar{M}=\lbrace 0.55,0.59,0.65,0.734\rbrace$ respectively. The dashed segments correspond to Minkowski embeddings, while the continuous segments correspond to black hole embeddings.}
\label{Condensate_TM}
\end{figure}

\begin{figure}
\begin{center}
\begin{tabular}{cc}
\includegraphics[width=0.45\textwidth]{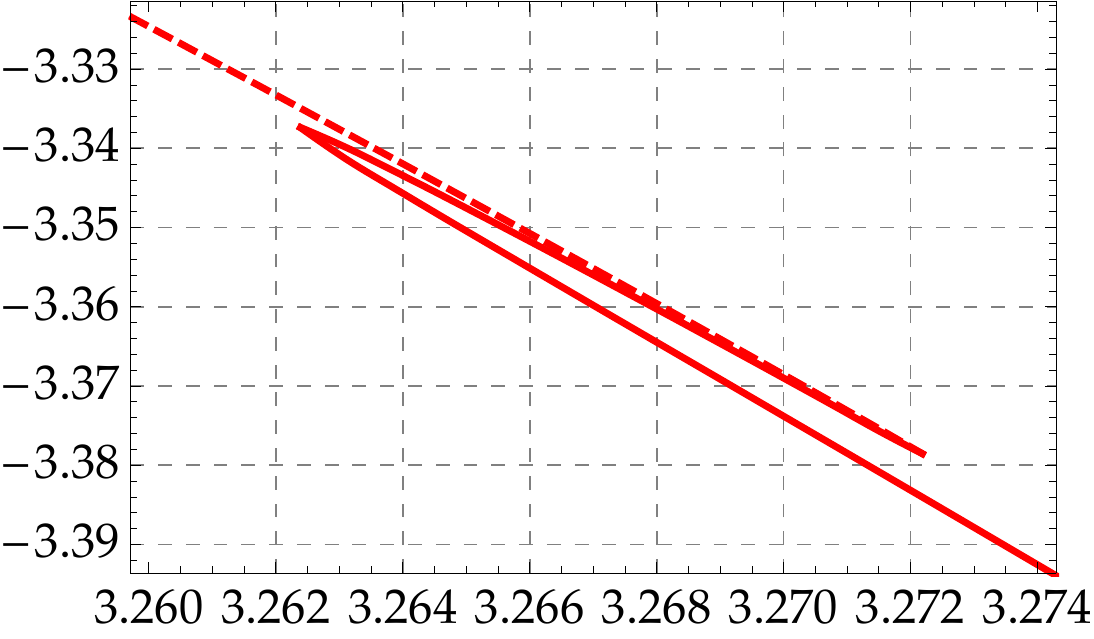} 
\qquad\qquad & 
\includegraphics[width=0.45\textwidth]{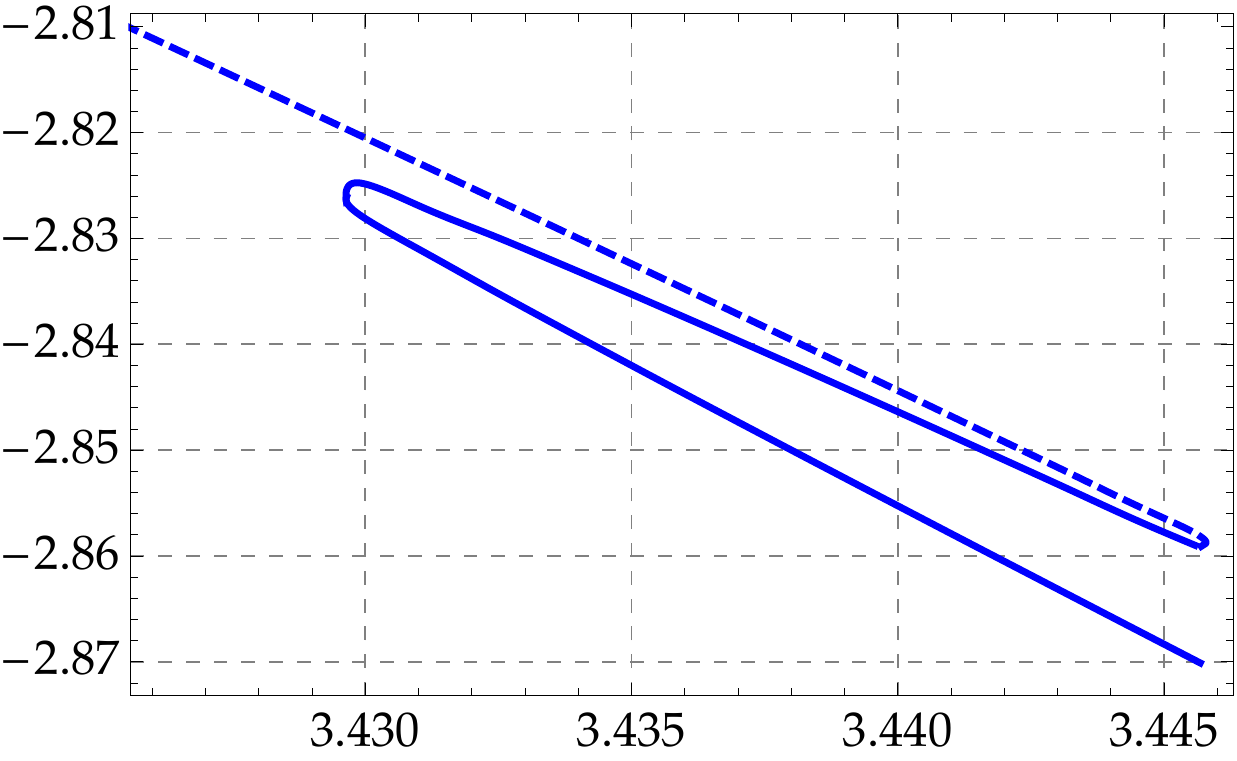}
\qquad
 \put(-450,70){$\frac{\langle \bar{q}q\rangle}{\kappa\bar{M}^{3}}$}
   \put(-250,-10){$\frac{b}{\bar{M}^{2}}$}
    \put(-220,70){$\frac{\langle \bar{q}q\rangle}{\kappa\bar{M}^{3}}$}
   \put(-18,-10){$\frac{b}{\bar{M}^{2}}$}
 \\
(a) & (b)\\
& \\
\includegraphics[width=0.45\textwidth]{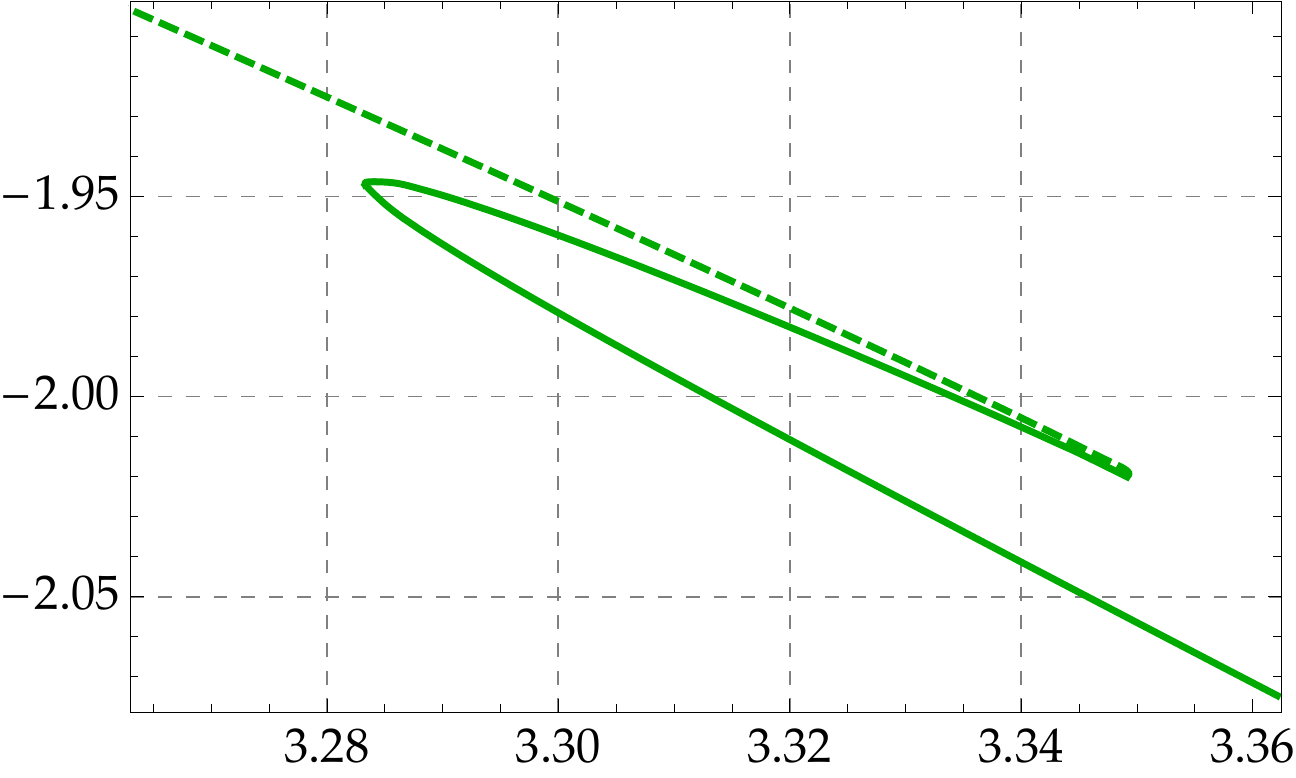} 
\qquad\qquad & 
\includegraphics[width=0.45\textwidth]{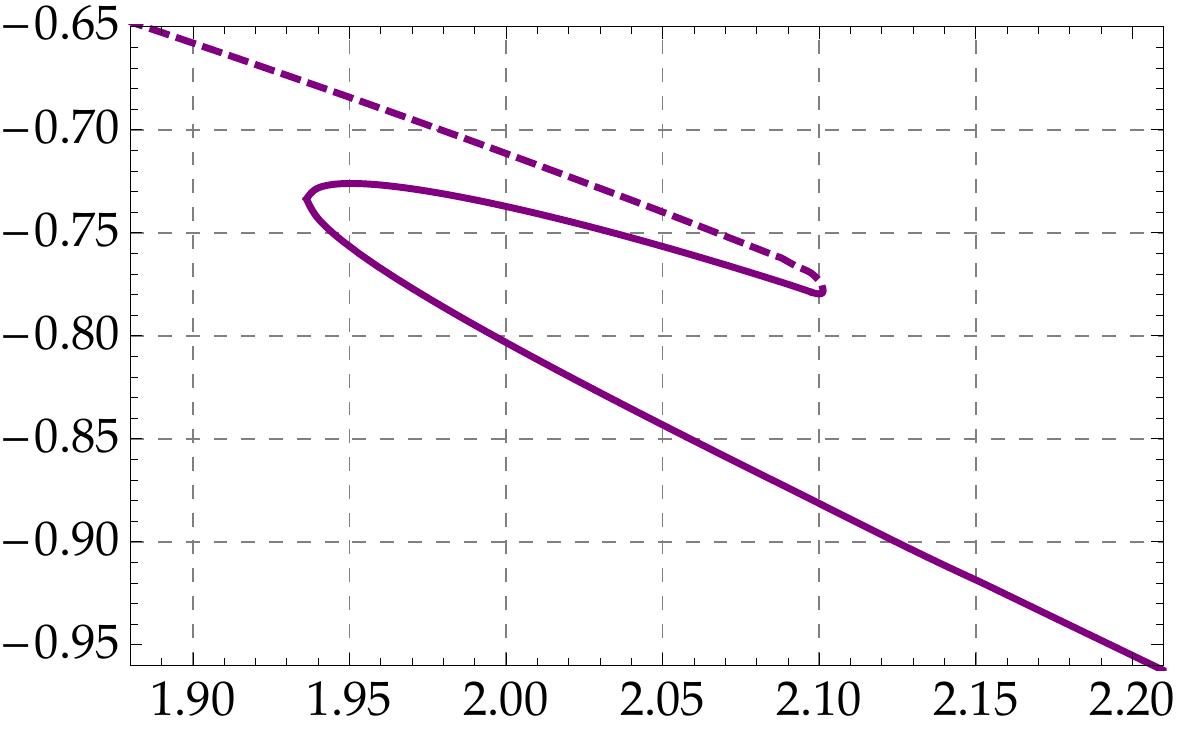}
\qquad
 \put(-450,70){$\frac{\langle \bar{q}q\rangle}{\kappa\bar{M}^{3}}$}
   \put(-250,-10){$\frac{b}{\bar{M}^{2}}$}
    \put(-220,70){$\frac{\langle \bar{q}q\rangle}{\kappa\bar{M}^{3}}$}
   \put(-18,-10){$\frac{b}{\bar{M}^{2}}$}
 \\
(c) & (d)
\end{tabular}
\end{center}
\caption{\small Zoom into the transition region between Minkowski and black hole embeddings for the quark condensate. The values of the temperature are $T/\bar{M}=$ 0.55 (a), 0.59 (b), 0.65 (c), and 0.734 (d). The dashed lines represent Minkowski embeddings, whereas the solid lines represent black hole embedding.}
\label{Condensate_TM_zoom}
\end{figure}

To conclude this subsection we show the quark condensate $\langle \bar{q}q\rangle/\kappa\bar{M}^{3}$ as a function of the magnetic field at fixed temperature in Fig. (\ref{Condensate_TM}). The values of $T/\bar{M}$ displayed are in the range in which the phase transition can be triggered by changing the magnetic field intensity. We can see that the magnitude of the quark condensate increases with the magnetic field for all the temperatures that we explored, which means that we are observing MC in regard to the quark condensate. This seems to be consistent with lattice results \cite{Bali:2011qj,Bali:2012zg,Endrodi:2019zrl}, as it is known that for heavy quarks MC is observed at low temperatures. Lastly, in Fig. (\ref{Condensate_TM_zoom}) we zoom-in near the region where the transition takes place. The shape of the curves signalize that this is a first order phase transition.
 
\subsection{Free energy density}
According to the holographic dictionary, the free energy of the system is related to the on-shell Euclidean D7-brane action \eqref{renormalized_action} through
\begin{equation}
\mathcal{F}=T S_{D7}.
\end{equation}
Note that because there is no dependence on the gauge theory directions, it is possible to factorize the infinite three-dimensional volume $\text{vol}(x)$ and a factor of $1/T$ coming from the integration over the Euclidean time from \eqref{renormalized_action}. While the remaining radial dependence in principle can be computed just by direct evaluation, for numerical purposes it is more convenient to rewrite \eqref{renormalized_action} as a finite integral, instead of two infinite integrals with a finite diference. 

To this end we express the counterterm action \eqref{counter_action} and the finite term \eqref{finite_term} using the near-boundary expansions of the fields \eqref{chi_boundary} and \eqref{r_expansions} evaluated at $r=r_{max}$. Given that $r_{max}$ is understood to be taken to the boundary, we can keep only the leading terms. The result can be organized as
\begin{equation}
\frac{T(S_{ct}+S_{f})}{2\pi^{2}T_{D7}N_{f}\text{vol}(x)}=\sum_{i=0}^{4}a_{i}r_{max}^{i}-\frac{b^{2}}{6}\log{r_{max}},
\end{equation}
where
\begin{eqnarray}
&& a_{4}=-\frac{1}{4},
\cr
&& a_{3}=-\frac{U_{1}}{2},
\cr
&& a_{2}=\frac{1}{24}(-9U_{1}^{2}+2\sqrt{6}\varphi_{0}+12m^{2}),
\cr
&& a_{1}=\frac{U_{1}}{24}(-3U_{1}^{2}+2\sqrt{6}\varphi_{0}+12m^{2}),
\cr
&& a_{0}=-\frac{U_{1}^{4}}{64}-\frac{U_{4}}{8}+\frac{U_{1}^{2}\varphi_{0}}{8\sqrt{6}}-\frac{\varphi_{0}^{2}}{12}-\frac{U_{1}^{2}m^{2}}{8}-\frac{m^{4}}{4}+mc-b^{2}\left(\frac{1}{12}+2C_{2}\right).
\label{a_i}
\end{eqnarray}
In order to create a finite integral, we rewrite the $r_{max}$ terms as an integral from $r_{min}$ to $r_{max}$ minus the lower integration limit terms\footnote{Note that $r_{min}=r_{h}$ for black hole embeddings, while $r_{min}=r_{i}$ for Minkowski embeddings.}. This gives
\begin{equation}
\frac{T(S_{ct}+S_{f})}{2\pi^{2}T_{D7}N_{f}\text{vol}(x)}=\int_{r_{min}}^{r_{max}}dr\mathcal{L}_{ct}+\sum_{i=0}^{4}a_{i}r_{min}^{i}-\frac{b^{2}}{6}\log{r_{min}},
\end{equation}
where we have denoted
\begin{equation}
\mathcal{L}_{ct}=4 a_{4}r^{3}+3 a_{3}r^{2}+2 a_{2}r+a_{1}-\frac{b^{2}}{6}\frac{1}{r}.
\label{Lct}
\end{equation}
In conclusion, the final expression for the free energy that we evaluate numerically is
\begin{equation}
\frac{\mathcal{F}}{\mathcal{N}}=\int_{r_{min}}^{r_{max}} dr \left(\mathcal{L}+\mathcal{L}_{ct}\right)+\sum_{i=0}^{4}a_{i}r_{min}^{i}-\frac{b^{2}}{6}\log{r_{min}},
\end{equation}
where we normalized our result with respect to
\begin{equation}
\mathcal{N}=2\pi^{2}T_{D7}N_{f}\text{vol}(x)=\frac{\text{vol}(x)}{8\pi^{4}}\lambda N_{c}N_{f}.
\label{mathN}
\end{equation}
The integrand $\mathcal{L}+\mathcal{L}_{ct}$ goes to zero in the limit $r_{max}\rightarrow\infty$ by construction, so the integral has been regularized. Notice that fixing a specific value for the free coefficient $C_{2}$ will change the dependence of the free energy on the magnetic field. If we work at a fixed field intensity hovewer, this term amounts only to an additive constant.

\begin{figure}[ht!]
 \centering
 \includegraphics[width=0.85\textwidth]{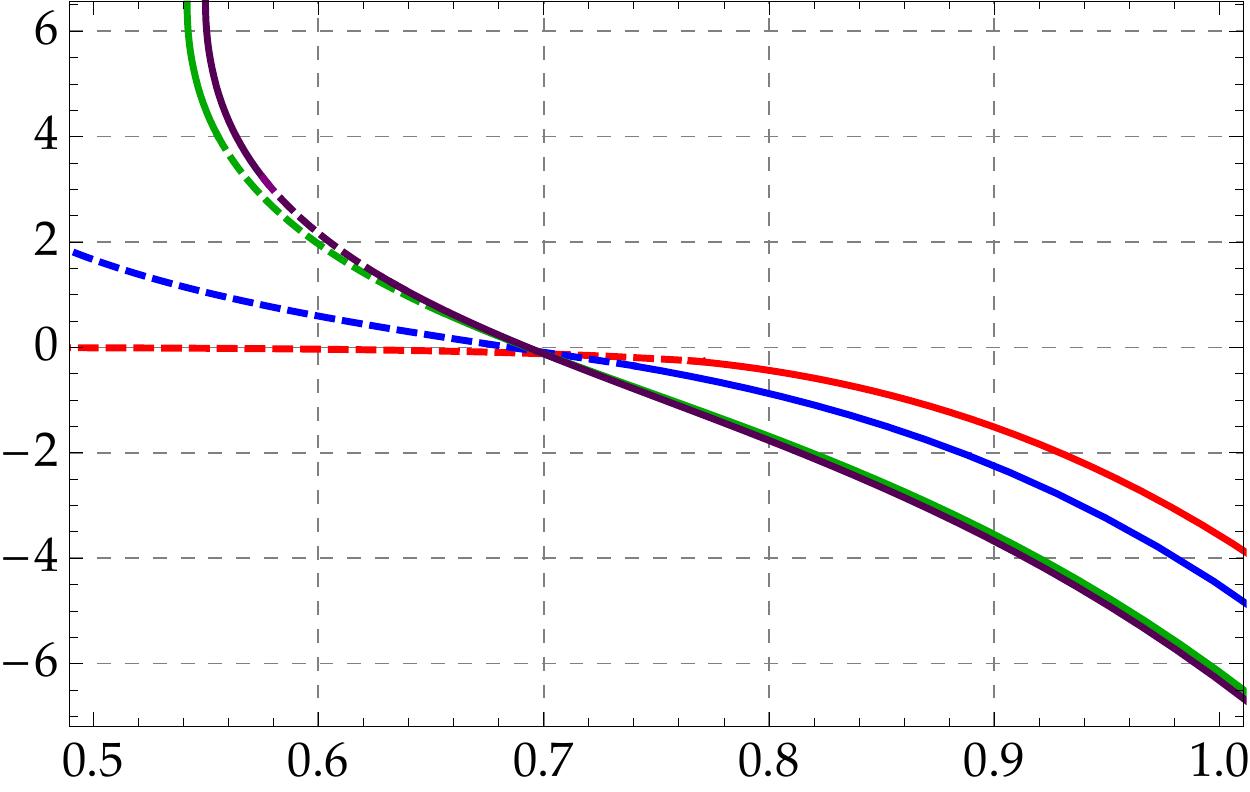}
 \put(0,-10){\Large $\frac{T}{\bar{M}}$}
 \put(-390,220){\Large $\frac{\mathcal{F}}{\mathcal{N}\bar{M}^{4}}$}
\caption{\small Free energy density $\mathcal{F}/\mathcal{N}\bar{M}^{4}$ as a function of $T/\bar{M}$. Red, blue, green, and purple curves (bottom to top on the left) correspond to $b/\bar{M}^{2}=\lbrace0,2,3.3,3.4\rbrace$ respectively. The dashed segments correspond to Minkowski embeddings, while the continuous segments correspond to black hole embeddings. The finite term is fixed at $C_{2}=0$.}
\label{Free_bM2}
\end{figure}

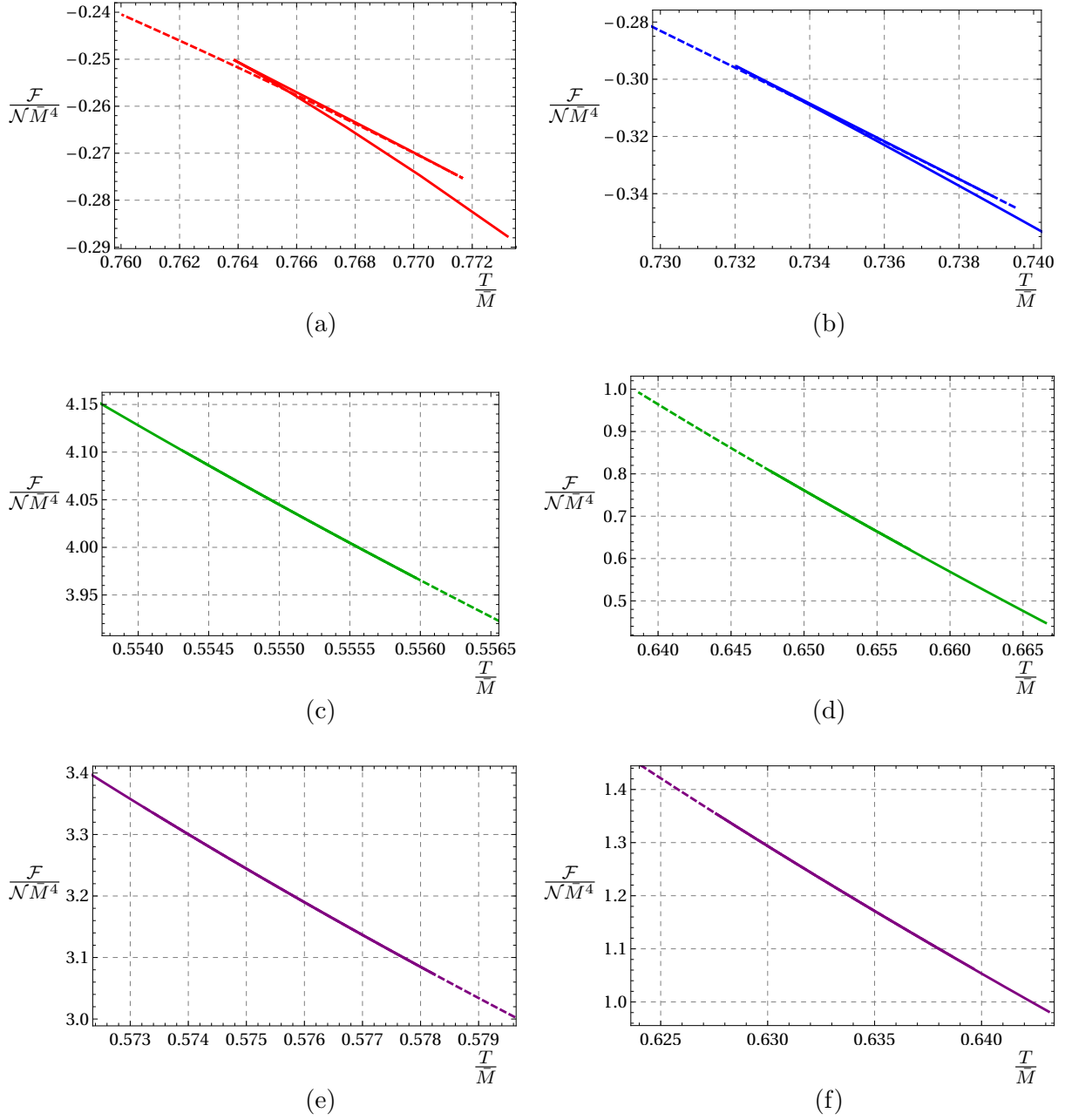
\begin{figure}
\begin{center}
\begin{tabular}{cc}
\includegraphics[width=0.45\textwidth]{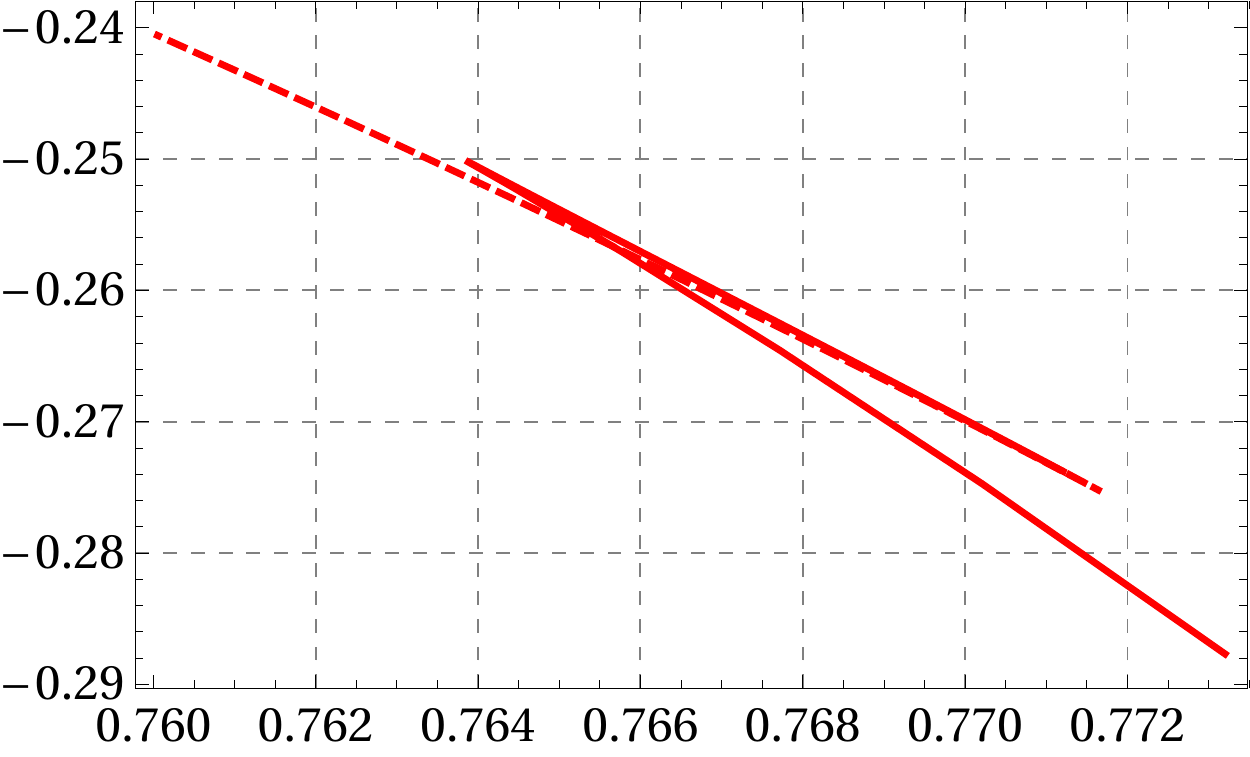} 
\qquad\qquad & 
\includegraphics[width=0.45\textwidth]{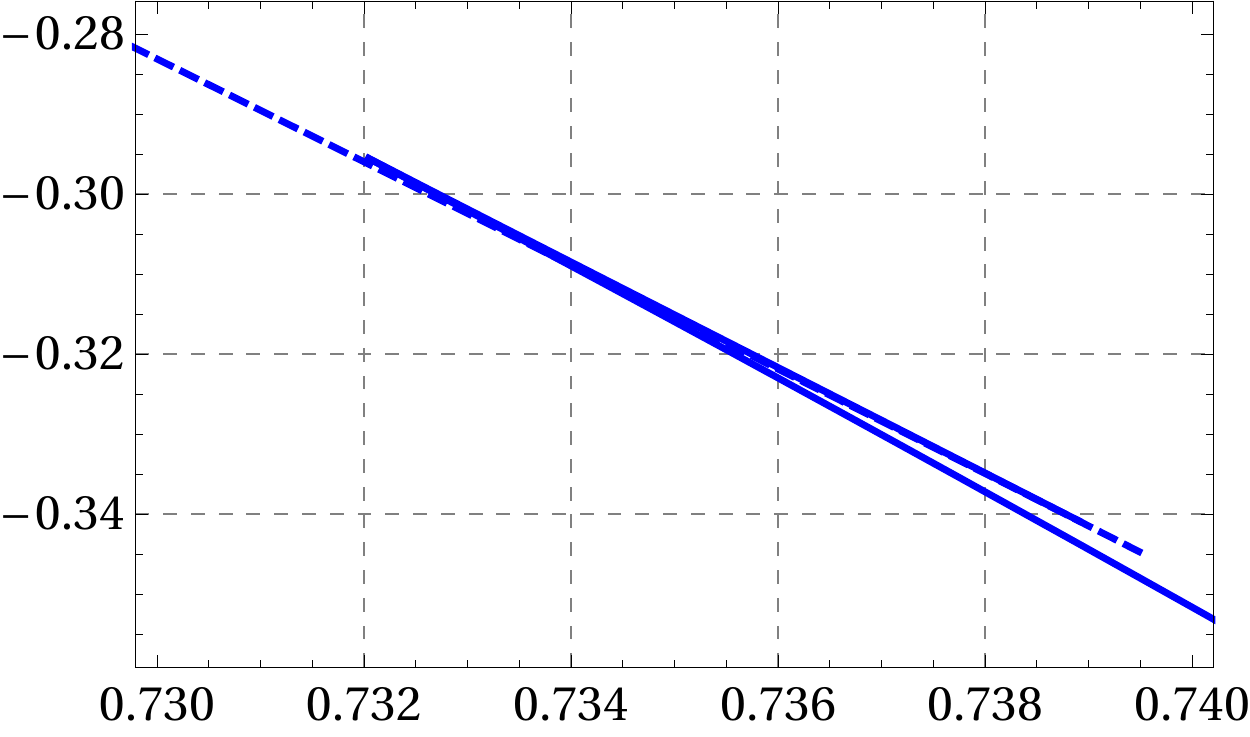}
\qquad
 \put(-450,70){$\frac{\mathcal{F}}{\mathcal{N}\bar{M}^{4}}$}
   \put(-250,-10){$\frac{T}{\bar{M}}$}
    \put(-218,70){$\frac{\mathcal{F}}{\mathcal{N}\bar{M}^{4}}$}
   \put(-18,-10){$\frac{T}{\bar{M}}$}
 \\
(a) & (b)\\
& \\
\includegraphics[width=0.45\textwidth]{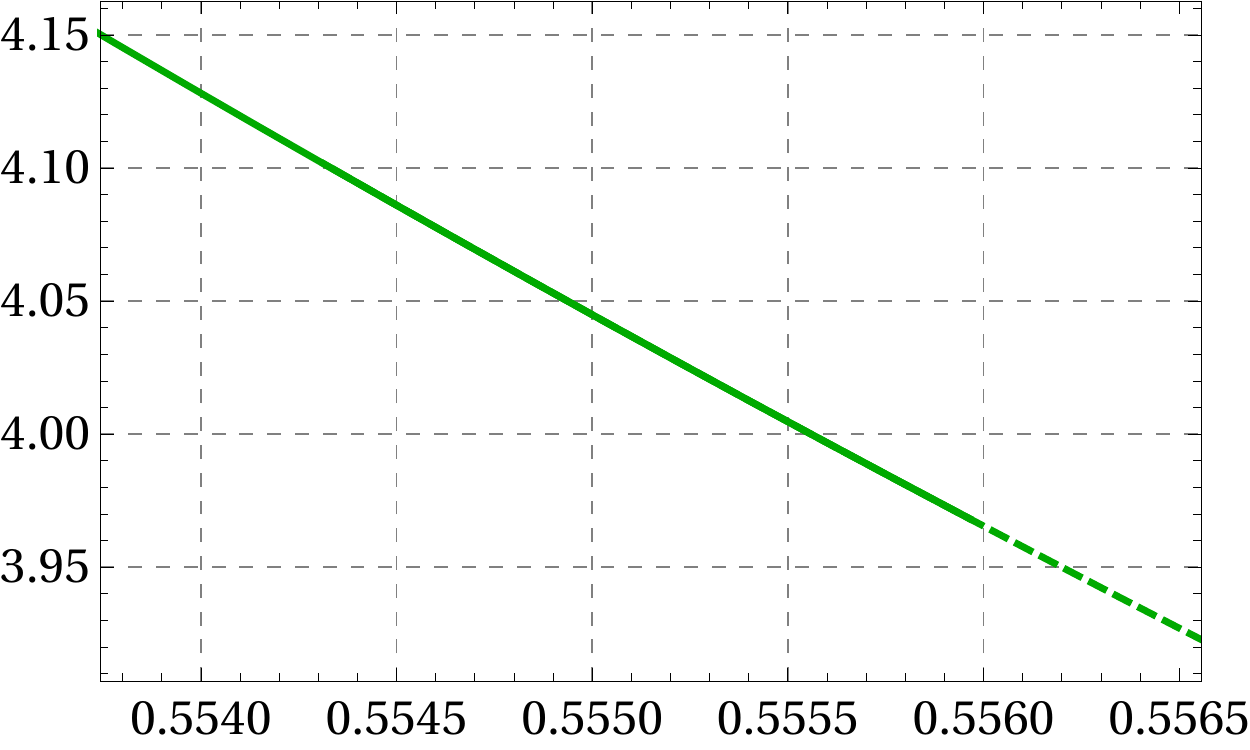} 
\qquad\qquad & 
\includegraphics[width=0.45\textwidth]{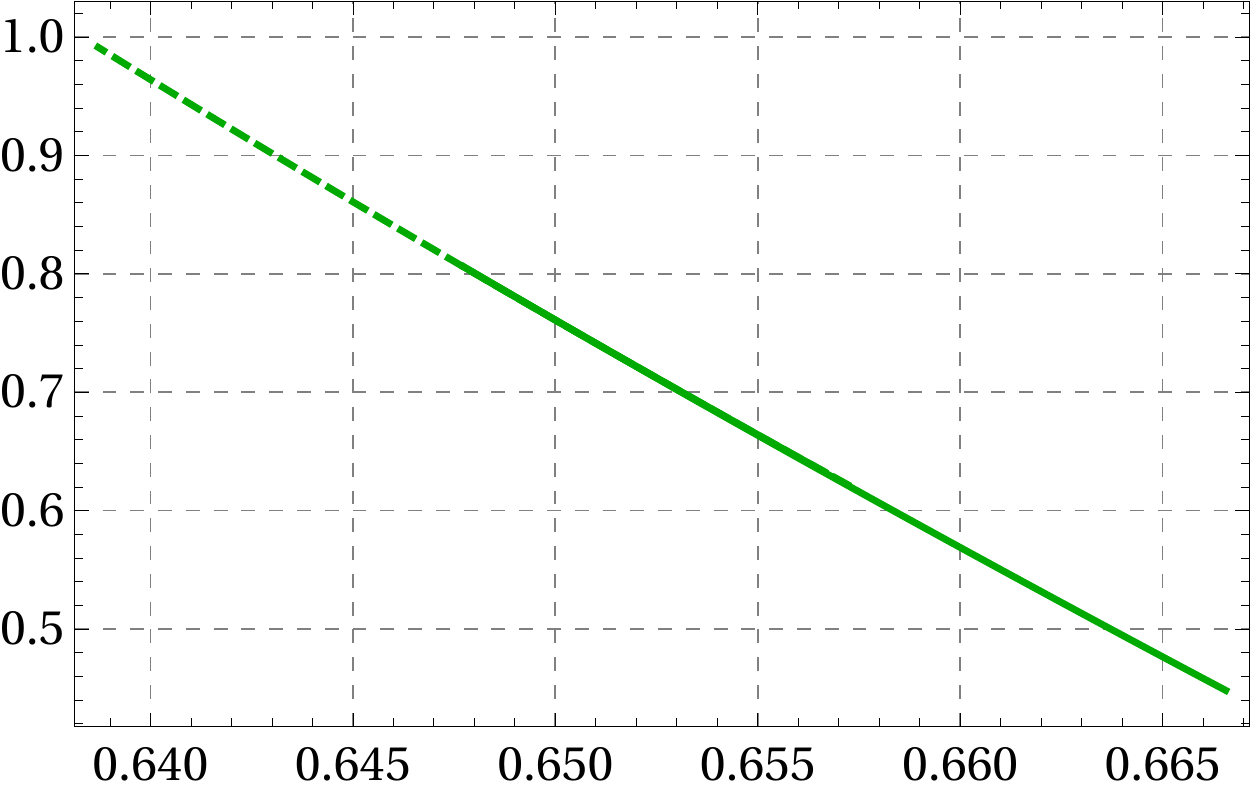}
\qquad
 \put(-450,70){$\frac{\mathcal{F}}{\mathcal{N}\bar{M}^{4}}$}
   \put(-250,-10){$\frac{T}{\bar{M}}$}
    \put(-220,70){$\frac{\mathcal{F}}{\mathcal{N}\bar{M}^{4}}$}
   \put(-18,-10){$\frac{T}{\bar{M}}$}
         \\
(c)& (d) \\
& \\
\includegraphics[width=0.45\textwidth]{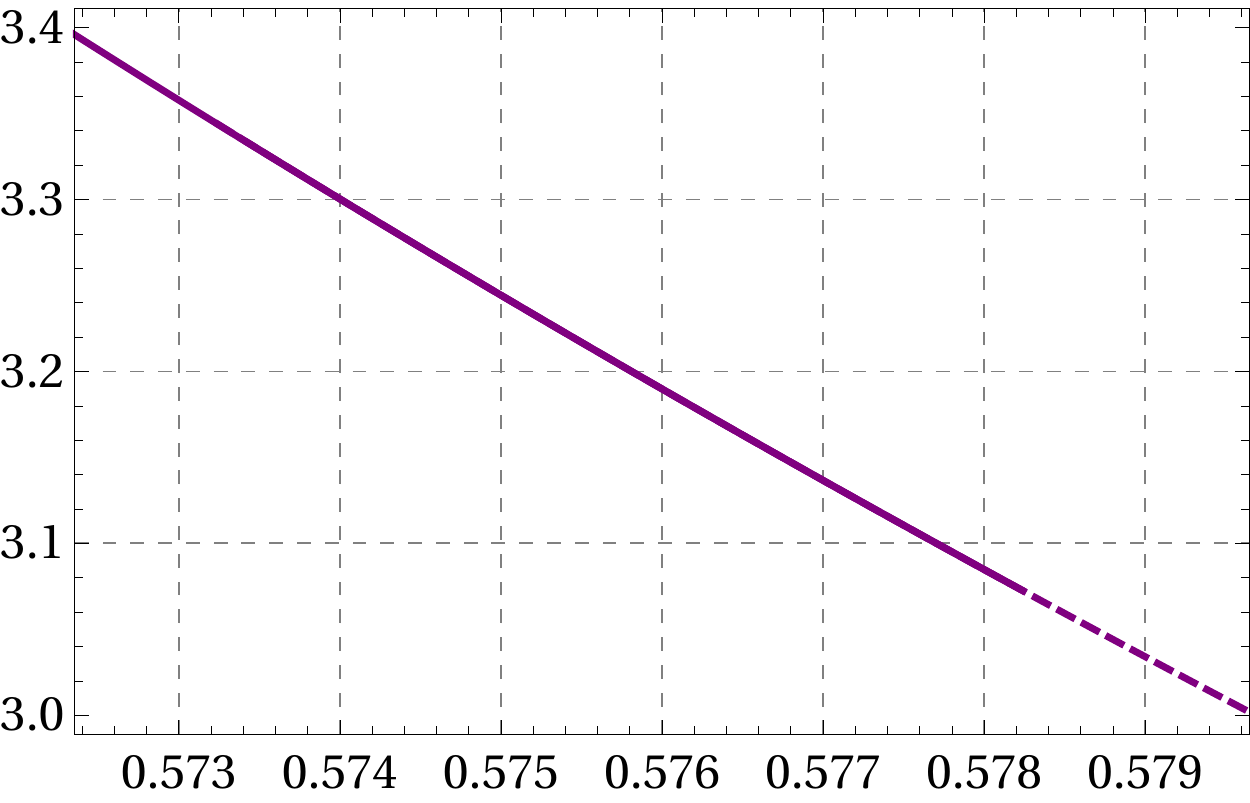} 
\qquad\qquad & 
\includegraphics[width=0.45\textwidth]{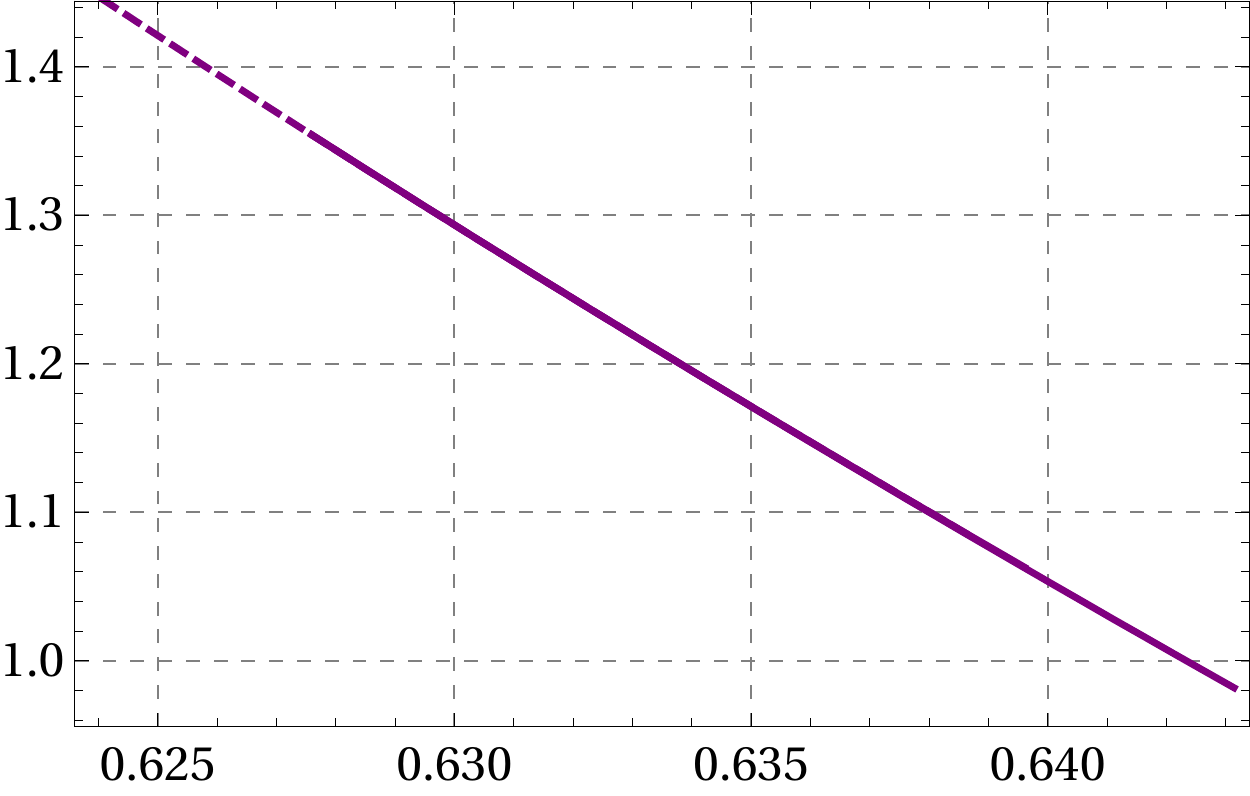}
\qquad
 \put(-450,70){$\frac{\mathcal{F}}{\mathcal{N}\bar{M}^{4}}$}
   \put(-250,-10){$\frac{T}{\bar{M}}$}
    \put(-220,70){$\frac{\mathcal{F}}{\mathcal{N}\bar{M}^{4}}$}
   \put(-18,-10){$\frac{T}{\bar{M}}$}
         \\
(e) & (f) 
\end{tabular}
\end{center}
\caption{\small Zoom into the transition region between Minkowski and black hole embeddings for the free energy density. The values of the magnetic field are $b/\bar{M}^{2}=$ 0 (a), 2 (b), 3.3 (c) and (d), and 3.4 (e) and (f). Dashed lines represent Minkowski embeddings, whereas solid lines represent black hole embedding.}
\label{Free_bM2_Zooms}
\end{figure}

In Fig. (\ref{Free_bM2}) we show $\mathcal{F}/\mathcal{N}\bar{M}^{4}$ as a function of $T/\bar{M}$ for different values for $b/\bar{M}^{2}$ at $C_{2}=0$. Note that the free energy seemingly diverges as $T/\bar{M}$ aproaches its minimum value for the given magnetic field, displaying the same behavior as the quark condensate. As in that case, the explanation is that the background itself is expected to undergo a phase transition at that point. In Fig. (\ref{Free_bM2_Zooms}) we zoom over the parameter region at which the transition takes place. For $b/\bar{M}^{2}=0$ the free energy displays the ``swallow tail" shape characteristic of a first order phase transition (Fig. (\ref{Free_bM2_Zooms}) (a)), which agrees with the findings in \cite{Mateos:2007vn}. The same behavior is observed for $b/\bar{M}^{2}=2$ (Fig. (\ref{Free_bM2_Zooms}) (b)). For $b/\bar{M}^{2}=3.3$ and $b/\bar{M}^{2}=3.4$ the the swallow tail seems to disappear both for the cold (Fig. (\ref{Free_bM2_Zooms}) (c) and (e)) and hot (Fig. (\ref{Free_bM2_Zooms}) (d) and (f)) phase transitions. However, from our analysis of the quark condensate we know that this is still a first order phase transition, even if the free energy difference between both phases is very small.

Unlike the quark condensate, the free energy does depend on the dimensionless quantity $\mu/\bar{M}$. However, changing $\mu/\bar{M}$ modifies the curves only by an additive constant, and so it does not change the dependence of the critical temperature on the magnetic field nor the order of the phase transition. The plots shown in Fig. (\ref{Free_bM2}) and (\ref{Free_bM2_Zooms}) are traced at $\mu/\bar{M}=1$.
 
\begin{figure}[ht!]
 \centering
 \includegraphics[width=0.75\textwidth]{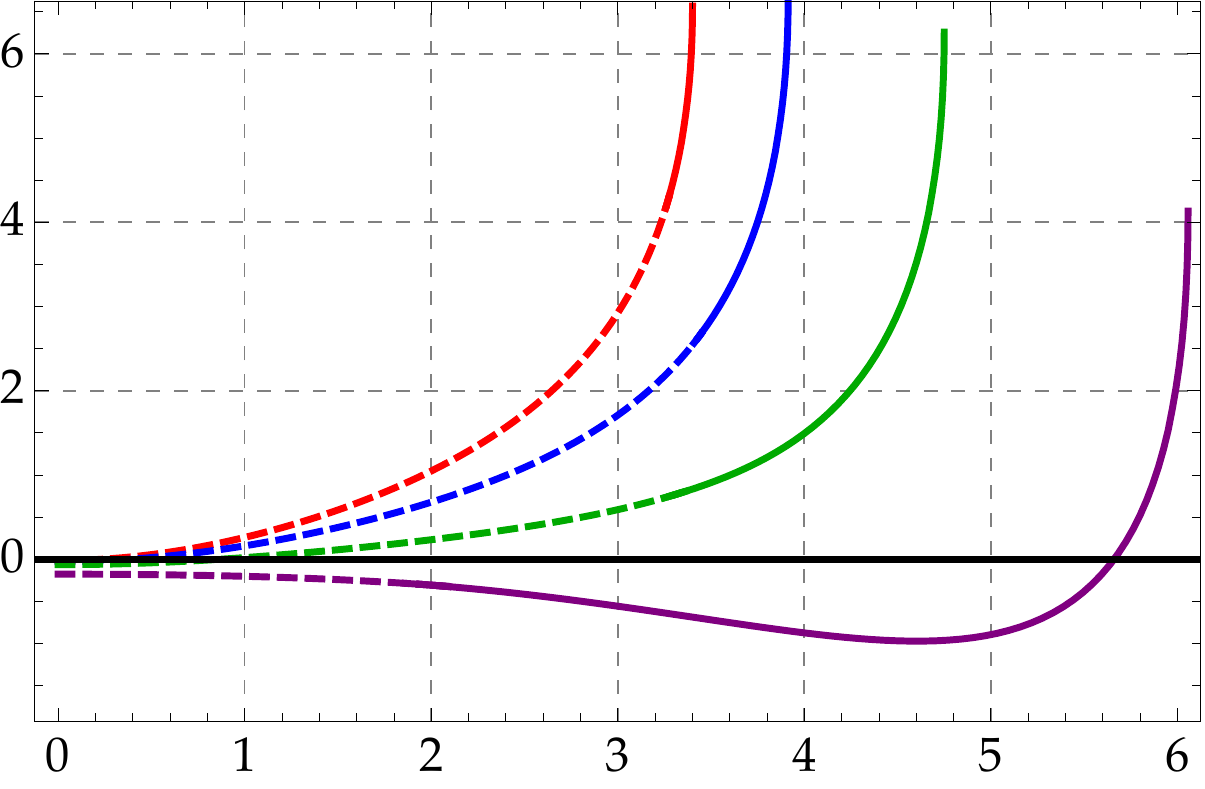}
 \put(-30,-10){\Large $\frac{b}{\bar{M}^{2}}$}
 \put(-353,200){\Large $\frac{\mathcal{F}}{\mathcal{N}\bar{M}^{4}}$}
\caption{\small Free energy $\mathcal{F}/\mathcal{N}\bar{M}^{4}$ as a function of $b/\bar{M}^{2}$. Red, blue, green, and purple curves (top to bottom) correspond to $T/\bar{M}=\lbrace 0.55,0.59,0.65,0.734\rbrace$ respectively. The dashed segments correspond to Minkowski embeddings, while the continuous segments correspond to black hole embeddings.}
\label{Free_TM}
\end{figure}

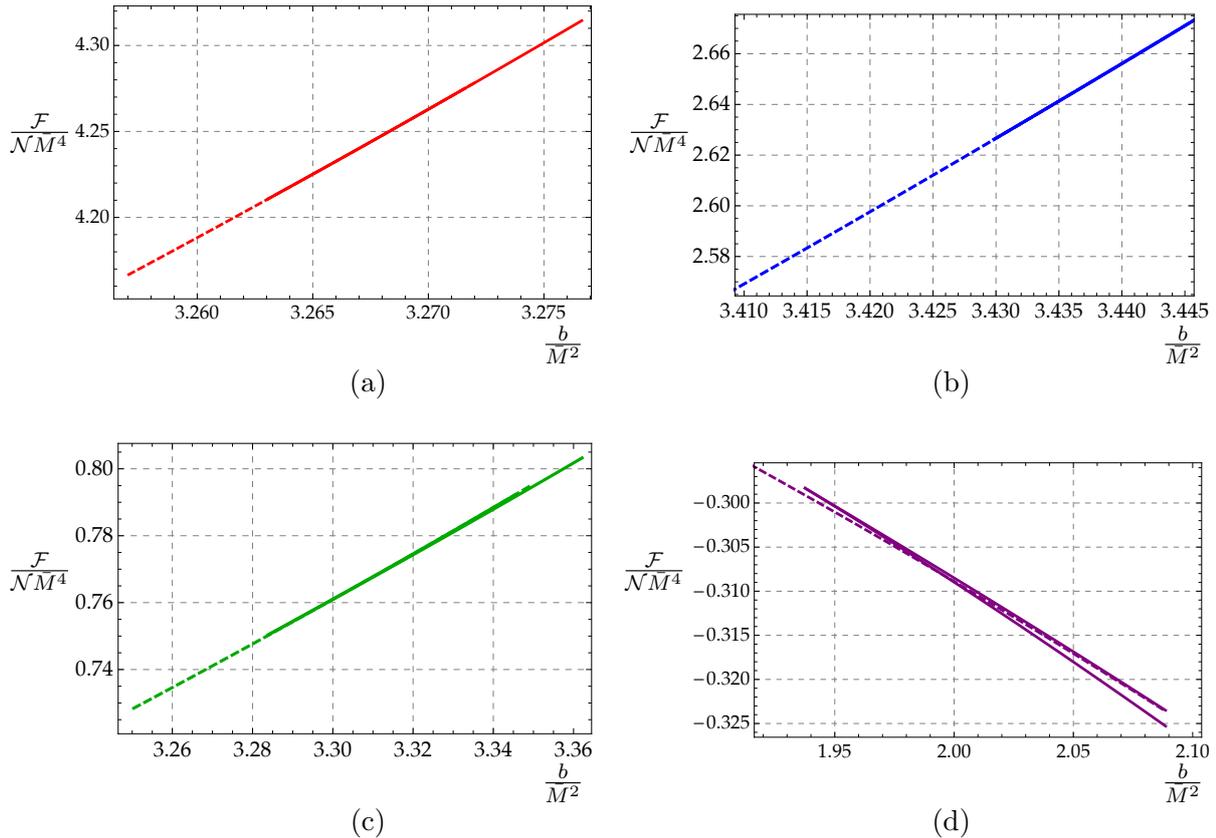
\begin{figure}
\begin{center}
\begin{tabular}{cc}
\includegraphics[width=0.45\textwidth]{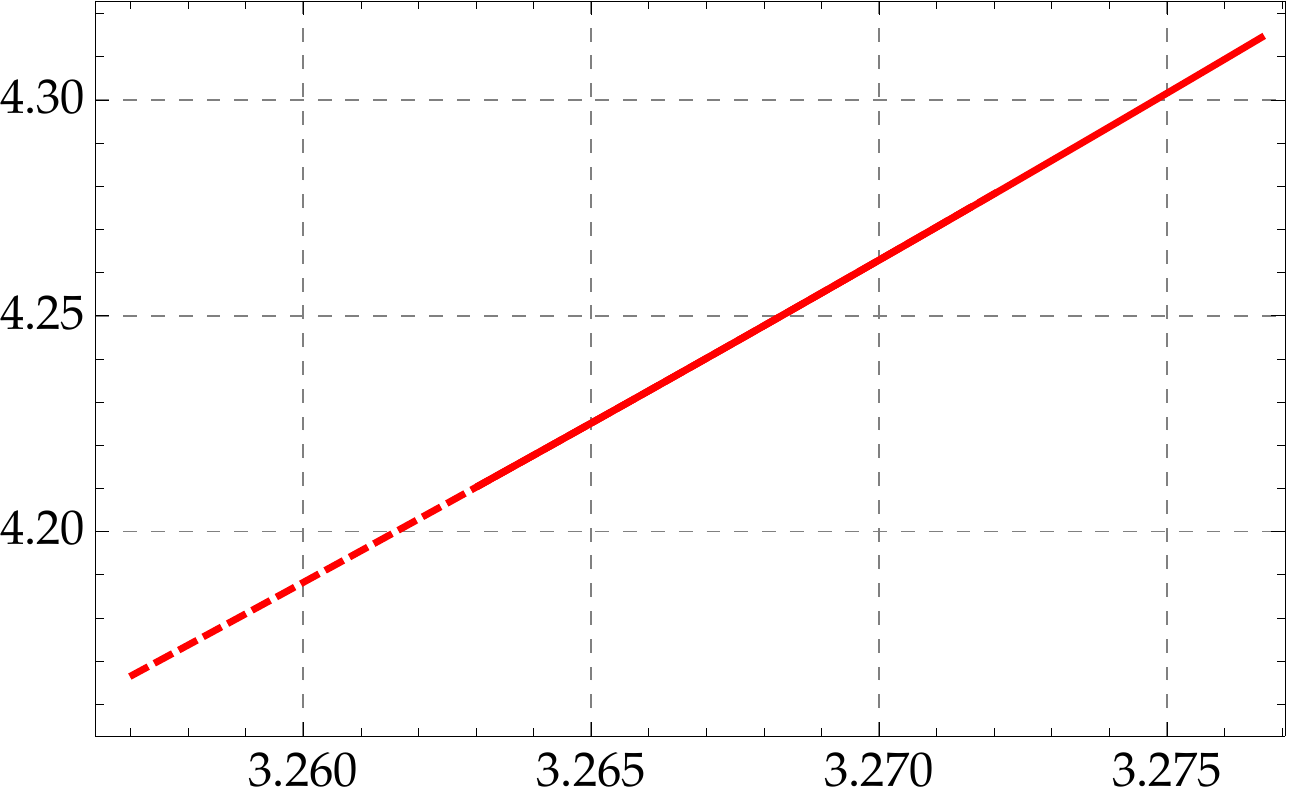} 
\qquad\qquad & 
\includegraphics[width=0.45\textwidth]{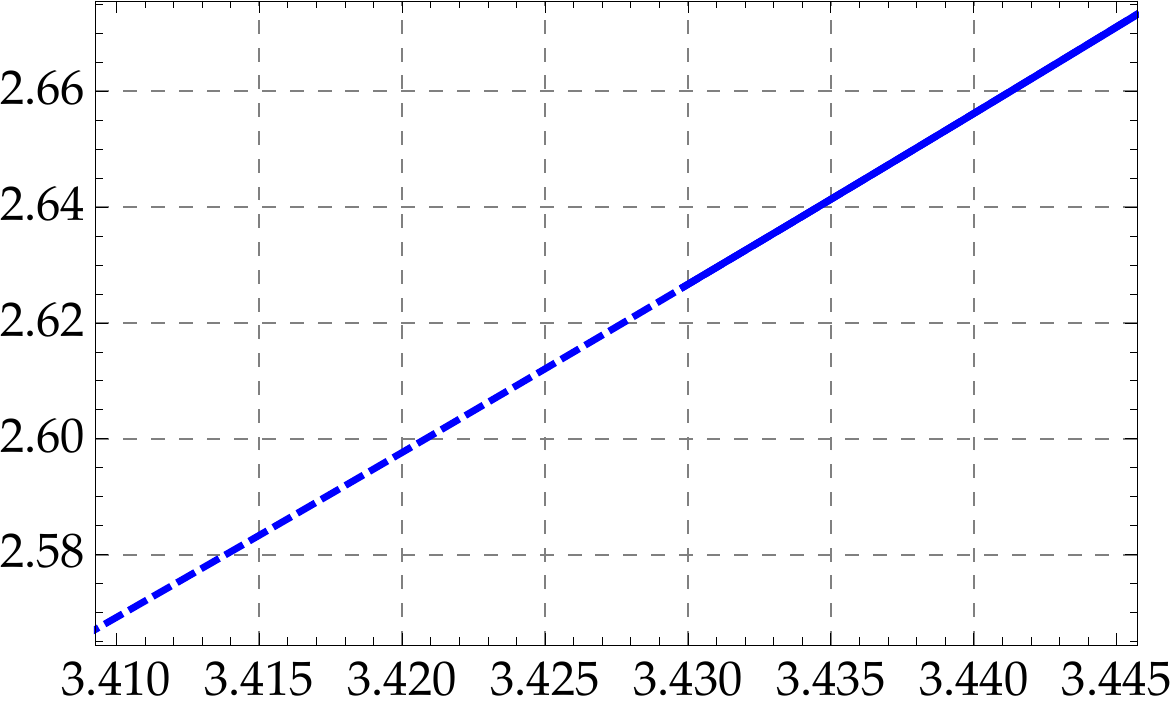}
\qquad
 \put(-450,70){$\frac{\mathcal{F}}{\mathcal{N}\bar{M}^{4}}$}
   \put(-250,-10){$\frac{b}{\bar{M}^{2}}$}
    \put(-218,70){$\frac{\mathcal{F}}{\mathcal{N}\bar{M}^{4}}$}
   \put(-18,-10){$\frac{b}{\bar{M}^{2}}$}
 \\
(a) & (b)\\
& \\
\includegraphics[width=0.45\textwidth]{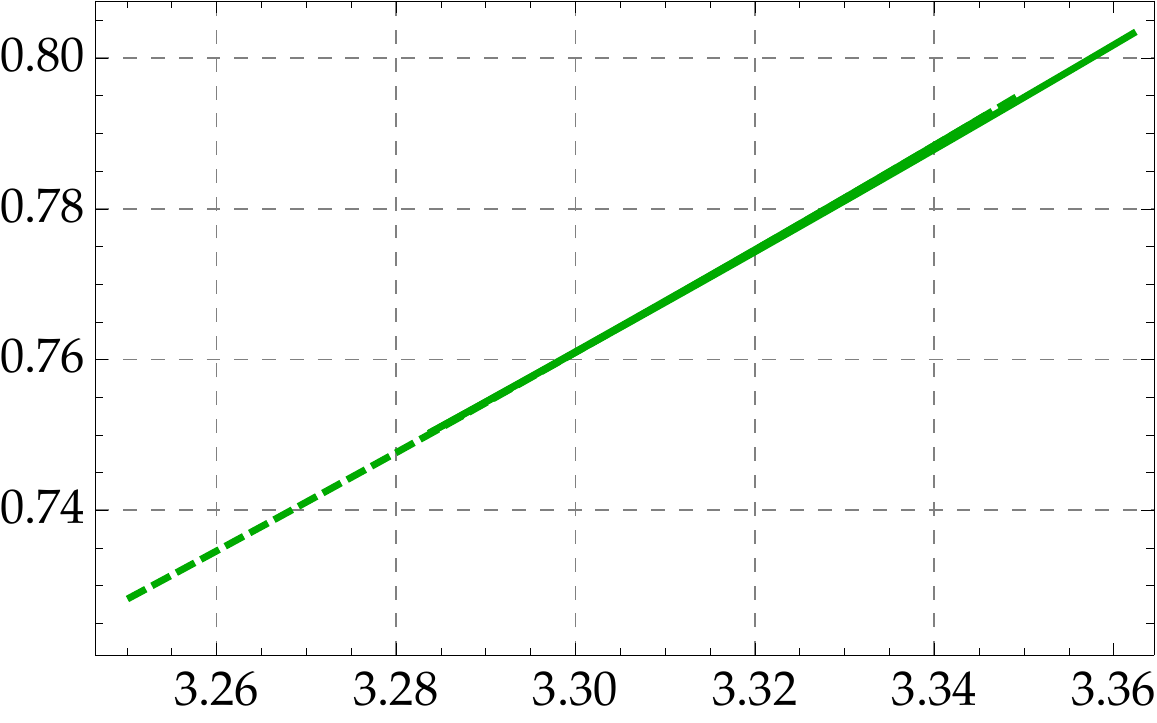} 
\qquad\qquad & 
\includegraphics[width=0.45\textwidth]{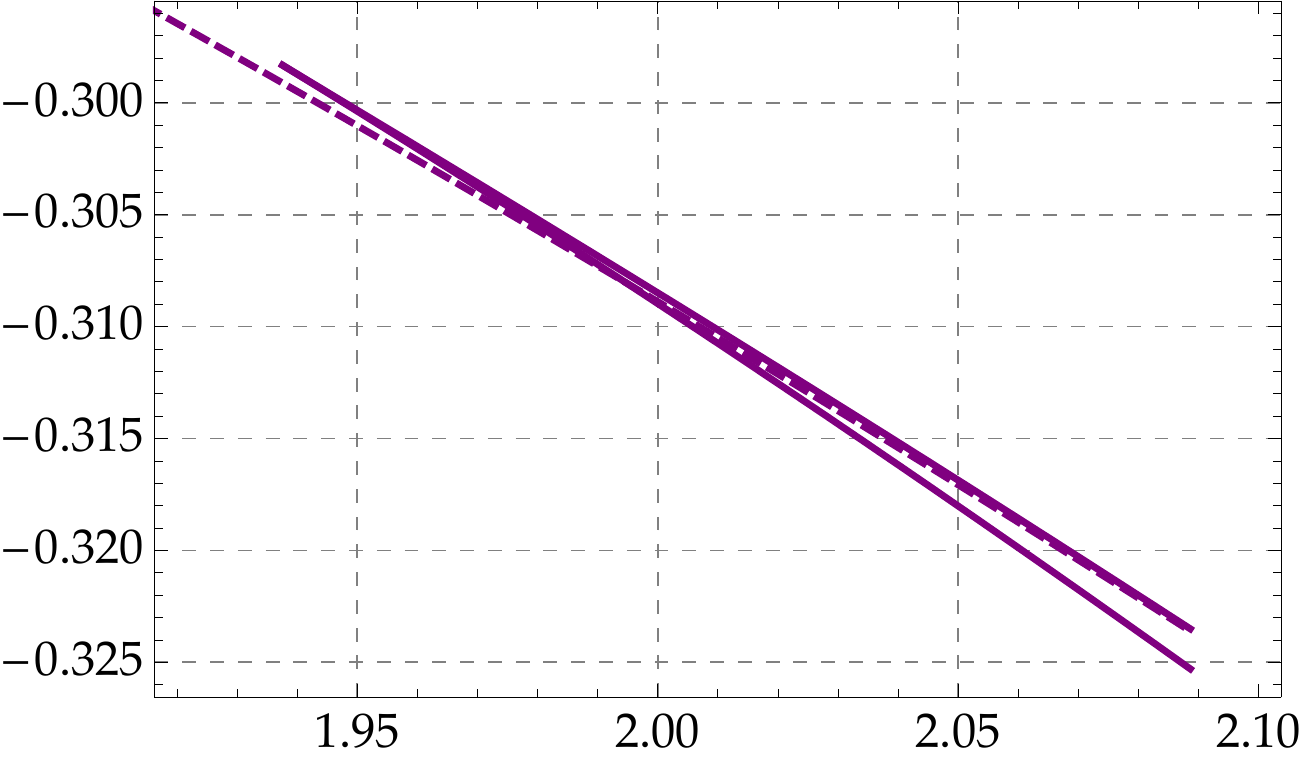}
\qquad
 \put(-450,70){$\frac{\mathcal{F}}{\mathcal{N}\bar{M}^{4}}$}
   \put(-250,-10){$\frac{b}{\bar{M}^{2}}$}
    \put(-220,70){$\frac{\mathcal{F}}{\mathcal{N}\bar{M}^{4}}$}
   \put(-18,-10){$\frac{b}{\bar{M}^{2}}$}
         \\
(c) & (d) 
\end{tabular}
\end{center}
\caption{\small Zoom into the transition region between Minkowski and black hole embeddings for the free energy density. The values of the temperature are $T/\bar{M}=$ 0.55 (a), 0.59 (b), 0.65 (c), and 0.734 (d). Dashed lines represent Minkowski embeddings, whereas solid lines represent black hole embeddings.}
\label{Free_TM_Zooms}
\end{figure}

We also show the free energy as a function of $b/\bar{M}^{2}$ for four different values of $T/\bar{M}$ in Fig.(\ref{Free_TM}). The values of $T/\bar{M}$ displayed are in the range in which the phase transition can be triggered by changing the magnetic field intensity. Again, we note that the free energy seemengly diverges as $b/\bar{M}^{2}$ aproaches its maximim value for the given temperature. In Fig. (\ref{Free_TM_Zooms}) we zoom over the parameter region at which the phase transition takes place. This, along with our findings for the quark condensate, confirms that the transition is of first order.

\subsection{Entropy and energy densities}
\begin{figure}[ht!]
 \centering
 \includegraphics[width=0.85\textwidth]{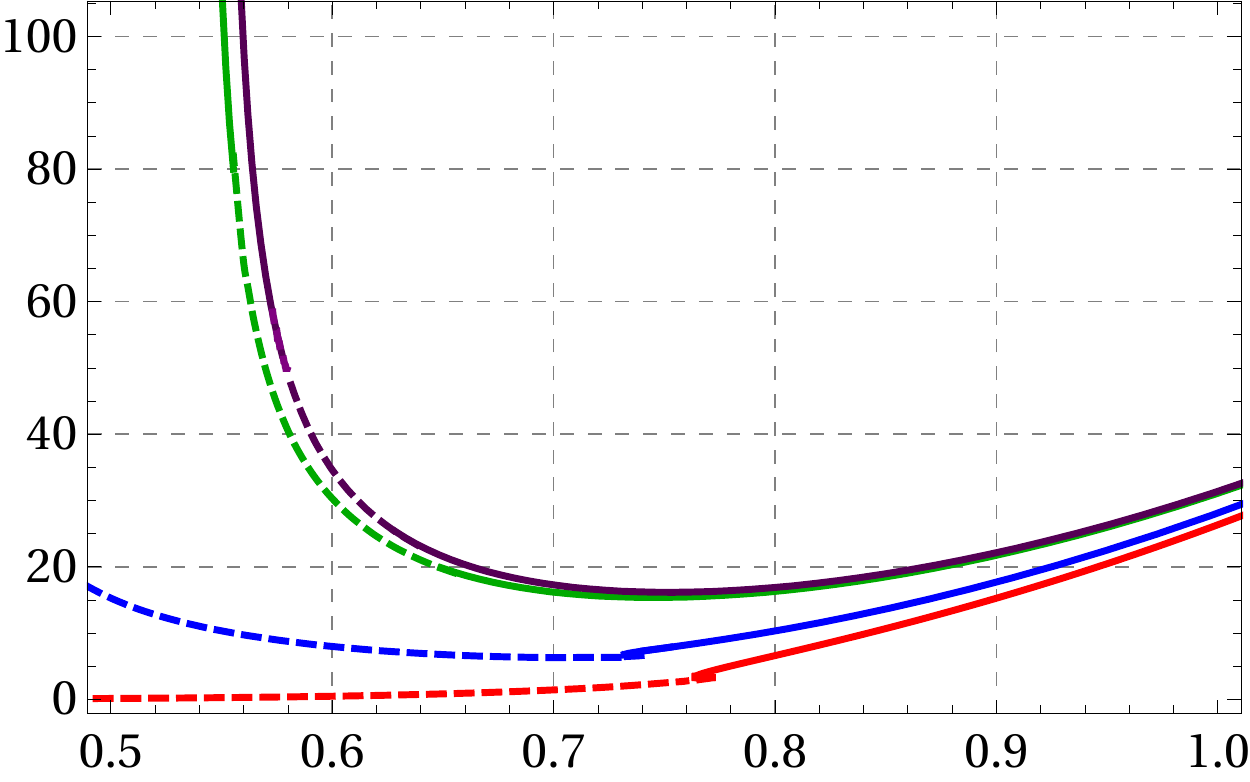}
 \put(0,-10){\Large $\frac{T}{\bar{M}}$}
 \put(-395,220){\Large $\frac{s}{\mathcal{N}\bar{M}^{3}}$}
\caption{\small Entropy density $s/\mathcal{N}\bar{M}^{3}$ as a function of $T/\bar{M}$. Red, blue, green, and purple curves (bottom to top) correspond to $b/\bar{M}^{2}=\lbrace0,2,3.3,3.4\rbrace$ respectively. The dashed segments correspond to Minkowski embeddings, while the continuous segments correspond to black hole embeddings.}
\label{Entropy_bM2}
\end{figure}

The entropy is given by the derivative of the free energy with respect to the temperature at fixed magnetic field and quark mass
\begin{equation}
s=-\left(\frac{\partial \mathcal{F}}{\partial T}\right)_{b,M_{q}}.
\label{entropy}
\end{equation}
As explained in Section \ref{Background}, we don't have direct numerical control over $b$ and $\bar{M}$, as both quantities need to be extracted once a numerical solution is known. In consequence, we computed the derivative in \eqref{entropy} numerically, as opossed to the analytical aproach followed in \citep{Mateos:2007vn}. We present the result of said numerical derivative in Fig. (\ref{Entropy_bM2}), where it can be seen that the general effect of the magnetic field is to increase the entropy, becoming infinite at the minimum temperature. It is important to note that the entropy, like the quark condensate, does not depend on the specific value of $\mu/\bar{M}$. This is also realized in the fact that the term proportional to $C_{2}$ in the free energy vanishes when taking the derivative \eqref{entropy}.

\begin{figure}[ht!]
 \centering
 \includegraphics[width=0.85\textwidth]{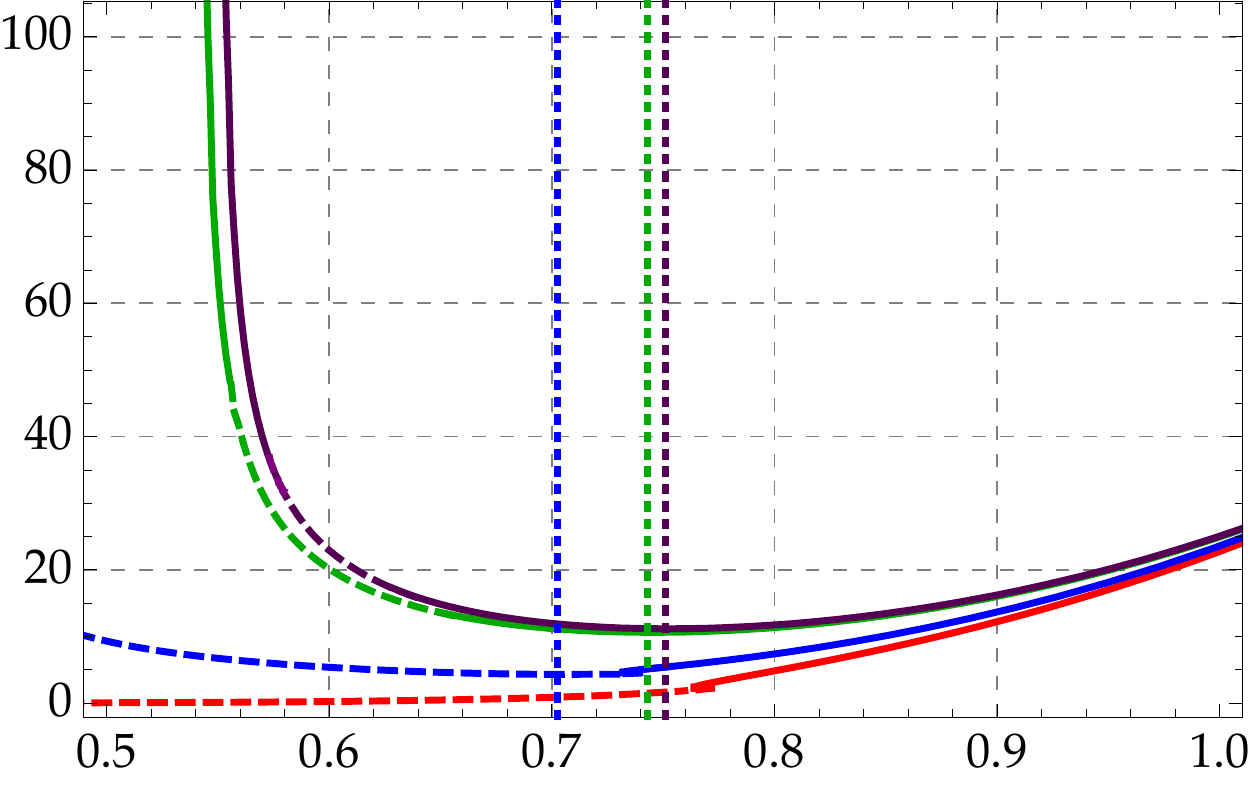}
 \put(0,-10){\Large $\frac{T}{\bar{M}}$}
 \put(-395,220){\Large $\frac{E}{\mathcal{N}\bar{M}^{4}}$}
\caption{\small Energy density $E/\mathcal{N}\bar{M}^{4}$ as a function of $T/\bar{M}$. Red, blue, green, and purple curves (bottom to top) correspond to $b/\bar{M}^{2}=\lbrace0,2,3.3,3.4\rbrace$ respectively. The dashed segments correspond to Minkowski embeddings, while the continuous segments correspond to black hole embeddings. The dotted vertical lines denote the temperature at which the specific heat becomes negative for each magnetic field other than zero ($b/\bar{M}^{2}=\lbrace 2,3.3,3.4\rbrace$ from left to right). The finite term is fixed at $C_{2}=0$.}
\label{Energy_bM2}
\end{figure}

\begin{figure}
\begin{center}
\begin{tabular}{cc}
\includegraphics[width=0.45\textwidth]{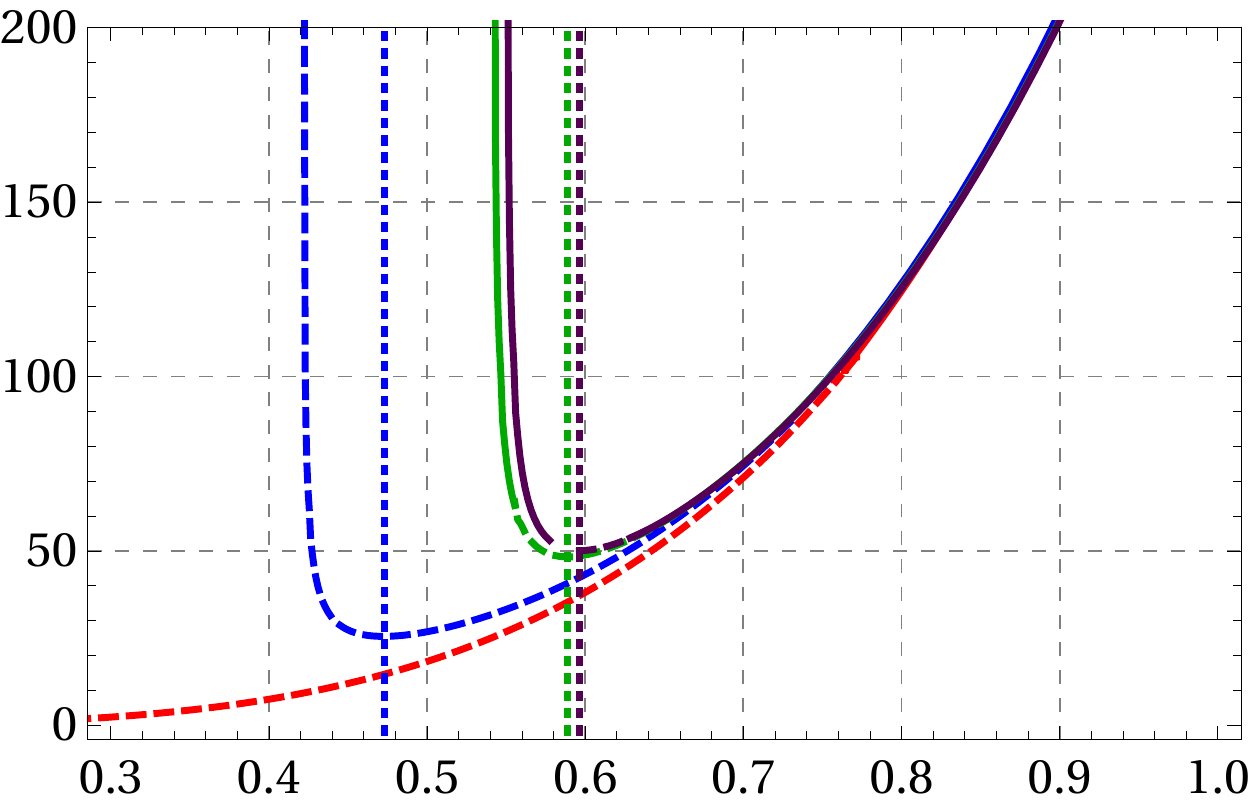} 
\qquad\qquad & 
\includegraphics[width=0.45\textwidth]{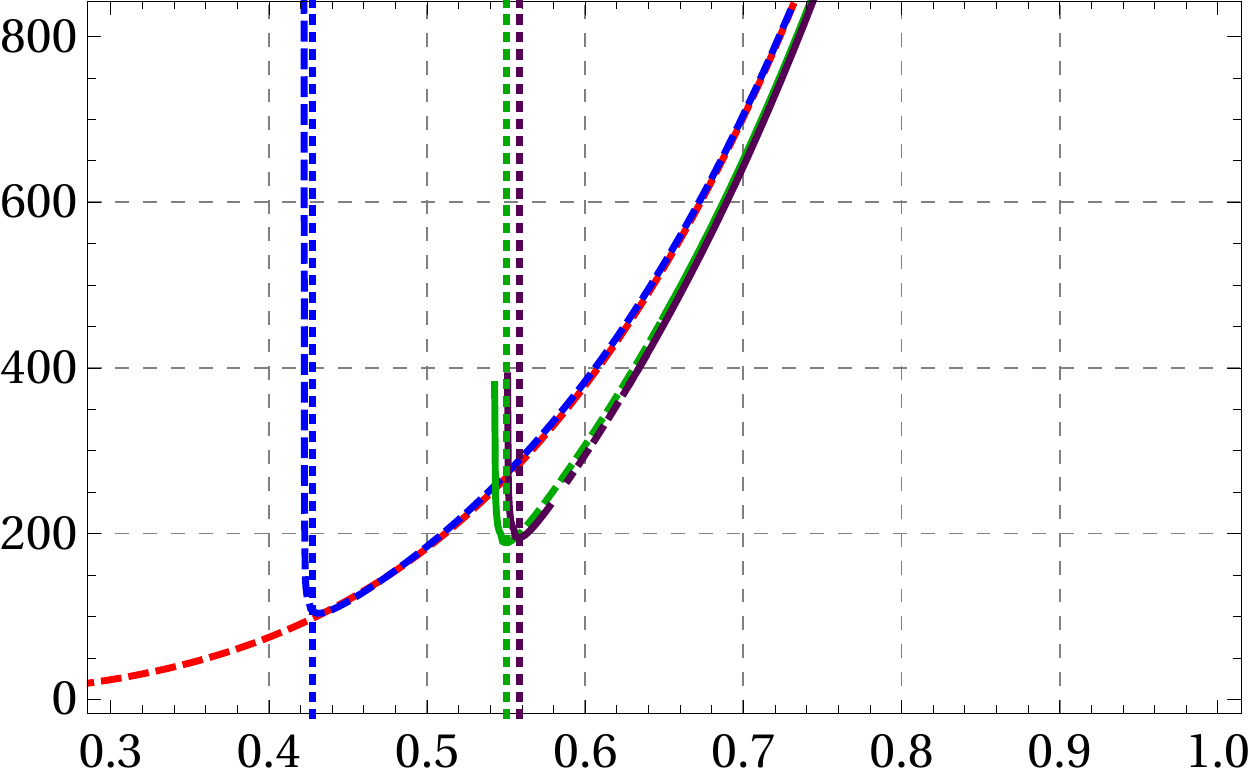}
\qquad
 \put(-450,70){$\frac{E_{total}}{\mathcal{N}\bar{M}^{4}}$}
   \put(-250,-10){$\frac{T}{\bar{M}}$}
    \put(-218,70){$\frac{E_{total}}{\mathcal{N}\bar{M}^{4}}$}
   \put(-18,-10){$\frac{T}{\bar{M}}$}
 \\
(a) & (b)
\end{tabular}
\end{center}
\caption{\small Total energy density $E_{total}/\mathcal{N}\bar{M}^{4}$ of the color and flavor degrees of freedom as a function of $T/\bar{M}$. Red, blue, green, and purple curves (bottom to top) correspond to $b/\bar{M}^{2}=\lbrace0,2,3.3,3.4\rbrace$ respectively. The dashed segments correspond to Minkowski embeddings, while the continuous segments correspond to black hole embeddings. The dotted vertical lines denote the temperature at which the specific heat becomes negative for each magnetic field other than zero ($b/\bar{M}^{2}=\lbrace 2,3.3,3.4\rbrace$ from left to right). (a) corresponds to $N_{c}/N_{f}=100$ while (b) corresponds to $N_{c}/N_{f}=1000$. In both cases we fixed $\lambda=100\pi^{2}$ and $C_{2}=0$.}
\label{TotalEnergy}
\end{figure}

With both the entropy and free energy at hand, we can compute the internal energy by using the simple thermodynamic identity
\begin{equation}
E=\mathcal{F}+Ts.
\label{energy}
\end{equation}
Just like with $\mathcal{F}$, we need to specify the value of $C_{2}$. In particular, the plots in Fig. (\ref{Energy_bM2}) were generated for $C_{2}=0$ and $\mu/\bar{M}=1$. We can also read the qualitative behavior of the specific heat from Fig. (\ref{Energy_bM2}), as it is given by
\begin{equation}
C=\left(\frac{\partial E}{\partial T}\right)_{b,M_{q}}.
\end{equation}
The main conclusion from this is that, as can be seen from the slope of the curves, the specific heat becomes negative at sufficiently low temperatures for any non-vanishing magnetic field. Note that this is independent of the renormalization scheme, as the finite term vanishes when taking the derivative at fixed $b$. The temperature at which the specific heat becomes negative is displayed in Fig. (\ref{Energy_bM2}) as a vertical line. This happens in either phase of the system, as for $b/\bar{M}^{2}=3.3$ and $b/\bar{M}^{2}=3.4$ cases the specific heat becomes negative in the black hole phase, while for $b/\bar{M}^{2}=2$ it occurs in the Minkowski phase. A negative specific heat signalizes a thermodynamic instability, and thus, perturbing the system while in any of this states will cause it to decay into a stable one. However, this affirmation is in regard of the entire system, being flavor and color degrees of freedom. In order to determine whether the system is thermodynamically unstable or not, we need to also consider the contribution of the adjoint fields to the specific heat. From the gravity perspective, this is achieved by including in Fig. (\ref{Energy_bM2}) the energy density of the background itself. We previously computed the energy density of our gravity configurations in \cite{Avila:2018hsi}, where all the details regarding the holographic renormalization are discussed. The energy density of the gravity backgrounds is given by
\begin{equation}
\frac{8\pi^{2}}{N_{c}^{2}\text{vol}(x)}E_{c}=-3U_{4}-\frac{1}{3}\varphi_{0}^{2}-2C_{2}b^{2},
\end{equation}
where $N_{c}$ is the number of colors. We show the total energy density of the system $E_{total}$ in Fig. (\ref{TotalEnergy}) for $N_{c}/N_{f}=100$ in (a) and $N_{c}/N_{f}=1000$ in (b), fixing $C_{2}=0$. We can see that as the limit $N_{f}\ll N_{c}$ is taken, the temperature at which the specific heat of the whole system becomes negative coincides with the minimum temperature at which the background itself exists for each given magnetic field. To provide the visualization above of the effect that the $N_{c}\rightarrow\infty$ limit has on the total energy we used $\lambda=100\pi^{2}$ when producing the plots in Fig. (\ref{TotalEnergy}). Identical plots would be obtained for other choices for $\lambda$ just at different values of $N_{c}$ as it approaches infinity.

\section{Meson spectrum}
\label{spectrum}
From the gauge theory perspective, the phase transition just described takes place between a discrete meson spectrum (Minkowski embeddings) and a gapless distribution of excitations (black hole embeddings). Holographically, the mesons are related to stable excitations over the D7-branes. Thus in order to study the meson spectrum, it is necessary to study the dynamics of perturbations over the Minkowski embeddings.  

Scalar perturbations are the ones related to vibrational modes of the D7-brane itself. A general excitation of this kind can be implemented as
\begin{equation}
\chi(r,t,\vec{x},\Sigma_{3})=\chi^{(0)}(r)+\epsilon \chi^{(1)}(r,t,\vec{x},\Sigma_{3}),
\label{chi_pert}
\end{equation}
\begin{equation}
\phi(r,t,\vec{x},\Sigma_{3})=\phi^{(0)}(r)+\epsilon \phi^{(1)}(r,t,\vec{x},\Sigma_{3}),
\label{phi_pert}
\end{equation}
where $\vec{x}$ denotes the gauge theory spatial directions $(x,y,z)$, $\Sigma_{3}$ denotes the coordinates of the 3-cycle $(\psi,\vartheta_{1},\vartheta_{2})$, $\epsilon\ll 1$ is a dimensionless parameter and $\chi^{(0)}$ and $\phi^{(0)}$ are the exact Minkowski solutions discussed earlier.

In order to study these perturbations we substitute \eqref{chi_pert} and \eqref{phi_pert} in the most general EOM coming from the DBI action and solve them at first order in $\epsilon$. This reveals that the $\chi^{(1)}$ and $\phi^{(1)}$ perturbations decouple and can be solved separately. We look for separable solutions of the form
\begin{equation}
\chi^{(1)}(r,t,\vec{x},\Sigma_{3})=\chi_{r}(r)e^{i(\omega t-\vec{x}\cdot\vec{k})}F(\Sigma_{3}),
\label{chi_sep}
\end{equation}
\begin{equation}
\phi^{(1)}(r,t,\vec{x},\Sigma_{3})=\phi_{r}(r)e^{i(\omega t-\vec{x}\cdot\vec{k})}G(\Sigma_{3}),
\label{phi_sep}
\end{equation}
where $\omega$ and $\vec{k}$ are dual to the energy and momentum of the mesons. Given that Lorentz invariance is broken by the finite temperature of the plasma, the notion of meson mass is frame dependent. We will work with the same convention as \cite{Mateos:2007vn} and define the meson mass in its rest-frame given by $\omega$ at $\vec{k}=0$. 

By substituting \eqref{chi_sep} and \eqref{phi_sep} in their respective equations of motion, we obtain two equations, one for $\chi_{r}$ and one for $\phi_{r}$, that only depend on the radial coordinate $r$, and a third equation that both $F(\Sigma_{3})$ and $G(\Sigma_{3})$ need to satisfy
\begin{equation}
\nabla_{S^{3}}^{2}F+\frac{2b^{2}x^{2}}{VX\Delta}(1-\chi^{(0)})(\partial_{\vartheta_{1}}^{2}F+2\partial_{\vartheta_{1}}\partial_{\vartheta_{2}}F+\partial_{\vartheta_{2}}^{2}F)=-\lambda F,
\label{angular_ec}
\end{equation}
where $\nabla_{S^{3}}^{2}$ denotes the Laplace operator on the 3-sphere. Note however that this equation not only depends on the 3-sphere coodinates, but also on the radial coordinate and one of the gauge theory coordinates by the presence of the second term. This was not a problem for vanishing magnetic field, as the second term on the right hand side of \eqref{angular_ec} is eliminated if $b=0$, and the expression reduces to Laplace equation on the 3-sphere as in \cite{Kruczenski:2003be,Mateos:2007vn,Hoyos:2006gb}. Thus the solutions of \eqref{angular_ec} for $b=0$ are the spherical harmonics $\mathcal{Y}^{l}_{m_{1},m_{2}}$ given by \cite{LachiezeRey:2005hs,Achour:2015zpa} 
\begin{equation}
\mathcal{Y}^{l}_{m_{1},m_{2}}=C^{l}_{m_{1},m_{2}}(e^{i\vartheta_{1}}\cos\psi)^{m_{2}+m_{1}}(e^{i\vartheta_{2}}\sin\psi)^{m_{2}-m_{1}}P^{(m_{2}-m_{1},m_{2}+m_{1})}_{\frac{l}{2}-m_{2}}(\cos 2\psi),
\label{spherical_harmonics}
\end{equation}
where $C^{l}_{m_{1},m_{2}}$ is a normalization constant and $P$ is a Jacobi polynomial. This spherical harmonics transform in the $(\frac{l}{2},\frac{l}{2})$ representation of $SO(4)$ and satisfy
\begin{equation}
\nabla_{S^{3}}^{2}\mathcal{Y}^{l}_{m_{1},m_{2}}=-l(l+2)\mathcal{Y}^{l}_{m_{1},m_{2}}.
\end{equation}
The quantum numbers $(l, m_{1}, m_{2})$ are such that $l\in\mathbb{N}$ and $|m_{1,2}|\leq l/2$ vary independently on integer steps, hence taking integer or half-integer values according to the parity of $l$. For any non-vanishing magnetic field both terms on the left hand side of equation \eqref{angular_ec} are present, and hence the only way that $F$, as a function of the angular coordinates exclusively, can be a solution is to satisfy
\begin{equation}
(\partial_{\vartheta_{1}}^{2}+2\partial_{\vartheta_{1}}\partial_{\vartheta_{2}}+\partial_{\vartheta_{2}}^{2})F=0
\label{constriccion}
\end{equation}
on top of Laplace equation. By plugging \eqref{spherical_harmonics} in \eqref{constriccion} we conclude that this can only happen if $F$ is a spherical harmonic with $m_{2}=0$. Given the range of values that this quantum number can take, this means that the presence of the magnetic field limits the angular momentum $l$ of the perturbations on the 3-sphere to take only even values.

The remaining equations\footnote{Both equations are rather lenghty and not very iluminating, so we will ommit them.} for $\chi_{r}$ and $\phi_{r}$ are two independent ordinary second order differential equations, which depend on the frequency $\omega$ and the angular momentum on the 3-sphere $l$. By solving said equations near the boundary we obtain that the behavior of a generic solution is of the form
\begin{equation}
\chi_{r}(r)=Ar^{-1+l}+Br^{-3-l},
\end{equation}
\begin{equation}
\phi_{r}(r)=Cr^{l}+Dr^{-2-l},
\end{equation}
where $A$, $B$, $C$, and $D$ are free coefficients that can be read once a particular numerical solution its known. We are looking for the solutions that remain normalizable when we take the limit $r\rightarrow\infty$, which are the ones for which $A=0$ and $C=0$ \cite{Kruczenski:2003be,Mateos:2007vn,Hoyos:2006gb}.

The numerical procedure to solve the equations for $\chi_{r}$ and $\phi_{r}$ is analogous to the one used in the previous sections. Given that the Minkowski embeddings are the ones for which $\theta=\pi/2$ for a $r_{i}>r_{h}$, the equation of motion is degenerated at $r_{i}$. Thus, we first solve the equations by a power series method around $r_{i}$
\begin{equation}
\chi_{r}(r)=\chi_{r_{i}}+\sum_{j=1}^{\infty}a_{j}(r-r_{i})^{j},
\label{chi_pert_exp}
\end{equation}
\begin{equation}
\phi_{r}(r)=\phi_{r_{i}}+\sum_{j=1}^{\infty}b_{j}(r-r_{i})^{j},
\label{phi_pert_exp}
\end{equation}
which allows to solve for the $a_{j}$ and $b_{j}$ coefficients in terms of the value that the metric functions, the scalar field and the unperturbed Minkowski embedding functions take at $r_{i}$, as well as the frequency $\omega$ and the angular momentum $l$. Using this as initial conditions we integrate the equations for $\chi_{r}$ and $\phi_{r}$ numerically from $r=r_{i}+\epsilon$ to the boundary at $r\rightarrow\infty$. For fixed $T$, $b$ and $\bar{M}$, we then look for the frequencies that corresponds to normalizable solutions. This discrete set $\omega_{n,l,m_{1},m_{2}}$ corresponds to the meson masses, where $n=0,1,2,...$ denotes the radial quantum number.

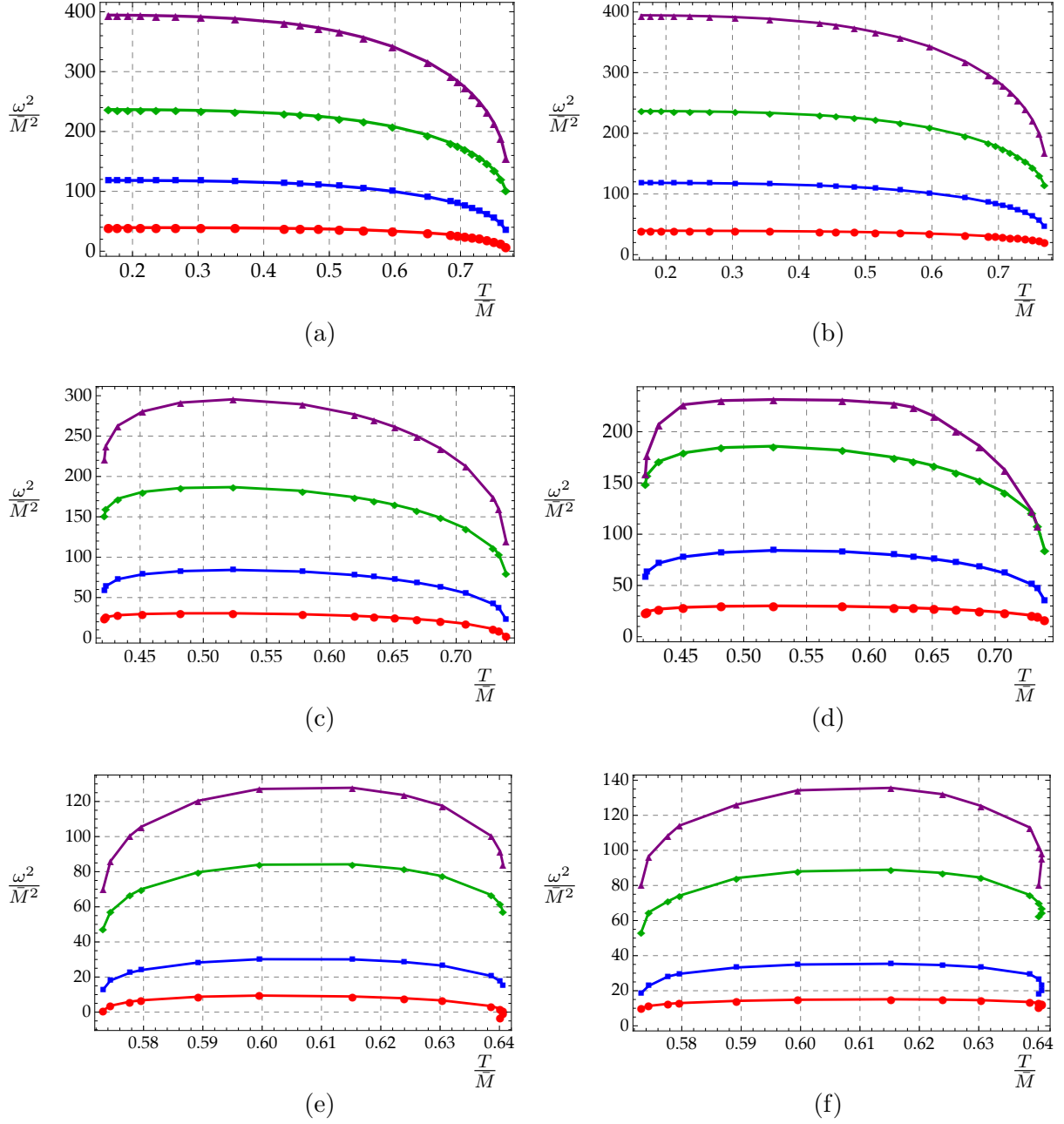
\begin{figure}
\begin{center}
\begin{tabular}{cc}
\includegraphics[width=0.45\textwidth]{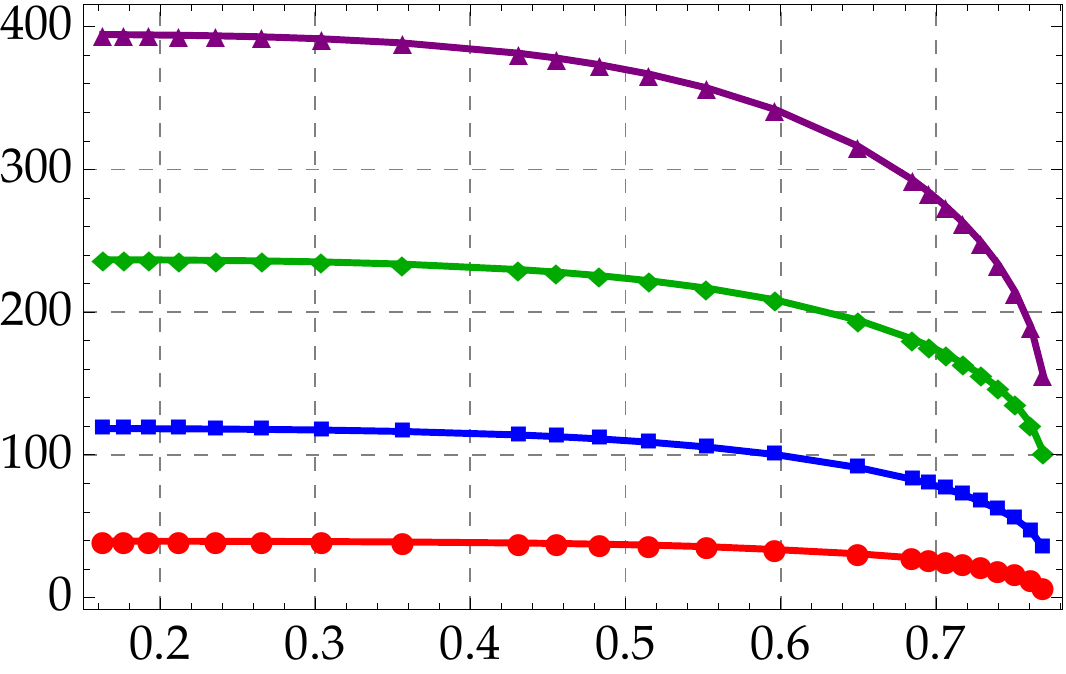} 
\qquad\qquad & 
\includegraphics[width=0.45\textwidth]{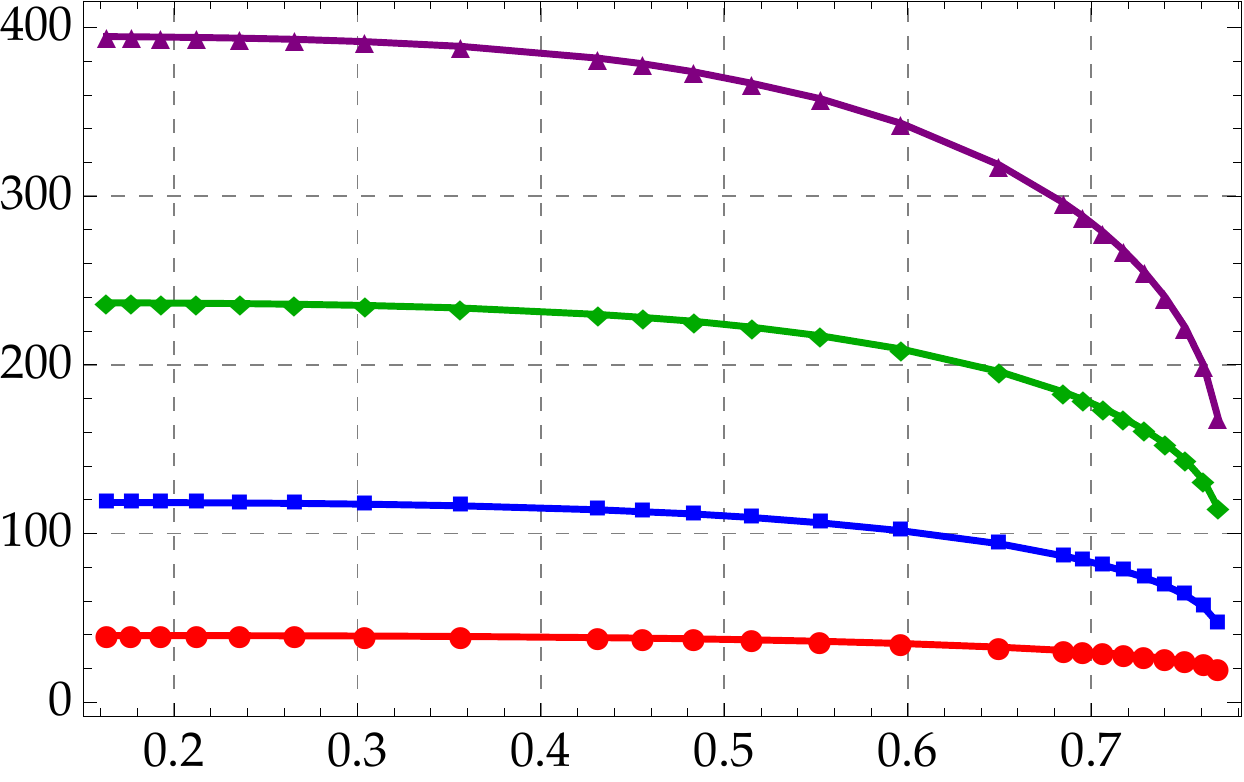}
\qquad
 \put(-450,70){$\frac{\omega^{2}}{\bar{M}^{2}}$}
   \put(-250,-10){$\frac{T}{\bar{M}}$}
    \put(-218,70){$\frac{\omega^{2}}{\bar{M}^{2}}$}
   \put(-18,-10){$\frac{T}{\bar{M}}$}
 \\
(a) & (b)\\
& \\
\includegraphics[width=0.45\textwidth]{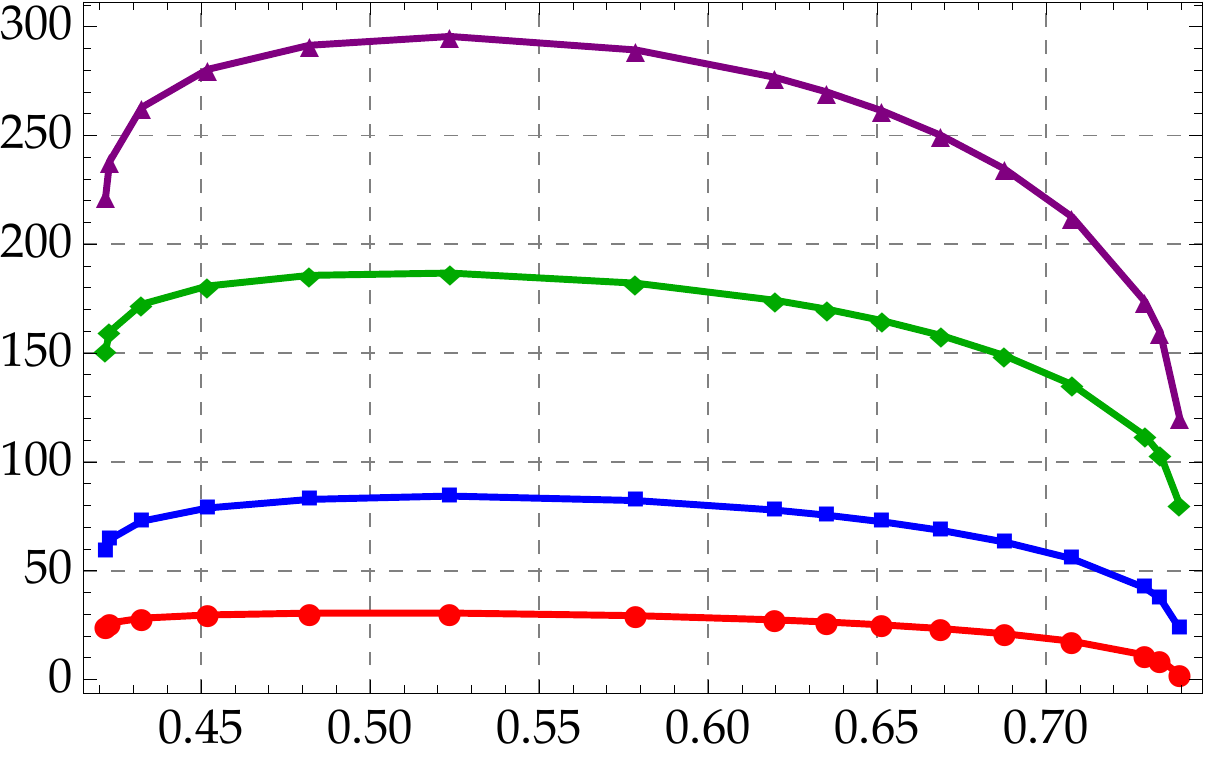} 
\qquad\qquad & 
\includegraphics[width=0.45\textwidth]{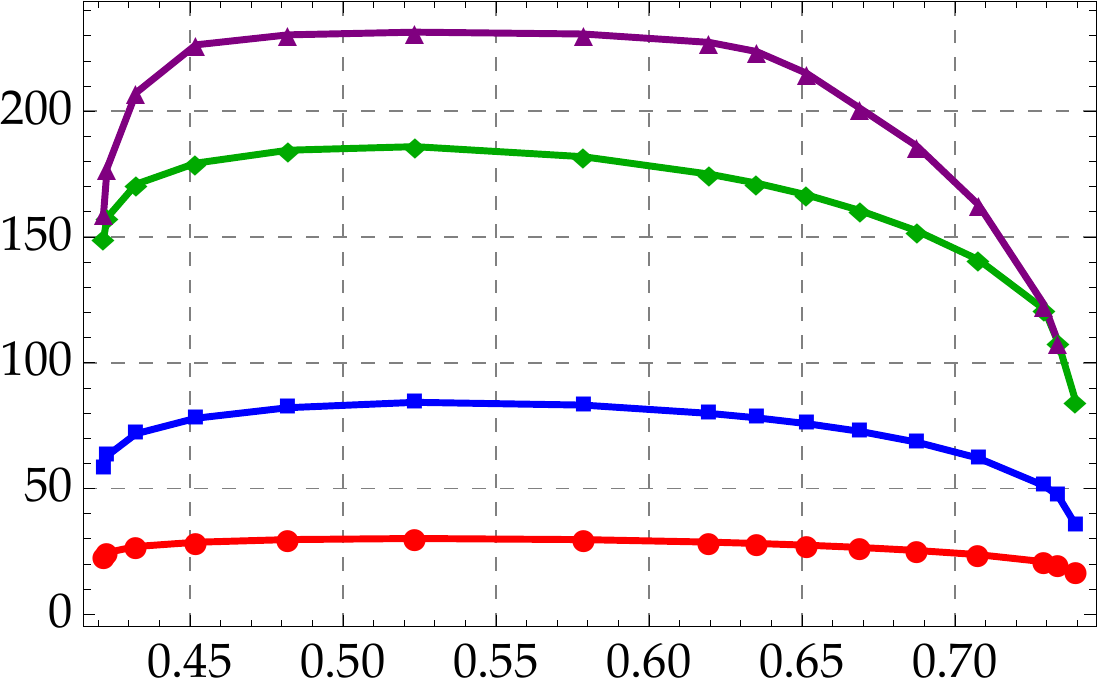}
\qquad
 \put(-450,70){$\frac{\omega^{2}}{\bar{M}^{2}}$}
   \put(-250,-10){$\frac{T}{\bar{M}}$}
    \put(-220,70){$\frac{\omega^{2}}{\bar{M}^{2}}$}
   \put(-18,-10){$\frac{T}{\bar{M}}$}
         \\
(c)& (d) \\
& \\
\includegraphics[width=0.45\textwidth]{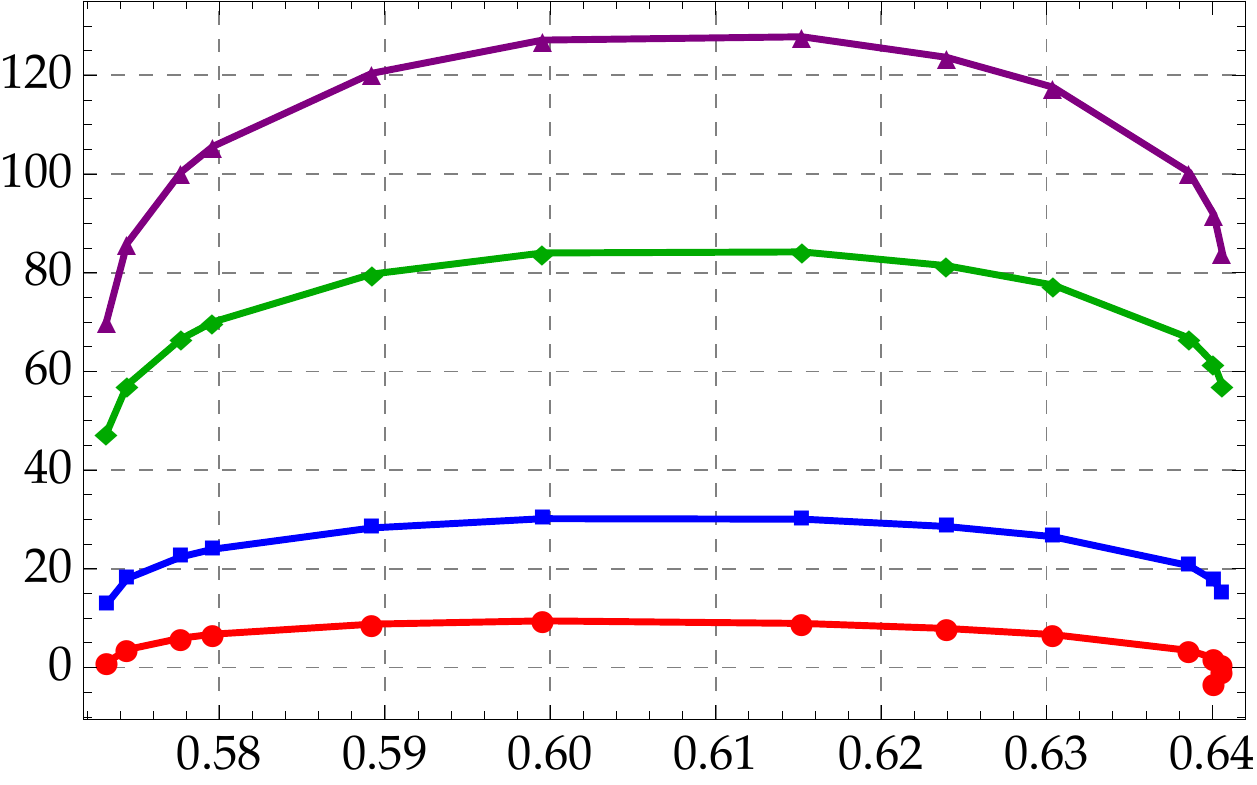} 
\qquad\qquad & 
\includegraphics[width=0.45\textwidth]{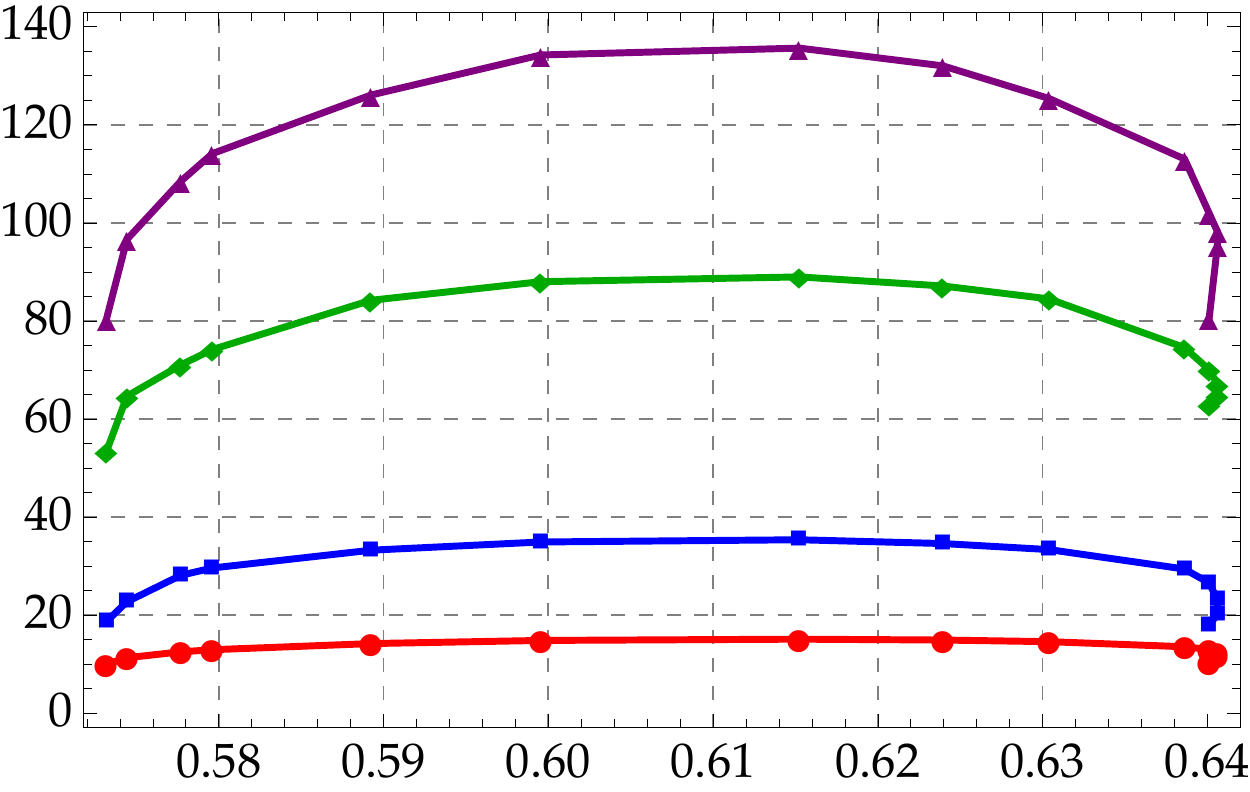}
\qquad
 \put(-450,70){$\frac{\omega^{2}}{\bar{M}^{2}}$}
   \put(-250,-10){$\frac{T}{\bar{M}}$}
    \put(-220,70){$\frac{\omega^{2}}{\bar{M}^{2}}$}
   \put(-18,-10){$\frac{T}{\bar{M}}$}
         \\
(e) & (f) 
\end{tabular}
\end{center}
\caption{\small Meson spectrum $\frac{\omega^{2}}{\bar{M}^{2}}$ for the $\chi$ (left column) and $\phi$ (right column) perturbations as a function of $T/\bar{M}$ for fixed magnetic field $b/\bar{M}^{2}=$ 0 (a) and (b), 2 (c) and (d), and 3.4 (e) and (f). The different colors correspond to different quantum numbers $(n,l)$ of the mesons, being from bottom to top red $(0,0)$, blue $(1,0)$, green $(0,2)$, and purple $(1,2)$.}
\label{Mesons_bM2}
\end{figure}

In Fig. (\ref{Mesons_bM2}) we show the meson spectrum $\omega^{2}$ normalized with respect to $\bar{M}^{2}$ for the $\chi$ (left column) and $\phi$ (right column) perturbations, as a function of $T/\bar{M}$ for fixed magnetic field and different values for the quantum numbers $n$ and $l$. Fig. (\ref{Mesons_bM2}) (a) and (b) display the case of vanishing magnetic field, which coincides with the values reported in \cite{Mateos:2007vn}. It can be seen how the mass gap goes to zero as we aproach the melting temperature, and also how the meson masses increase as the temperature is reduced, asymptoting a constant value as $T/\bar{M}\rightarrow 0$. Fig. (\ref{Mesons_bM2}) (c) and (d) show the results for $b/\bar{M}^{2}=2$. As we aproach the melting temperature $T_{hot}/\bar{M}$ the meson masses go to zero, just like for vanishing magnetic field. However, as we reduce the temperature and aproach its minimum value the meson masses start decreasing without asymptoting to a specific value. Fig. (\ref{Mesons_bM2}) (e) and (f) display the results for $b/\bar{M}^{2}=3.4$, for which there are two melting temperatures, $T_{hot}/\bar{M}$ and $T_{cold}/\bar{M}$. We can see that as we aproach both temperatures the meson masses go to zero. 

\begin{figure}
\begin{center}
\begin{tabular}{cc}
\includegraphics[width=0.45\textwidth]{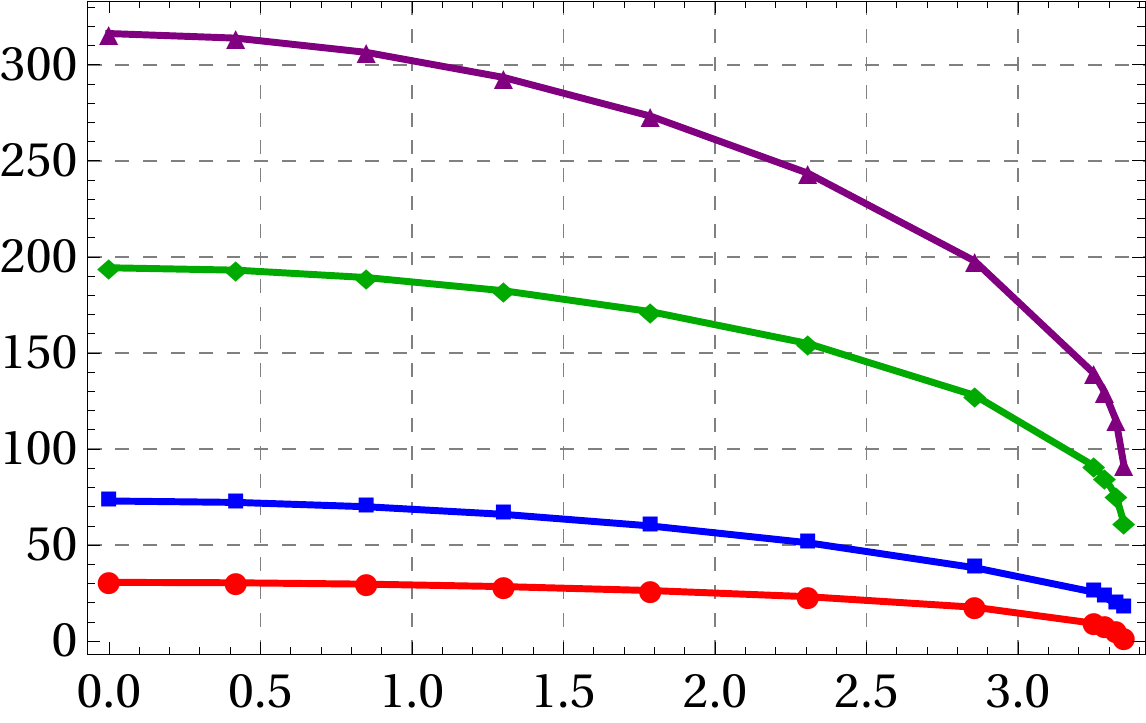} 
\qquad\qquad & 
\includegraphics[width=0.45\textwidth]{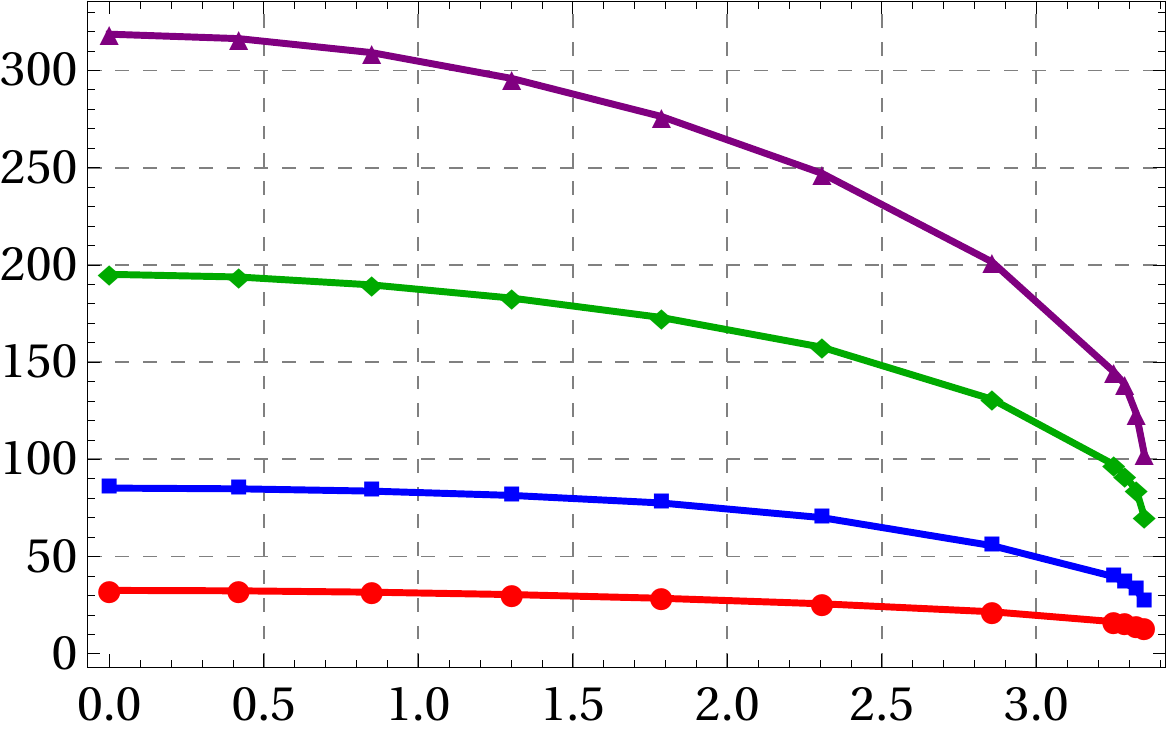}
\qquad
 \put(-450,70){$\frac{\omega^{2}}{\bar{M}^{2}}$}
   \put(-250,-10){$\frac{b}{\bar{M}^{2}}$}
    \put(-218,70){$\frac{\omega^{2}}{\bar{M}^{2}}$}
   \put(-18,-10){$\frac{b}{\bar{M}^{2}}$}
 \\
(a) & (b)\\
\end{tabular}
\end{center}
\caption{\small Meson spectrum $\frac{\omega^{2}}{\bar{M}^{2}}$ for the $\chi$ (a) and $\phi$ (b) perturbations as a function of $b/\bar{M}^{2}$. In both cases the temperature is fixed at $T/\bar{M}=0.65$. The different colors correspond to different quantum numbers $(n,l)$ of the mesons, being from bottom to top red $(0,0)$, blue $(1,0)$, green $(0,2)$, and purple $(1,2)$.}
\label{Mesons_TM}
\end{figure}

We close this section by showing the meson spectrum as a function of $b/\bar{M}^{2}$ at fixed temperature in Fig. (\ref{Mesons_TM}). While we specifically present the results for $T/\bar{M}=0.65$, we checked that the qualitative behavior is the same for any temperature in the range in which the phase transition can take place. We can see how the mass gap aproaches zero as the magnetic field goes to its critical value, confirming that the mesons can be melted by means of the magnetic field alone. Additionally, we recover the results from \cite{Mateos:2007vn} for vanishing magnetic field.

\section{Discussion}
\label{discussion}

In this paper we employed holographic methods to study the effects of an external magnetic field over flavor degrees of freedom. While previously this magnetic field had been introduced as an excitation over the world volume of the probe D7-branes \cite{Filev:2007gb,Albash:2007bk,Erdmenger:2007bn}, here we followed a different aproach and considered a family of solutions in which the magnetic field backreacts in the geometry itself. Notably, the results obtained with our setup are very different, and even completely opposite when comparable, to the ones obtained in \cite{Filev:2007gb,Albash:2007bk,Erdmenger:2007bn}.

Some of our main results are conveniently summarized in the phase diagram in Fig.(\ref{Phase_Diagram}), where we can see that the presence of the magnetic field induces a very rich thermodynamic behavior. While previously \cite{Avila:2019pua} we had only observed IMC for meson dissociation, we see now that the physics of the system is much more interesting, as the dependence of the phase transition on the magnetic field is highly non-trivial. 

One of the main novelties is that, for some magnetic field intensities, mesons can be melted by decreasing the temperature of the plasma. This can be seen explicitly from our thermodynamic analysis, as there is now a first order phase transition at low temperatures in addition to the expected first order phase transition at high temperatures. While for the first hot temperature the effect of the magnetic field is the one of IMC, we observe MC for the cold critical temperature. This is reminiscent of the effect of the magnetic field on the quark condensate for light quarks \cite{Bali:2011qj,Bali:2012zg,Endrodi:2019zrl}, as there are some temperatures for which both behaviors can be observed. However, it is important to remark that this behavior is not only a consequence of the magnetic field, but of its interplay with a scalar field that saturates the BF bound. It is possible that for low temperatures the energy of the whole system, consisting of the scalar field and the quarks, is such that it is unfavorable to have quark-antiquark bound states. In order to properly check that this is the case one would need to isolate the energy associated with the scalar field from the stress-energy tensor of the background solution. Nonetheless, this is not a trivial matter given the intricate interplay between the metric, scalar and magnetic field.

We would also like to remark that what we study here is neither a confinement/deconfinement transition nor a chiral symmetry breaking transition. This is because the existence of a horizon implies that the adjoint degrees of freedom are deconfined and chiral symmetry is broken as in both phases of our system the expectation value of the chiral condensate is different from zero for finite values of $b/\bar{M}^{2}$ and $T/\bar{M}$. In fact, lattice calculations \cite{Asakawa:2003re,Datta:2003ww,Hatsuda:2005nw} show that the temperature at which the mesons dissociate is higher than that at which deconfinement happens, thus it is expected that the transition we study does not coincide with any of the former two. 

Another important result is that, for a certain range of temperatures, the mesons can be melted by just changing the intensity of the magnetic field. We first observed this in \cite{Avila:2019pua}, where we named this phenomenom magnetic meson melting. For sufficiently high temperatures the mesons are melted regardless of the magnetic field. The opposite behavior is observed from the setup in \cite{Albash:2007bk,Erdmenger:2007bn}, where it was imposible to melt the mesons for magnetic field intensities above a certain critical value.

From our results for the quark condensate in Fig. (\ref{Condensate_TM}), we can see that the magnitude of the quark condensate increases with the magnetic field for the temperatures that we explored. Thus we conclude that we observe MC in regard to chiral symmetry. This seems to be consistent with lattice calculations \cite{Bali:2011qj,Bali:2012zg,Endrodi:2019zrl}, which predict that for sufficientely low temperatures and heavy quarks, MC is observed.

By performing a thermodynamic analysis we showed that the phase transition is of first order at both critical temperatures. We also found that for any non-vanishing magnetic field intensity there exists a temperature below which the specific heat of the flavor degrees of freedom becomes negative. However, the specific heat of the whole system, consisting of flavor and color degrees of freedom, remains positive for all the temperatures accesible with our setup as the limit $N_{f}\ll N_{c}$ is taken.

By studying perturbations of the D7-brane position we computed the spectrum of scalar mesons. We showed that in the case of the Minkowski embeddings the meson spectrum is discrete, as it was known for zero magnetic field \cite{Kruczenski:2003be,Mateos:2007vn,Hoyos:2006gb}. However, we found that a non-vanishing magnetic field has many important effects over the meson spectrum. The first one is that it reduces the number of possible excited states on the 3-sphere by imposing that the quantum number $m_{2}$ has to vanish for any $b\neq 0$. The second one is that the magnetic field reduces the meson masses for all $T/\bar{M}$ and any quantum number, as shown in Fig. (\ref{Mesons_TM}). For the values of $T/\bar{M}$ for which the transition can be triggered using the magnetic field we showed that the meson masses, including its mass gap, tend to zero as the magnetic field reaches its melting value.

Our results for the meson spectrum are in agreement with those found in \cite{Ayala:2018zat}, where a one-loop calculation is done in a linear sigma model to find that the mass of the neutral pion is reduced by the application of a magnetic field. The same behavior was observed using lattice calculations in \cite{Bali:2017ian} for the same pion. The spectrum of the $\eta'$ meson was already computed in the holographic context \cite{Zayakin:2008cy}. However, their calculation indicates an increase in the mass with the magnetic field and the authors concluded that it should be corrected by taking backreaction effects into account. Similar corrections should be applied to \cite{Evans:2010xs}, where the melting transition is studied out of equilibrium. Given that the magnetic field backreacts on the geometry of our construction, we provide such correction, showing that the effect of the magnetic field is the inverse to the one reported in \cite{Zayakin:2008cy}.

It is important to remark that our gravitational setup does not allow us to explore all values of the temperature and the magnetic field, as shown in Fig. (\ref{Phase_Diagram}). However, our thermodynamic analysis shows that the quark condensate, free energy, entropy, and internal energy, diverge as we aproach the critical value of $b/T^{2}$, showing that an interesting thermodynamic behavior is expected. 
\section{Acknowledgments}

We acknowledge partial financial support from PAPIIT IN113618, UNAM. We thank Omar del Pilar and Aslan Garc\'ia for their help with the execution of numerical computations. We also thank David Mateos, Alejandro Ayala and Maria Elena Tejeda-Yeomans for useful discussion and comments.

\appendix

\section{General truncation ansatz}
\label{App10D}
The family of solutions to 10-dimensional SUGRA IIB we found is part of the general truncation ansatz presented in \cite{Cvetic:1999xp}. In this apprendix we will show how to obtain it and the family of solutions from \cite{DHoker:2009mmn} from this general setting. We will use the notation employed in \cite{Cvetic:1999xp}, which is sligthly different than the one used in the main text. We will remind the reader when necessary.

The five-dimensional family of solutions presented in \cite{Cvetic:1999xp} can be uplifted to a ten-dimensional family of solutions to the SUGRA IIB equations of motion by considering the metric and self-dual five-form as the only non-vanishing fields. The truncation ansatz for the ten-dimensional metric is given by
\begin{equation}
ds_{10}^{2}=\Delta^{\frac{1}{2}}ds_{5}^{2}+\frac{L^{2}}{\Delta^{\frac{1}{2}}}\sum_{i=1}^{3}X_{i}^{-1}\left(d\mu_{i}^{2}+\mu_{i}^{2}\left(d\phi_{i}+\frac{A^{i}}{L}\right)^{2}\right),
\label{met_10}
\end{equation}
where $L$ is the AdS radius, $ds_{5}^{2}$ denotes the line element of a 5-dimensional non-compact spacetime, $A^{i}$ are three independent Maxwell fields living in the non-compact part of the geometry, $\mu_{i}$ are parametrized by two angular coordinates $\{\theta,\psi\}$ as
\begin{equation}
\mu_{1}=\sin\theta, \qquad \mu_{2}=\cos\theta\sin\psi, \qquad \mu_{3}=\cos\theta\cos\psi,
\end{equation}
while $\Delta$ is given by
\begin{equation}
\Delta=\sum_{i=1}^{3}X_{i}\mu_{i}^{2},
\end{equation}
with
\begin{equation}
X_{i}=e^{-\frac{1}{2}\vec{a}_{i}\cdot\vec{\varphi}}, \qquad \vec{a}_{i}=(a_{i}^{(1)},a_{i}^{(2)}),\qquad \vec{a_{i}}\cdot\vec{a_{j}}=4\delta_{ij}-\frac{4}{3},\qquad \text{y} \qquad \vec{\varphi}=(\varphi_{1},\varphi_{2}),
\label{truncA}
\end{equation}
$\varphi_{1}$ and $\varphi_{2}$ are two independent scalar fields living in the non-compact part of the spacetime. Meanwhile the 5-form is given by
\begin{equation}
F_{5}=G_{5}+\star G_{5},
\label{5form}
\end{equation}
where $\star$ denotes the Hodge dual associated with \eqref{met_10} and
\begin{equation}
\begin{split}
G_{5}=&\frac{2}{L}\sum_{i=1}^{3}(X_{i}^{2}\mu_{i}^{2}-\Delta X_{i})\epsilon_{5}-\frac{L}{2}\sum_{i=1}^{3}X_{i}^{-1}\bar{\star}dX_{i}\wedge d(\mu_{i}^{2})\\&+\frac{L^{2}}{2}\sum_{i=1}^{3}X_{i}^{2}d(\mu_{i}^{2})\wedge\left(d\phi_{i}+\frac{A^{i}}{L}\right)\wedge\bar{\star}F^{i},
\end{split}
\end{equation}
where $\epsilon_{5}$ and $\bar{\star}$ are the volume form and the Hodge dual associated with the five-dimensional metric, and $F^{i}=dA^{i}$. Note that $G_{5}$ reduces to $\epsilon_{5}$ when the scalar and Maxwell fields are taken equal to zero.

The family of solution we found can be recovered from the general truncation ansatz by taking
\begin{equation}
\frac{2}{\sqrt{3}}\varphi_{2}=2\varphi_{1}=\varphi, \qquad A^{1}=0, \qquad A^{2}=A^{3}=\sqrt{2}A,
\label{particular1}
\end{equation}
and the vectors $\vec{a}_{i}$ as
\begin{equation}
\vec{a}_{1}=\left(\frac{2}{\sqrt{6}},\sqrt{2}\right), \qquad \vec{a}_{2}=\left(\frac{2}{\sqrt{6}},-\sqrt{2}\right), \qquad \vec{a}_{3}=\left(-\frac{4}{\sqrt{6}},0\right).
\label{particular2}
\end{equation}
This in turn implies that
\begin{equation}
X=X_{2}=X_{3}=e^{\frac{1}{\sqrt{6}}\varphi}, \qquad X_{1}=X^{-2}.
\end{equation} 

With this choice the line element \eqref{met_10} simplifies to
\begin{equation}
ds_{10}^{2}=\Delta^{\frac{1}{2}}ds_{5}^{2}+\frac{L^{2}}{\Delta^{\frac{1}{2}}}\left[X\Delta d\theta^{2}+X^{2}\sin^{2}\theta d\phi_{1}^{2}+X^{-1}\cos^{2}\theta d\Sigma_{3}^{2}(A)\right],
\end{equation}
where
\begin{equation}
d\Sigma_{3}^{2}(A)=d\psi^{2}+\sin^{2}\psi\left(d\phi_{2}+\sqrt{2}\frac{A}{L}\right)^{2} +\cos^{2}\psi\left(d\phi_{3}+\sqrt{2}\frac{A}{L}\right)^{2},
\end{equation}
which is precisely the line element \eqref{metric_10} from the main text after renaming the angular coordinates $\{\phi,\psi,\vartheta_{1},\vartheta_{2}\}\rightarrow\{\phi_{1},\psi,\phi_{2},\phi_{3}\}$. As it is explained in the main text, this metric permits an easy embedding of a D7-brane because it doesn't depend on $\phi_{1}$ and the direction that it represents remains orthogonal to the rest of the spacetime. This allows to put the D7-brane at a fixed position in $\phi_{1}$ while it extends along the AdS directions and wraps the 3-cycle, which gets maximal for $\theta=0$. The family of solutions presented in \cite{DHoker:2009mmn} doesn't possesses this characteristics.

The family of solutions from \cite{DHoker:2009mmn} is also part of the general truncation ansatz described above. It can be recovered by taking
\begin{equation}
\varphi_{2}=\varphi_{1}=0, \qquad A^{1}=A^{2}=A^{3}=2A/\sqrt{3},
\end{equation}
which implies that $X_{i}=1$ and $\Delta=1$ and thus the 10D metric is given by
\begin{equation}
ds_{10}^{2}=ds_{5}^{2}+L^{2}\left[d\theta^{2}+\sin^{2}\theta\left(d\phi_{1}+\frac{2}{L\sqrt{3}}A\right)^{2}+\cos^{2}\theta d\Sigma_{3}^{2}(A)\right],
\end{equation}
where
\begin{equation}
d\Sigma_{3}^{2}(A)=d\psi^{2}+\sin^{2}\psi\left(d\phi_{2}+\frac{2}{L\sqrt{3}}A\right)^{2} +\cos^{2}\psi\left(d\phi_{3}+\frac{2}{L\sqrt{3}}A\right)^{2}.
\end{equation}
Note that, crucially, the $\phi_{1}$ coordinate is no longer orthogonal to the rest of the spacetime and thus the D7-brane cannot lie at constant $\phi_{1}$ unless the embedding profile depends on the gauge theory coordinates.

\section{Boundary expansions of the background fields}
\label{AppB}

Given that the background solution is naturally computed numerically using the $r$ coordinate defined in \eqref{metric}, for many purposes it is convenient to know how a generic solution behaves at the boundary at $r\rightarrow\infty$. This can be obtained by solving the equations of motion \eqref{EOM_fondo} by a power series method around the boundary. The only restrictions we impose is that the metric asymptotes exactly the metric of $AdS_{5}$ and the non-normalizable mode of the scalar field is turned off. The result reads
\begin{eqnarray}
&& U(r)=r^{2}+U_{1}r+\frac{U_{1}^{2}}{4}+\frac{1}{r^{2}}\left(U_{4}-\frac{2}{3}b^{2}\log{r}\right)+\mathcal{O}\left(\frac{1}{r^{4}}\right),
\cr
&& V(r)=r^{2}+U_{1}r+\frac{U_{1}^{2}}{4}+\frac{1}{r^{2}}\left(-\frac{1}{2}W_{4}-\frac{1}{6}\varphi_{0}^{2}+\frac{1}{3}b^{2}\log{r}\right)+\mathcal{O}\left(\frac{1}{r^{4}}\right),
\cr
&& W(r)=r^{2}+U_{1}r+\frac{U_{1}^{2}}{4}+\frac{1}{r^{2}}\left(W_{4}-\frac{2}{3}b^{2}\log{r}\right)+\mathcal{O}\left(\frac{1}{r^{4}}\right),
\cr
&& \varphi(r)=\frac{\varphi_{0}}{r^{2}}-\frac{U_{1}\varphi_{0}}{r^{3}}+\frac{1}{12r^{4}}\left(-2\sqrt{6}b^{2}+\varphi_{0}(9U_{1}^{2}-\sqrt{6}\varphi_{0})\right)+\mathcal{O}\left(\frac{1}{r^{5}}\right),
\label{r_expansions}
\end{eqnarray}
where $U_{1}$, $U_{4}$, $W_{4}$ and $\varphi_{0}$ are coefficients that are not fixed by the equations of motion, but can be read as functions of the magnetic field intensity $b$ and the temperature $T$ once a particular numerical solution is known. As explained in detail in \cite{Avila:2018hsi}, $\varphi_{0}$ is dual to the VEV of the scalar operator $\langle\mathcal{O}_{\varphi}\rangle$ while $U_{4}$ and $W_{4}$ are both related to the stress-energy tensor. 
\section{Holographic renormalization}
\label{AppA}

As it is commonly the case in the gauge/gravity correspondence, the Euclidean DBI action \eqref{Euclidean_DBI} suffers from long-distance divergences when evaluated on-shell, making the variational principle ill defined. There exists a systematic method to deal with this issue, known as holographic renormalization, that has been studied extensively \cite{Skenderis:2002wp,Bianchi:2001kw}. The first step is to write the near-boundary behavior of all the fields that enter in the Euclidean action, which is more conveniently done in the Fefferman-Graham coordinate where the five-dimensional metric \eqref{metric} takes the form
\begin{equation}
ds_{5}^{2}=\frac{du^{2}}{u^{2}}+\gamma_{ij}(u)dx^{i}dx^{j}=\frac{1}{u^{2}}\left(du^{2}+g_{ij}(u)dx^{i}dx^{j}\right),
\label{metric_5_FG}
\end{equation}

Next we solve the equations of motion \eqref{EOM_fondo} by a power series method around the boundary, located in this particular coordinate at $u=0$. The resulting expansions for $g, \varphi,$ and $F$ are given by
\begin{eqnarray}
&& g_{ij}(u)={g_{ij}}_{(0)}+({g_{ij}}_{(4)}+h_{ij}\log{u}+H_{ij}\log^{2}{u})u^{4}+\mathcal{O}(u^{6}),
\cr
&& \varphi(u)=u^{2}(\varphi_{(0)}+\psi_{(0)}\log{u}+(\varphi_{(2)}+\psi_{(2)}\log{u}+ \Psi_{(2)}\log^{2}{u})u^{2})+\mathcal{O}(u^{6})
\cr
&& F_{u\nu}=0, \qquad F_{ij}=F_{ij}(t,x,y,z),
\label{Fondo_FG}
\end{eqnarray}
with
\begin{eqnarray}
&& \Psi_{(2)}=-\frac{\psi_{(0)}^{2}}{2\sqrt{6}},
\cr
&& \psi_{(2)}=\frac{\psi_{(0)}}{\sqrt{6}}(\psi_{(0)}-\varphi_{(0)}),
\cr
&& \varphi_{(2)}=-\frac{1}{\sqrt{6}}\left(\frac{1}{2}F_{ik}F_{jl}{g^{ij}}_{(0)}{g^{kl}}_{(0)}+\frac{1}{2}\varphi_{(0)}^{2}+\frac{3}{4}\psi_{(0)}^{2}-\varphi_{(0)}\psi_{(0)}\right),
\cr
&& {g^{ij}}_{(0)}{g_{ij}}_{(4)}=\frac{1}{12}F_{ik}F_{jl}{g^{ij}}_{(0)}{g^{kl}}_{(0)}-\frac{1}{3}\varphi_{(0)}^{2}-\frac{1}{24}\psi_{(0)}^{2},
\cr
&& h_{ij}=\frac{1}{4}{g_{ij}}_{(0)}F_{nk}F_{ml}{g^{nm}}_{(0)}{g^{kl}}_{(0)}-F_{ik}F_{jl}{g^{kl}}_{(0)}-\frac{1}{6}{g_{ij}}_{(0)}\varphi_{(0)}\psi_{(0)},
\cr
&& H_{ij}=-\frac{1}{12}{g_{ij}}_{(0)}\psi_{(0)}^{2},
\label{solucion_FG}
\end{eqnarray}
and any not listed coefficient up to the specified order is equal to zero. Doing the same for the D7-brane profile $\chi(u)$ we obtain
\begin{equation}
\chi(u)=u\chi_{(0)}+u^{3}\chi_{(2)}+\mathcal{O}(u^{5}),
\label{chi_FG}
\end{equation}
where $\chi_{(0)}$ and $\chi_{(2)}$ are both free coefficients not fixed by the equations of motion.

The next step is to evaluate the Euclidean DBI action \eqref{Euclidean_DBI} on the solutions \eqref{Fondo_FG} and \eqref{chi_FG}. After that we integrate from a radial cut-off $\epsilon$ to an arbitrary $u_{\text{max}}$, which even if bigger than $\epsilon$, is still close to the boundary and remains fixed. This allows us to collect all the divergent terms in the limit $\epsilon\rightarrow 0$, which can be organized as follows:
\begin{equation}
\frac{S_{E}}{2\pi^{2}T_{D7}N_{f}}=\int d^{4}x \sqrt{g_{(0)}}\left(a_{(1)}+a_{(2)}+a_{(3)}+b_{(1)}+b_{(2)}+ c_{(1)}+c_{(2)}\right),
\label{div_action}
\end{equation}
where
\begin{equation}
a_{(1)}=\frac{1}{4\epsilon^{4}}, \qquad a_{(2)}=-\frac{1}{2}\frac{\chi_{(0)}^{2}}{\epsilon^{2}}, \qquad a_{(3)}=-\frac{1}{8}F_{ik}F_{jl}{g^{ij}}_{(0)}{g^{kl}}_{(0)}\log{\epsilon},
\end{equation}
are independent of the scalar field, while
\begin{equation}
b_{(1)}=\frac{1}{12}(\psi_{(0)}^{2}\log^{2}{\epsilon}+2\varphi_{(0)}\psi_{(0)}\log{\epsilon}),
\qquad
b_{(2)}=-\frac{1}{2\sqrt{6}}(\varphi_{(0)}+\psi_{(0)}\log{\epsilon})\frac{1}{\epsilon^{2}}, 
\end{equation}
are divergent terms that only depend on $\varphi$, and
\begin{equation}
c_{(1)}=-\frac{5}{48}\psi_{(0)}^{2}\log{\epsilon},\qquad c_{(2)}=-\frac{1}{4\sqrt{6}}\frac{\psi_{(0)}}{\epsilon^{2}},
\end{equation}
are depend only on the source $\psi_{(0)}$. In order to properly substract this divergent terms from the action, it is necessary to invert the series for all the fields, and express the $a_{(i)}$, $b_{(i)}$, and $c_{(i)}$ terms in a covariant manner. To this end we first note that up to leading order in $\epsilon$
\begin{equation}
\epsilon^{2}\gamma_{ij}={g_{ij}}_{(0)}+\ldots, \qquad \frac{\gamma^{ij}}{\epsilon^{2}}={g^{ij}}_{(0)}+\ldots, \qquad
\frac{\chi}{\epsilon}=\chi_{(0)}+\ldots,
\end{equation}
where the dots denote terms that vanishes in the limit $\epsilon\rightarrow 0$. This allows us to rewrite the $a_{(i)}$ terms in a covariant manner as
\begin{equation}
a_{(1)}=\frac{1}{4\epsilon^{4}},\qquad a_{(2)}=-\frac{\chi^{2}}{2\epsilon^{4}}, \qquad a_{(3)}=-F_{ij}F^{ij}\frac{\log{\epsilon}}{8\epsilon^{4}},
\end{equation}
where the indexes are raised and lowered using the boundary metric $\gamma_{ij}$. 

Let us turn now to the $b_{(i)}$ terms. Given that both $\varphi_{(0)}$ and $\psi_{(0)}$ appear at the same order in the $\epsilon$ expansion of $\varphi$, the inversions need to be performed carefully. To leading order we have
\begin{equation}
\frac{\varphi}{\epsilon^{2}}=\varphi_{(0)}+\psi_{(0)}\log{\epsilon}+\ldots,
\end{equation}
from where
\begin{equation}
\frac{\varphi^{2}}{\epsilon^{4}}=\varphi_{(0)}^{2}+2\varphi_{(0)}\psi_{(0)}\log{\epsilon}+ \psi_{(0)}^{2}\log^{2}{\epsilon}+\ldots,
\label{varphi1}
\end{equation}
and discarding the finite term in the limit $\epsilon\rightarrow 0$ we conclude that
\begin{equation}
b_{(1)}=\frac{\varphi^{2}}{12\epsilon^{4}}.
\end{equation}
Note however that if the source term $\psi_{(0)}$ vanishes, the finite term we just discarded is the only contribution. Thus, for vanishing source, $b_{(1)}$ gives only a finite term which is fixed for $\varphi_{(0)}\neq 0$. For $b_{(2)}$, from \eqref{solucion_FG} in \eqref{Fondo_FG}
\begin{equation}
\frac{1}{\epsilon^{2}}(\varphi_{(0)}+\psi_{(0)}\log{\epsilon})=\frac{\varphi}{\epsilon^{4}}-\frac{1}{\sqrt{6}}\psi_{(0)}^{2}\log{\epsilon}+\frac{1}{2\sqrt{6}}\log{\epsilon}(2\varphi_{(0)}\psi_{(0)}+ \psi_{(0)}^{2}\log{\epsilon})-\varphi_{(2)}+\ldots,
\end{equation}
and then using \eqref{varphi1}
\begin{equation}
\frac{1}{\epsilon^{4}}\left(\varphi-\frac{\varphi^{2}}{\sqrt{6}\log{\epsilon}}+\frac{\varphi^{2}}{2\sqrt{6}}\right)=\frac{1}{\epsilon^{2}}(\varphi_{(0)}+\psi_{(0)}\log{\epsilon})+\varphi_{(2)}+\ldots.
\end{equation}
Discarding the finite term $\varphi_{(2)}$ we conclude that
\begin{equation}
b_{(2)}=\frac{1}{\epsilon^{4}}\left(-\frac{\varphi}{2\sqrt{6}}+\frac{\varphi^{2}}{12\log{\epsilon}}-\frac{\varphi^{2}}{24}\right).
\end{equation}
It turns out that for $\psi_{(0)}=0$ the term that goes like $1/\log{\epsilon}$ vanishes and once again the $\varphi^{2}$ term gives only a finite contribution. Even if our case of interest is the one with no source for the scalar field, we conducted the analysis up to this point not setting $\psi_{(0)}$ to zero because it induces a fixed a finite counterterm that should be taken into account also in the sourceless case. Nothing similar will happen for the remaining $c_{(i)}$ terms, which vanish identically for $\psi_{(0)}=0$, so from here onwards, as in the main text, we will limit ourselves to that case.

Finally, by using \eqref{Fondo_FG} and \eqref{solucion_FG} the expansion of the determinant of the boundary metric is found to be
\begin{equation}
\sqrt{g_{(0)}}=\epsilon^{4}\sqrt{\gamma}\left(1+\frac{1}{6}\varphi^{2}\right).
\end{equation}
With this ingredients we can write \eqref{div_action} in a covariant manner, and thus the counterterms are given by
\begin{equation}
\frac{S_{ct}}{2\pi^{2}T_{D7}N_{f}}=-\int d^{4}x\sqrt{\gamma}\left(\frac{1}{4}-\frac{1}{2}\chi^{2}-\frac{1}{8}F_{ij}F^{ij}\log{\epsilon}-\frac{1}{2\sqrt{6}}\varphi+\frac{1}{12}\varphi^{2}\right),
\end{equation}
where any term that goes to zero in the limit $\epsilon\rightarrow 0$ was discarded.
\section{Computation of the quark condensate}
\label{AppC}
The quark condensate is given by the variation of the action with respect to the quark mass \cite{Kruczenski:2003uq,Skenderis:2002wp}
\begin{equation}
\langle\bar{q}q\rangle=\frac{1}{\sqrt{g_{(0)}}}\left(\frac{\delta S_{D7}}{\delta M_{q}}\right)_{M_{q}=0}=\pi\sqrt{\frac{2}{\lambda}}\frac{1}{\sqrt{g_{(0)}}}\left(\frac{\delta S_{D7}}{\delta m}\right)_{m=0},
\end{equation}
where we have used \eqref{quark_mass} to express everything in terms of the parameter $m$. This can be rewritten as a variation with respect to the embedding profile as
\begin{equation}
\langle\bar{q}q\rangle=\pi\sqrt{\frac{2}{\lambda}}\lim_{\epsilon\rightarrow 0}\left(\frac{1}{\epsilon^{3}}\frac{1}{\sqrt{\gamma}}\frac{\delta S_{D7}}{\delta\chi}\right).
\end{equation}
Using \eqref{renormalized_action} and the equations of motion \eqref{EOM_chi} we can compute this variation explicitly in terms of the boundary values of the fields as
\begin{equation}
\frac{1}{\sqrt{\gamma}}\frac{\delta S_{D7}}{\delta\chi}=-2\pi^{2}T_{D7}N_{f}\left(\frac{1}{\sqrt{\gamma}}\frac{\partial\mathcal{L}}{\partial\chi'}-\chi+\chi^{3})\right)_{u=\epsilon},
\end{equation}
and thus after inserting the FG expansions of all the background fields \eqref{Fondo_FG} and the embedding profile \eqref{chi_FG} we conclude that the quark condensate is given by
\begin{equation}
\langle\bar{q}q\rangle=-4\pi^{3}T_{D7}N_{f}\sqrt{\frac{2}{\lambda}}\chi_{(2)}=-\frac{1}{2^{3/2}\pi^{3}}\sqrt{\lambda}N_{c}N_{f}\chi_{(2)}.
\label{condensate_FG}
\end{equation}
Finally, in order to evaluate this expression numerically more easily it is convenient to express the FG coefficient $\chi_{(2)}$ in terms of $m$ and $c$ appearing in the $r$-coordinate expansion of $\chi$ \eqref{chi_boundary}. To that end we use the fact that both radial coordinates are given in terms of each other near the boundary as \cite{Avila:2018hsi}
\begin{eqnarray}
&& u(r)=\frac{1}{r}-\frac{U_{1}}{2r^{2}}+\frac{U_{1}^{2}}{4r^{3}}-\frac{U_{1}^{3}}{8r^{4}}+\frac{1}{r^{5}}\left(\frac{1}{48}(b^{2}+3U_{1}^{4}-6U_{4})+\frac{1}{12}b^{2}\log{r}\right)+\mathcal{O}\left(\frac{1}{r^{6}}\right),
\cr
&& r(u)=\frac{1}{u}-\frac{U_{1}}{2}+u^{3}\left(\frac{1}{48}(b^{2}-6U_{4})-\frac{1}{12}b^{2}\log{u}\right)+\mathcal{O}(u^{5}),\label{u2r}
\end{eqnarray}
which permits to write $\chi$ in terms of the $u$-coordinate
\begin{equation}
\chi=m u+\left(c-\frac{1}{4}mU_{1}^{2}\right)u^{3}.
\label{chi_u}
\end{equation}
By comparing \eqref{chi_u} to \eqref{chi_FG} we can read that the relation between coefficients is
\begin{equation}
\chi_{(0)}=m, \qquad \chi_{(2)}=c-\frac{1}{4}mU_{1}^{2}.
\label{rel_coeff}
\end{equation}
After substitution of \eqref{rel_coeff} in \eqref{condensate_FG}, \eqref{condensate} is readily derived.

\end{document}